\documentclass[12pt]{article}

\usepackage{arxiv}
\usepackage[utf8]{inputenc} 
\usepackage[T1]{fontenc}    
\usepackage[tbtags]{amsmath}
\usepackage{bm}
\usepackage{graphicx}
\graphicspath{ {./Figures/} }
\usepackage{caption}
\usepackage{subcaption}
\usepackage{cool}
\usepackage[bottom]{footmisc}
\usepackage{titlesec}
\usepackage{enumerate}
\usepackage[round]{natbib}
\usepackage{hyperref}
\usepackage[capitalise]{cleveref}
\usepackage{tikz}
\usepackage{pgfplots}
\pgfplotsset{compat=1.3}
\titleformat*{\subsubsection}{\normalsize\itshape}
\usepackage{rotating}
\usepackage{array}
\usepackage[font=small]{caption}
\usepackage{chngpage}
\usepackage[misc]{ifsym}

\title{Foundations and their practical implications for the constitutive coefficients of poromechanical dual-continuum models}

\author{
  Mark Ashworth\\
  Institute of GeoEnergy Engineering\\
  Heriot-Watt Unversity, Edinburgh\\
  \texttt{ma174@hw.ac.uk} \\
   \And
  Florian Doster\\
  Institute of GeoEnergy Engineering\\
  Heriot-Watt Unversity, Edinburgh\\
  \texttt{f.doster@hw.ac.uk} \\
}

\date{}

\newlength\figureheight
\newlength\figurewidth

\begin{document}
\maketitle
\captionsetup[figure]{labelfont={bf},labelformat={default},labelsep=period,name={Figure}}
\setlength{\abovedisplayskip}{-5pt}

\begin{abstract}
A dual-continuum model can offer a practical approach to understanding first order behaviours of poromechanically coupled multiscale systems. To close the governing equations, constitutive equations with models to calculate effective constitutive coefficients are required. Several coefficient models have been proposed within the literature. However, a holistic overview of the different modelling concepts is still missing. To address this we first compare and contrast the dominant models existing within the literature. In terms of the constitutive relations themselves, early relations were indirectly postulated that implicitly neglected the effect of the mechanical interaction arising between continuum pressures. Further, recent users of complete constitutive systems that include inter-continuum pressure coupling have explicitly neglected these couplings as a means of providing direct relations between composite and constituent properties, and to simplify coefficient models. Within the framework of micromechanics we show heuristically that these explicit decouplings are in fact coincident with bounds on the effective parameters themselves. Depending on the formulation, these bounds correspond to end-member states of isostress or isostrain. We show the impacts of using constitutive coefficient models, decoupling assumptions and parameter bounds on poromehcanical behaviours using analytical solutions for a 2D model problem. Based on the findings herein we offer recommendations for how and when to use different coefficient modelling concepts.\\
\end{abstract}

\section{Introduction}
Many natural and manufactured geomaterials exhibit strong heterogeneities in their material properties owing to the existence of porous constituents at various length scales. Examples of multiscale systems that are commonly encountered in subsurface operations include fissured or fractured rock and soil aggregates (\citealt{Warren1963}; \citealt{Kazemi1976}; \citealt{Nelson2001}; \citealt{Gerke2006}; \citealt{Koliji2008}; \citealt{Romero2011}). Modelling of such materials is invaluable in understanding how these systems behave in response to extraneous activities. In general, modelling can be done using either explicit (e.g. discrete fracture matrix models) or implicit methods (e.g. continuum approaches) (\citealt{Berre2018}).  
   
With respect to fractured systems, using explicit methods can be computationally prohibitive at large scales (\citealt{Karimi-Fard2006}; \citealt{Gong2007}; \citealt{Garipov2016}). Additionally, explicit methods may require data (e.g. spatial data) that is not obtainable without direct access (\citealt{Berkowitz2002}; \citealt{Blessent2014}). In cases where field-scale modelling of multiscale systems is required, implicit representations are then often preferred. The most common type of implicit model is the dual-continuum (or double-porosity) model, originally attributed to \cite{Barenblatt1960}. In this, the dual-material is considered as the superposition of two overlapping continua, which communicate through a mass transfer term. Continua are defined on the basis of their material properties. For example, fractures (or inter-aggregate pores) generally have high permeabilities and poor storage capabilities, vice-versa the matrix. Although less detailed than their explicit counterparts, dual-continuum models can provide practical and valuable insight into the first order behaviours of multiscale systems. Further, use of these models is desirable due to the low number of fitting parameters that allow for efficient calibration to historical data.

Multiscale systems can also exhibit strong coupling between deformation and fluid flow, and vice-versa. This phenomenon is known as poromechanical coupling (\citealt{Rutqvist2003}), and is described by the well established poromechanical theory (see for example \citealt{Biot1941,Biot1977}; \citealt{Detournay1995}; \citealt{Coussy1995,Coussy2004}; \citealt{Wang2000}; \citealt{DeBoer2012}; \citealt{Cheng2016}).  

\cite{Aifantis1977,Aifantis1979} and \cite{Wilson1982} were the first to introduce the generalised notion of deformation within the dual-continuum setting. Further offerings then came from \cite{Elsworth1992}, \cite{Lewis1997} and \cite{Bai1999}. However, all of these models implicitly neglected the effects of coupling between pressures of different pore domains due to their postulation of the form of the constitutive equations. The absence of these pressure couplings was shown to give unphysical responses by \cite{Khalili2003}. Specifically, the author's results showed discontinuous pressure jumps in the matrix and fracture continua that were incompatible with the prescribed boundary conditions. The cause of the observations made by \cite{Khalili2003} still remains an open question. 

Additional models in which the constitutive equations included inter-continuum pressure coupling were introduced by \cite{Berryman1995}, \cite{Tuncay1996a}, \cite{Loret1999}, \cite{Berryman2002}, \cite{Berryman2002a},  \cite{Khalili1996} and \cite{Khalili2008}. The difference between these presentations comes in the way that the authors choose to calculate the constitutive coefficients that govern a dual-continuum's poromechanical behaviour. For example, some authors implicitly assume the high permeability continuum to be all void space (e.g. \citealt{Khalili1996}), whilst others allow for an intrinsic phase stiffness (e.g. \citealt{Berryman2002a}). Models (referred to as coefficient models herein) used to calculate the constitutive coefficients are required due to the potential difficulty in measuring these properties experimentally. Whilst various coefficient models exist within literature there is still no general guideline for how and when to use them. 

More recent users of these later constitutive/coefficient models have explicitly decoupled pore domain pressures when expressing the constitutive relations in terms of stress and continuum pressures (pure stiffness setting) (see for example \citealt{Nguyen2010}; \citealt{Kim2012}; \citealt{Mehrabian2014}; \citealt{Mehrabian2018}). This has been done as a form of non-algebraic closure and to provide explicit relations between composite and constituent properties resulting in simplified coefficient models. However, such decoupling assumptions have been made without discussing the origin and sensitivities that may arise as a result. 

The aim of this paper is to formulate a set of recommendations for how and when to use different constitutive modelling concepts. In doing we show the impacts of making implicit and explicit decoupling assumptions. In the case of the latter we use heuristic arguments from micromechanics to show that these assumptions are coincident with bounds on the effective parameters themselves.  

\begin{figure}[h]
\centering
\includegraphics[scale = 0.7]{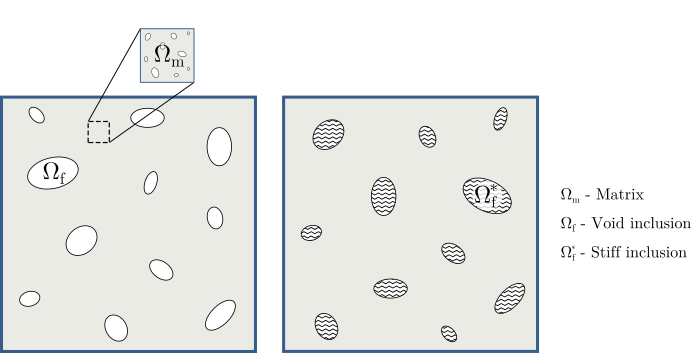}
\caption{Examples of matrix-void inclusion (left) and matrix-stiff inclusion (right) composites. Fractures can be considered as either depending on modelling assumptions.} \label{fig:1}
\end{figure}

We structure the paper as follows. In Section 2 we introduce the governing and constitutive equations pertinent to double-porosity materials. For the latter set of equations we support their form using arguments from the energy approach to poromechanics (\citealt{Coussy2004}). Section 3 presents the most prevalent modelling approaches for calculating the effective poromechanical coefficients. Section 4 details the origins of explicit assumptions made on constitutive/coefficient models within the framework of micromechanics. From here we offer upscaling recommendations for constituent moduli when composite moduli may not be available. In Section 5 we use analytical solutions to the double-porosity Mandel problem to explore the physical implications, and relevance, of different coefficient models and decoupling assumptions. We conclude by way of offering recommendations for how and when to use coefficient models in light of (a) intrinsic fracture stiffness effects and (b) pressure decoupling assumptions between pore domains. Throughout this paper our reference multiscale material is that of a naturally fractured system. Such systems can be considered as void space inclusion composites or stiff inclusion composites (\cref{fig:1}).     

We note that work has been done on the determination of effective properties of multiple-porosity materials via homogenisation methods (e.g. \citealt{Berryman2006} and \citealt{Levin2012}). However, equivalent continuum models can fail to provide insight into processes occurring at the different porosity scales due to use of an averaged flow field (\citealt{Berre2018}). In contrast, this work is concerned with double-porosity materials for which two distinct flow fields exist. Upscaling of such flow fields for inelastic materials has been addressed by periodic homogenisation (\citealt{Arbogast1990}), but such a treatment for deformable materials is, to the best of the authors knowledge, still missing. Given this context, the introduction of the phenomological approaches described herein for the determination of constitutive coefficients is desirable due to their ease of use, and resulting explicit relations to underling properties.    

\section{Double-porosity mathematical model}
We present the balance equations and constitutive laws for the dual-continuum system within the macroscopic framework of \cite{Coussy1995,Coussy2004}. The dual system is considered as the superposition of two overlapping poroelastic continua. Elastic deformation of each continuum is thus implied. Quantities denoted by $m$ and $f$ refer to matrix and fracture continua respectively. It is assumed that the poroelastic double-porosity material is isotropic, and is saturated by a slightly compressible fluid which can undergo isothermal flow. Under the assumptions of quasi-static deformations and infinitesimal transformations the momentum balance for the dual medium recovered as

\begin{equation}
\nabla\cdot\bm{\sigma} + \rho\bm{g} = \overline{\gamma}, \label{eqn:1}
\end{equation}

\noindent
where $\bm{\sigma}$ is the Cauchy stress tensor, $\bm{g}$ is the gravity vector, $\rho=\rho_s(1-\phi)+\rho_l\phi$ is the density of the bulk medium, $\rho_s$ is the intrinsic density of the solid matrix, $\rho_l$ is the intrinsic density of the fluid, and $\phi$ is the Lagrangian porosity. This property is defined as the ratio of the current pore volume, $\Omega_p$, to the bulk volume of the undeformed configuration, $\Omega^0$, where superscript $0$ denotes measurement at reference conditions. Assuming small perturbations in Lagrangian porosity, and solid and fluid densities, allows us to take these quantities at reference conditions where necessary. In keeping with convention, stress is taken as positive in the tensile direction. Finally, $\overline{\gamma}$ represents the momentum transfer arising as a result of the mass transfer between the two pore continua. Often $\overline{\gamma}$ is assumed to be negligible with respect to the other force density terms (\citealt{Elsworth1992}; \citealt{Pao2002}; \citealt{Fornells2007}; \citealt{Kim2012}). 

Next we introduce the linearised strain tensor given according to the strain-displacement compatibility relation

\begin{equation}
\bm{\epsilon} = \nabla^{sym}\bm{u} = \frac{1}{2}(\nabla\bm{u} + \nabla^{\top}\bm{u}), \label{eqn:2}
\end{equation}

\noindent
where $\bm{u}$ denotes the displacement vector.

The inter-continuum momentum transfer is given by

\begin{equation}
\overline{\gamma} = \sum\limits_{\alpha = m,f} \gamma_{\alpha}\bm{v}_{l,\alpha}, \label{eqn:3}
\end{equation}

\noindent
where $\gamma_{\alpha}$ $[\alpha = m,f]$ is the rate of mass transfer from pore continuum $\alpha$ to pore continuum $\beta$, and $\bm{v}_{l,\alpha}$ is the absolute fluid velocity within each pore continuum. The rate of mass transfer is conservative between the matrix and the fractures thus 

\begin{equation}
\sum\limits_{\alpha = m,f} \gamma_\alpha = 0. \label{eqn:4} 
\end{equation}

\noindent
The absolute fluid velocity, $\bm{v}_{l,\alpha}$, is related to the volume flux, $\bm{q}_\alpha$, within each continuum by

\begin{equation}                           
\bm{q}_\alpha = \phi_\alpha\tilde{\bm{v}_{l,\alpha}} = \phi_\alpha(\bm{v}_{l,\alpha} - \bm{v}_s), \label{eqn:5}
\end{equation}

\noindent
where $\tilde{\bm{v}_{l,\alpha}}$ is the relative fluid velocity, $\bm{v}_s$ is the velocity of the solid matrix, and $\phi_\alpha$ is the Lagrangian porosity associated to each continuum. This is defined as

\begin{equation}
\phi_{\alpha}= \frac{\Omega_{p,\alpha}}{\Omega^0}, \text{ such that }  \phi = \sum\limits_{\alpha = m,f}\phi_\alpha, \label{eqn:6}
\end{equation}   

\noindent
where $\Omega_{p,\alpha}$ is the current pore volume of continuum $\alpha$. The volume flux for each pore continuum is then given by Darcy's law

\begin{equation}
\bm{q}_\alpha = \frac{-\bm{k}_\alpha}{\mu_l}(\nabla p_\alpha-\rho^0_l\bm{g}), \label{eqn:7}
\end{equation}

\noindent
where $\bm{k}_\alpha$ and $p_\alpha$ denote the permeability tensor and fluid pressures associated with pore continuum $\alpha$. 

The balance of fluid mass for each continuum is then given as 

\begin{equation}
\pderiv{m_{l,\alpha}}{t} + \rho^0_l\nabla\cdot\bm{q}_\alpha = \gamma_\alpha, \label{eqn:8}
\end{equation}

\noindent
where $m_{l,\alpha} = \rho_l\phi_\alpha$ is the fluid mass content of continuum $\alpha$. 

We require constitutive laws to provide closure to the model. In early presentations, constitutive relations were indirectly postulated in which inter-continuum pressures were implicitly decoupled (\citealt{Aifantis1977,Aifantis1979}; \citealt{Wilson1982}; \citealt{Elsworth1992}, \citealt{Lewis1997}; \citealt{Bai1999}). To provide more rigour to the form of the constitutive equations we make use of the energy approach to poromechanics under the assumption of infinitesimal strain theory (\citealt{Coussy2004}). It can be shown from a purely macroscopic approach (\citealt{Coussy2004}) or via micromechanical considerations (\citealt{Dormieux2006}), that the increment in strain work density, $\text{d}W_s$, on the skeleton due to the loading triplet ($\text{d}\bm{\epsilon},\text{ d}p_m,\text{ d}p_f$) can be expressed as

\begin{equation}
\text{d}W_s = \bm{\sigma}\text{d}\bm{\epsilon} + p_m\text{d}\phi_m + p_f\text{d}\phi_f. \label{eqn:9}
\end{equation}

\noindent 
Due to elasticity, our system is non-dissipative and thus the skeletal strain energy is stored entirely as an elastic potential

\begin{equation}
\text{d}W_s = \text{d}\Psi_s, \label{eqn:10}
\end{equation}

\noindent
where $\Psi_s$ denotes the Helmholtz free energy of the skeleton, and from which it follows 

\begin{equation}
\bm{\sigma}\text{d}\bm{\epsilon} + p_m\text{d}\phi_m + p_f\text{d}\phi_f - \text{d}\Psi_s = 0. \label{eqn:11}
\end{equation}

\noindent
\cref{eqn:11} is a trivial extension of the skeleton free energy expression for single porosity materials, and is indeed analogous (and identical) to the expression for the multiphase fluid single porosity poromechanical problem (see for example \citealt{Coussy2004}). Now introducing the following Legendre transform 

\begin{equation}
F_s = \Psi_s - p_m\phi_m - p_f\phi_f, \label{eqn:12}
\end{equation}

\noindent
into \cref{eqn:11} results in

\begin{equation}
\bm{\sigma}\text{d}\bm{\epsilon} - \phi_m \text{d}p_f - \phi_f \text{d}p_f - \text{d}F_s = 0. \label{eqn:13}
\end{equation}

\noindent
Next, it is useful to decompose the stress and strain tensors by way of their volumetric and deviatoric parts, 

\begin{align}
\bm{\sigma} = \sigma\bm{1} + \bm{\sigma}_d, \label{eqn:14} \\
\bm{\epsilon} = \frac{1}{3}\epsilon\bm{1} + \bm{\epsilon}_d, \label{eqn:15}
\end{align}

\noindent
where $\sigma = \frac{1}{3}\text{tr}(\bm{\sigma})$ is the mean stress, $\bm{\sigma}_d$ is the deviatoric component of the stress tensor, $\epsilon = \text{tr}(\bm{\epsilon})$ is the volumetric strain, and $\bm{\epsilon}_d$ is the deviatoric component of the total strain tensor. Making use of the stress and strain decompositions from \crefrange{eqn:14}{eqn:15} in \cref{eqn:13}, the state equations for double-porosity poroelasticity are then given as

\begin{equation}
\sigma = \frac{\partial F_s}{\partial \epsilon} \text{;} \quad \bm{\sigma}_d = \frac{\partial F_s}{\partial \bm{\epsilon}_d} \text{;} \quad \phi_m = \frac{\partial F_s}{\partial p_m} \text{;} \quad \phi_f = \frac{\partial F_s}{\partial p_f}. \label{eqn:16}
\end{equation}

\noindent
Applying \cref{eqn:16} to \cref{eqn:13}, and making use of the Maxwell symmetry relations which arise naturally from \cref{eqn:16} whilst also assuming isotropy of the material, we arrive at the constitutive equations for a linear isotropic poroelastic dual-continuum 

\begin{gather}
\text{d}\sigma = K_{dr}\epsilon - b_m\text{d}p_m - b_f\text{d}p_f, \label{eqn:17} \\
\text{d}\phi_m = b_m\epsilon + \frac{1}{N_{m}}\text{d}p_m + \frac{1}{Q}\text{d}p_f, \label{eqn:18} \\
\text{d}\phi_f = b_f\epsilon + \frac{1}{Q}\text{d}p_m + \frac{1}{N_{f}}\text{d}p_f. \label{eqn:19} \\
\text{d}\bm{\sigma}_d = 2G\bm{\epsilon}_d \label{eqn:20}
\end{gather}

\noindent
where parameters $K_{dr}$ and $G$ are the drained bulk and shear moduli of the dual-medium respectively (\citealt{Coussy2004}). Coefficients $b_\alpha$ can be thought of as effective Biot coefficients, and relate changes in effective Lagrangian porosity to skeletal straining under drained conditions. Coefficients $\frac{1}{N_{\alpha}}$ relate changes in the Lagrangian porosity of continuum $\alpha$ to changes in fluid pressure of the same medium, whilst the skeleton remains constrained and fluid pressure in continuum $\beta$ remains constant. Finally, $\frac{1}{Q}$ is a coupling coefficient that relates changes in the Lagrangian porosity of continuum $\alpha$ and pressure changes in continuum $\beta$.

In poromechanics it is common to formulate the constitutive equations in terms of the fluid mass content such that

\begin{equation}
\frac{\text{d}m_{l,\alpha}}{\rho_l} = \text{d}\phi_{\alpha} + \phi_\alpha\frac{\text{d}\rho_l}{\rho_l}\equiv \text{d}\phi_{\alpha} + \phi_\alpha \frac{\text{d}p_\alpha}{K_l}, \label{eqn:21}
\end{equation} 

\noindent
where $K_l$ is the fluid compressibility given by

\begin{equation}
\frac{1}{K_l} = \frac{1}{\rho_l}\frac{\text{d}\rho_l}{\text{d}p}. \label{eqn:22}
\end{equation}  

\noindent
With \cref{eqn:21} we can express \crefrange{eqn:18}{eqn:19} as

\begin{gather}
\text{d}\xi_m =  b_m\epsilon + \frac{1}{M_m}\text{d}p_m + \frac{1}{Q}\text{d}p_f, \label{eqn:23} \\
\text{d}\xi_f = b_f\epsilon + \frac{1}{Q}\text{d}p_m + \frac{1}{M_f}\text{d}p_f, \label{eqn:24}
\end{gather}

\noindent
where 

\begin{equation}
\text{d}\xi_\alpha = \frac{\text{d}m_{l,\alpha}}{\rho^0_l}, \label{eqn:25}
\end{equation}

\noindent
Comparison between \crefrange{eqn:18}{eqn:19} and \crefrange{eqn:23}{eqn:24} gives the additional relation 

\begin{equation}
\frac{1}{M_\alpha} = \frac{1}{N_{\alpha}} + \frac{\phi^0_\alpha}{K_l}. \label{eqn:26}
\end{equation}  

\noindent
Under long-term drainage conditions the double-porosity model must be able to reduce to the well known single-porosity model. This provides us with the following compatibility relations

\begin{gather}
b = b_m + b_f = 1 - \frac{K_{dr}}{K_s}, \label{eqn:27}  \\
\frac{1}{M_{\alpha}} + \frac{1}{Q} = \frac{b_\alpha-\phi^0_{\alpha}}{K_s} + \frac{\phi^0_{\alpha}}{K_l}. \label{eqn:28}
\end{gather}

\noindent
where $b$ is the single-porosity Biot coefficient (\citealt{Berryman1995}). 

A final constitutive equation is required to describe the mass transfer rate between the two pore continua. In accordance with \cite{Barenblatt1960} and \cite{Warren1963} the simplest model for mass transfer between the two continua is given by

\begin{equation}
\gamma_m = \eta\frac{\rho^0_l\overline{k}}{\mu_l}(p_f - p_m), \qquad \gamma_f = \eta\frac{\rho^0_l\overline{k}}{\mu_l}(p_m - p_f), \label{eqn:29}
\end{equation}

\noindent
where $\overline{k}$ is the interface permeability, which here is taken as the matrix permeability (\citealt{Barenblatt1960}; \citealt{Choo2015}), and $\eta$ is the shape factor. The first order nature of \cref{eqn:29} may in some cases over-simplify the physics of inter-continuum mass transfer. This should be taken into consideration when working with the dual-continuum paradigm.  

We make use of an analytical based shape factor as introduced in \cite{Lim1995}. For an isotropic material in two dimensions $\eta$ is defined as     

\begin{equation}
\eta = \frac{2\pi^2}{d^2}, \label{eqn:30}
\end{equation}

\noindent
where $d$ denotes the average spacing between the fractures. 

\section{Models of constitutive coefficients}
We require substantiation of the constitutive coefficients in \cref{eqn:17} and \crefrange{eqn:23}{eqn:24}. One option is direct measurement of these effective parameters. However, this approach is predicted to be non-trivial for dual-continua. For example, isolating matrix and fracture contributions would be challenging. 

An alternative option would be to calculate the effective parameters with models that use more accessible quantities. In the following we compare three modelling concepts that make use of different properties for the calculation of the constitutive parameters: 

\begin{enumerate}[(i)]
\item \cite{Khalili1996} - Constituent mechanical properties, assuming the high permeability, low storage continuum is all void space (no intrinsic fracture properties),
\item \cite{Borja2009} - Constituent pore fractions, assuming the high permeability, low storage continuum is all void space,
\item \cite{Berryman2002a} - Constituent mechanical properties, including intrinsic fracture properties.
\end{enumerate}

\noindent
We recognise these models, to the best of our ability, as the most dominant within the literature. They have been used in the works of \cite{Khalili2000}, \cite{Callari2000}, \cite{Pao2002}, \cite{Fornells2007}, \cite{Taron2009}, \cite{Kim2012}, \cite{Mehrabian2014}, \cite{Choo2015}, \cite{Choo2016} and \cite{Wang2018}. It should be stressed that in most cases the modelling concepts introduced in this section build on, or are aligned with, previous works and concepts introduced by \cite{Aifantis1977,Aifantis1979}, \cite{Wilson1982}, \cite{Elsworth1992}, \cite{Berryman1995}, \cite{Tuncay1995, Tuncay1996a}, \cite{Loret1999} and \cite{Dormieux2006} to name but a few. These works should thus by borne in mind in the section to follow. 

\subsection{Model approaches and assumptions}
\subsubsection{\cite{Khalili1996}}
The authors take a top down approach, postulating macroscopic balance laws and an effective stress expression. Closure for the model equations therein is then sought through thought experiments that isolate volumetric changes of the constituents. Superposition due to linearity, and Betti's reciprocal work theorem finally allow for recovery of the macroscopic behaviour in terms of constituent responses. In doing, expressions for the constitutive coefficients are identified.  

\cite{Khalili1996} implicitly assume that the fracture phase is all void space. Additionally, the following assumption is also made: $b_m\phi^0_f = b_f\phi^0_m$. This relation is restrictive, and is later removed in \cite{Khalili2003} and \cite{Khalili2003a}, due to the resulting compatibility enforced between bulk moduli (\citealt{Loret1999}). We present coefficient models derived by the authors without this assumption (\cref{tab:1}). The coefficient models from \cite{Khalili1996} (\cref{tab:1}), are then consistent with results from \cite{Berryman1995}, \cite{Loret1999}, and \cite{Dormieux2006}. 

\begin{table}[h]
\small
\centering
\begin{adjustwidth}{-1.2cm}{-1.2cm}
\begin{tabular} { >{\centering\arraybackslash}m{3cm} | c | c | c | c | c}
& $b_f$ & $b_m$ & $\frac{1}{M_f}$ & $\frac{1}{Q}$ & $\frac{1}{M_m}$ \\ 
\hline \hline \\ [-0.8em]
\cite{Khalili1996} & $1-\frac{K_{dr}}{K_m}$ & $\frac{K_{dr}}{K_m}-\frac{K_{dr}}{K_s}$ & $\frac{\phi^0_f}{K_l} + \frac{b_f - \phi^0_f}{K_m}$ & $\frac{b_f - \phi^0_f}{K_s} - \frac{b_f - \phi^0_f}{K_m}$ & $\frac{\phi^0_m}{K_l} + \frac{b_m - \phi^0_m}{K_s} - \frac{1}{Q}$ \\ [0.5em]
\cite{Borja2009}$\text{}^+$ & $\psi^0_f b$ & $\psi^0_m b$ & $\frac{\phi^0_f}{K_l}$ & $0$ & $\frac{\phi^0_m}{K_l}$ \\ [0.5em]
\cite{Berryman2002a}$\text{}^{++}$ & $b^*_f\frac{K_{dr}-K_m}{K_f-K_m}$ & $b^*_m\frac{K_{dr}-K_f}{K_m-K_f}$ & $A_{33} - \frac{(b_f)^2}{K_{dr}}$ & $A_{23} - \frac{b_mb_f}{K_{dr}}$ & $A_{22} - \frac{(b_m)^2}{K_{dr}}$ 
\end{tabular}
\end{adjustwidth}
\caption{poromechanical material cofficients by author. $\text{}^+$ Assuming $\pderiv{\psi_\alpha}{t}\approx0$. $\text{}^{++}$ Coefficients $A_{22}$, $A_{23}$ and $A_{33}$ are given in \cref{tab:2}. Note that our expressions for coefficient models from \cite{Khalili1996} are slightly different to their original presentation. This is done to highlight the similarities between expressions for $b_f$ and $\frac{1}{M_f}$ and single porosity equivalents, $b$ and $\frac{1}{M}$ (see for example \cite{Coussy2004}).} \label{tab:1}
\end{table}
   
\subsubsection{\cite{Borja2009}} 
In \cite{Borja2009}, the authors consider the evolution of internal energy density and derive a thermodynamically consistent effective stress law. Aggregate material is used as their reference material. The void space assumption is thus implicit. We present the effective stress expression from \cite{Borja2009} for an isotropic single phase dual-continuum system as follows 

\begin{equation}
\sigma' = \text{d}\sigma + \sum\limits_{\alpha=m,f}\psi_\alpha b\text{d}p_\alpha, \label{eqn:31}
\end{equation}  

\noindent
where $\sigma'=K_{dr}\epsilon$, and $\psi_\alpha$ denotes the pore fraction of continuum $\alpha$ such that

\begin{equation}
\psi_\alpha=\frac{\varphi_\alpha}{1-\varphi_s}. \label{eqn:32}
\end{equation}  

\noindent
Notation $\varphi_\alpha$ is the Eulerian porosity of continuum $\alpha$. This is defined as the ratio of the current pore volume, $\Omega_{p,\alpha}$, to the bulk volume of the current (deformed) configuration, $\Omega$. Notation $\varphi_s=\frac{\Omega_s}{\Omega}$ is the current volume fraction of the solid phase. In the limit of infinitesimal transformations, $\varphi_\alpha\approx\phi_\alpha$, from which, together with small perturbations in Lagrangian porosity, follows $\psi_\alpha\approx\psi^0_\alpha$. 

Comparison of \cref{eqn:31} with \cref{eqn:17} leads to the following relations for effective Biot coefficients, $b_m = \psi^0_m b$ and $b_f = \psi^0_f b$. 

\cite{Borja2009} identify the requirement for a constitutive expression for $\psi_\alpha$ based on energy conjugacy with $p_\alpha$ (their Eq. (76)). However, explicit constitutive equations for $\psi_\alpha$ remain, to the best of the current authors' knowledge, an open question. We do note, however, that \cite{Borja2016} develop a framework that allows for the tracking of pore fraction evolutions numerically. 

As an initial approach to deriving an algebraic expression for variations in $\psi_\alpha$ we first consider the following mass balance equation given in \cite{Borja2009},

\begin{equation}
b_\alpha\pderiv{\epsilon}{t} + (1-\phi_s)\pderiv{\psi_\alpha}{t} + \frac{\phi^0_\alpha}{K_l}\pderiv{p_\alpha}{t}+\frac{\bm{q}_\alpha}{K_l}\nabla p_\alpha + \nabla\cdot\bm{q}_\alpha  = \frac{1}{\rho^0_l} \gamma_\alpha. \label{eqn:33}
\end{equation}

\noindent
We can derive an alternative form of the mass balance here by substitution of \cref{eqn:23} or \cref{eqn:24}, together with Darcy's law, into \cref{eqn:8} such that  

\begin{equation}
b_\alpha\pderiv{\epsilon}{t} + (\frac{1}{N_{\alpha}}+\frac{\phi^0_\alpha}{K_l})\pderiv{p_\alpha}{t} + \frac{1}{Q}\pderiv{p_\beta}{t} + \nabla\cdot\bm{q}_\alpha = \frac{1}{\rho^0_l} \gamma_\alpha,  \label{eqn:34} \\
\end{equation}

\noindent
where we have used \cref{eqn:26} to decompose $\frac{1}{M_\alpha}$. 

Under the assumption of small perturbations in fluid density, the fourth term on the left-hand side of \cref{eqn:33} can be neglected. Comparing \cref{eqn:33} and \cref{eqn:34} leads to the following identity

\begin{equation}
(1-\phi_s)\pderiv{\psi_\alpha}{t} = \frac{1}{N_\alpha}\pderiv{p_\alpha}{t} + \frac{1}{Q}\pderiv{p_\beta}{t} \label{eqn:35}
\end{equation}

\noindent
In the works of \cite{Choo2015} and \cite{Choo2016}, the authors achieve non-algebraic closure of the mass balance equations for each continuum by assuming $\pderiv{\psi_\alpha}{t}\approx0$. This assumption is equivalent to

\begin{equation}
\frac{1}{N_\alpha} = \frac{1}{Q} = 0 \quad \text{or} \quad \pderiv{p_\alpha}{t} = -\frac{N_\alpha}{Q} \pderiv{p_\beta}{t}, \label{eqn:36}
\end{equation}

\noindent
in \cref{eqn:35}. We discard the latter relation in \cref{eqn:36} as it is overly restrictive with respect to fluid exchange between matrix and fractures. Hence we identify the closure condition $\pderiv{\psi_\alpha}{t}\approx0$ with values for material coefficients $\frac{1}{N_\alpha}$ and $\frac{1}{Q}$ of zero. 

A summary of the coefficient models from \cite{Borja2009}, under the explicit assumption $\pderiv{\psi_\alpha}{t}\approx0$, mapped to the constitutive model of \cref{eqn:17} and \crefrange{eqn:23}{eqn:24} is shown in \cref{tab:1}.

\subsubsection{\cite{Berryman2002a}} 
The motivation behind the approach  by \cite{Berryman2002a} is to formulate coefficient models using intrinsic fracture properties. Therefore, contrary to the previous models, no assumption is made on the values of $\phi^*_f$ and $K_f$. Inclusion of intrinsic fracture properties is concurrent with the fracture continuum having an associated stiffness. 

The authors use a top down approach whose starting point is the macroscopic constitutive model written within a pure stiffness setting ($\sigma$ and $p_\alpha$ as primary variables; contrary to conventional poromechanical modelling),

\begin{equation}
\begin{pmatrix}
\epsilon\\
\text{d}\xi_m\\
\text{d}\xi_f  
\end{pmatrix}
=\begin{pmatrix}
A_{11} & A_{12} & A_{13}\\
A_{21} & A_{22} & A_{23}\\
A_{31} & A_{32} & A_{33}  
\end{pmatrix}
\begin{pmatrix}
\text{d}\sigma\\
\text{d}p_m\\
\text{d}p_f 
\end{pmatrix}, \label{eqn:37}
\end{equation}

\noindent
where the goal is then to find expressions for coefficients $A_{11}-A_{33}$. Symmetry of the coefficient matrix in \cref{eqn:37} is implied. A comparison of \cref{eqn:37} with \cref{eqn:17} and \crefrange{eqn:23}{eqn:24} reveals the following relations

\begin{gather}
K_{dr} = \frac{1}{A_{11}}, \qquad b_m = \frac{A_{12}}{A_{11}}, \qquad b_f = \frac{A_{13}}{A_{11}}, \nonumber\\
\frac{1}{M_m} = A_{22}-\frac{\left(A_{12}\right)^2}{A_{11}}, \qquad \frac{1}{Q} = A_{23}-\frac{A_{12}A_{13}}{A_{11}}, \qquad \frac{1}{M_f} = A_{33}-\frac{\left(A_{13}\right)^2}{A_{11}}. \label{eqn:38}
\end{gather}  

\noindent
To identify expressions for the parameters in \cref{eqn:37}, \cite{Berryman2002a} considers scenarios of uniform expansion (or contraction). These scenearios are equivalent to asking whether we can find variations in uniform stress, $\text{d}\sigma=\text{d}\sigma_m=d\text{d}\sigma_f$, and variations in pore pressures, $\text{d}p_m$ and $\text{d}p_f$, such that $\epsilon=\epsilon_m=\epsilon_f$ (\citealt{Berryman2002a}) As a result the authors are able to relate \cref{eqn:37} to intrinsic constituent equations. The microscale equations used in \cite{Berryman2002a} are postulated based on the assumption that each constituent belonging to the dual-medium behaves as a Gassmann-material. That is, constituent solid phases are homogeneous and isotropic (\citealt{Cheng2016}). In the case of a Gassmann-material, the constitutive equations for a material $\alpha$ can be written as   

\begin{equation}
\begin{pmatrix}
\epsilon_\alpha\\
\text{d}\xi_\alpha
\end{pmatrix} 
= \frac{1}{K_\alpha} \begin{pmatrix}
1 & b^*_\alpha\\
v_\alpha b^*_\alpha & \frac{v_\alpha b^*_\alpha}{B^*_\alpha}
\end{pmatrix}
\begin{pmatrix}
\text{d}\sigma_\alpha \\
\text{d}p_\alpha
\end{pmatrix}, \label{eqn:39}
\end{equation}       

\noindent
where $v_\alpha$ denotes the volume fraction of material $\alpha$, and poroelastic coefficients intrinsic to a material $\alpha$ have been denoted by superscript $^*$. Volume fractions for matrix and fracture materials are given as 

\begin{equation}
v_m = \frac{\Omega^0_m}{\Omega^0}, \quad  v_f =  \frac{\Omega^0_f}{\Omega^0}, \label{eqn:40}
\end{equation}

\noindent
where $\Omega^0_m$ and $\Omega^0_f$ are the volumes of matrix and fracture continua at reference conditions respectively. Under the void space assumption $v_f=\phi^0_f$. The intrinsic Biot and Skempton coefficients, $b^*_\alpha$ and $B^*_\alpha$ for material $\alpha$ respectively, are

\begin{gather}
b^*_\alpha = 1 - \frac{K_\alpha}{K^{\alpha}_s}, \nonumber\\
B^*_\alpha = \frac{b^*_\alpha M^*_\alpha}{K_\alpha + (b^*_\alpha)^2M^*_\alpha} \text{  where  }
\frac{1}{M^*_\alpha} = \frac{\phi^{*}_\alpha}{K_l} + \frac{b^*_\alpha - \phi^{*}_\alpha}{K^{\alpha}_s}. \label{eqn:41}
\end{gather}

\noindent
where $K^\alpha_s$ is the solid grain modulus of the intact material, and $M^*_\alpha$ and $\phi^*_\alpha$ are the intrinsic Biot modulus and intrinsic porosity (measurement at reference conditions is implied) respectively.  

Expressions for $A_{11}-A_{33}$ are finally recovered (\cref{tab:2}), using the uniform expansion/contraction thought experiments described above. With the relations in \cref{eqn:38} we get material coefficient formulations pertaining to the conventional mixed compliance setting ($\epsilon$ and $p_\alpha$ as primary variables) (\cref{tab:1}).

As a final note on the \cite{Berryman2002a} coefficient models, recent users have explicitly assumed  $A_{23}=A_{32}=0$ as a closure condition to generalise the dual-continuum system to a multi-continuum one (\citealt{Kim2012}; \citealt{Mehrabian2014}; \citealt{Mehrabian2018}), providing explicit relations between material properties and simplifying the coefficient models. It is still unclear how this assumption may affect the system.     

\begin{table}[h]
\small
\centering
\begin{tabular} {l | l }
Coefficient & \cite{Berryman2002a} Formulation \\ 
\hline \hline \\ [-0.8em]
$A_{11}$ & \Large{$\frac{1}{K_{dr}}$} \\[0.5em]
$A_{12}$ & \Large{$\frac{b^*_m}{K_m}\frac{1-K_f/K_{dr}}{1-K_f/K_m}$} \\ [0.5em]
$A_{13}$ & \Large{$\frac{b^*_f}{K_f}\frac{1-K_m/K_{dr}}{1-K_m/K_f}$} \\[0.5em]
$A_{22}$ &  \Large{$\frac{v_mb^*_m}{B^*_mK_m}-(\frac{b^*_m}{1-K_m/K_f})^2\lbrace\frac{v_m}{K_m}+\frac{v_f}{K_f}-\frac{1}{K_{dr}}\rbrace$} \\[0.5em]
$A_{23}$ &  \Large{$\frac{K_mK_fb^*_mb^*_f}{(K_f-K_m)^2}\lbrace\frac{v_m}{K_m}+\frac{v_f}{K_f}-\frac{1}{K_{dr}}\rbrace$} \\[0.5em]
$A_{33}$ &  \Large{$\frac{v_fb^*_f}{B^*_fK_f}-(\frac{b^*_f}{1-K_f/K_m})^2\lbrace\frac{v_m}{K_m}+\frac{v_f}{K_f}-\frac{1}{K_{dr}}\rbrace$} 
\end{tabular}
\caption{\cite{Berryman2002a} material coefficient formulations for the pure stiffness constitutive model.} \label{tab:2}
\end{table}

\subsubsection{In sum} 
Coefficient models from \cite{Khalili1996} and \cite{Borja2009} both make an underlying void space assumption for the high permeability, low storage continuum. Models from \cite{Borja2009} make use of continuum pore fractions, but still require a final closure relationship for the evolution of each pore fraction. Finally, models from \cite{Berryman2002a} make no underlying assumption on the intrinsic porosity, and thus the stiffness of the high permeability, low storage continuum. 

In terms of pressure decoupling assumptions we have two types. The first are implicit assumptions for which the constitutive relations are postulated without inter-continuum pressure coupling. The second are explicit assumptions for which the full constitutive model is the starting point (or the requirement for constitutive expressions are at least identified in the case of \cite{Borja2009}). The explicit decoupling assumptions are then made so as to provide relations between material properties and to simplify coefficient models (e.g. due to non-algebraic closure). However, the physical justifications and/or implications of these assumptions still remain an open question. 

It is of interest to investigate how differences in coefficient models, in addition to deceoupling assumptions, may impact poromechanical behaviour. This is pursued in the remaining sections. 

\section{Micromechanics of dual-continua} 
To establish the physical implications of explicit decoupling assumptions such as taking $A_{23}=A_{32}=0$ we make use of the theoretical framework of micromechancis. Micromechanics is used as a tool to relate the macroscopic behaviour and properties of a composite to those of its underlying constituents (microscale) (\citealt{Nemat1993}). In the following we consider the dual-medium as a composite of two (poro-) elastic materials with each material having its own intrinsic constitutive model.

\subsection{Effective elastic properties}
We define effective properties over a representative elementary volume (REV) taken from a macroscopic body that is at least an order of magnitude greater in size than the REV itself. The scale requirements for the identification of such an REV can be found in \cite{Bear2012}. In addition to the material isotropy assumption, we assume the dual-material to be statistically homogeneous. In its most simple interpretation, statistical homogeneity implies that averages taken over any REV from a large composite body, approach averages taken over the whole body itself. Averages are thus positionally invariant and we can recover the classical volume averaging expression accordingly   

\begin{equation}
\overline{\bm{f}} = \frac{1}{\Omega}\int_{\Omega}\bm{f}(\bm{x})\text{ } dV, \label{eqn:42}
\end{equation}    

\noindent
where $\bm{f}$ is an arbitrary field and $\Omega$ is the volume of \textit{any} REV within a large macroscopic body.

With the averaging operation defined we proceed to formulate the effective elastic property problem. In doing we make use of the works of \cite{Hill1963} and \cite{Hashin1972}. Additionally we assume that the composite is drained and thus make no distinction between effective and total stress fields. 

Starting at the microscale, for which we have assumed linear elasticity, the micro stress and strain fields, $\bm{s}(\bm{x})$ and $\bm{e}(\bm{x})$ respectively, are related through the drained fourth order stiffness and compliance tensors, $\bm{C}^*_{dr}(\bm{x})$ and $\bm{S}_{dr}^*(\bm{x})$ respectively, by 

\begin{gather}
\bm{s}(\bm{x}) = \bm{C}_{dr}^*(\bm{x})\bm{e}(\bm{x}), \label{eqn:43} \\
\bm{e}(\bm{x}) = \bm{S}_{dr}^*(\bm{x})\bm{s}(\bm{x}). \label{eqn:44}
\end{gather}

\noindent
It can be shown that macroscopic stress and strain tensors are related to their underlying fields through the volume averaging operator (see for example \citealt{Hashin1972}) such that,

\begin{gather}
\bm{\sigma} = \bm{\overline{s}} = v_m\overline{\bm{s}}_m + v_f\overline{\bm{s}}_f, \label{eqn:45} \\
\bm{\epsilon} = \bm{\overline{e}} = v_m\overline{\bm{e}}_m + v_f\overline{\bm{e}}_f. \label{eqn:46}
\end{gather}

\noindent
where $\overline{\bm{e}}_\alpha=\Omega^{-1}_\alpha\int_{\Omega_\alpha}e_\alpha(x)\text{ }dV$ (resp. $\overline{\bm{s}}_\alpha$). Substitution of \crefrange{eqn:43}{eqn:44} in \crefrange{eqn:45}{eqn:46}, and making use of the decompositions in \crefrange{eqn:14}{eqn:15} due to the isotropy of the dual-medium leads to

\begin{gather}
\sigma = K_{dr}\epsilon = v_mK_m\overline{e}_m + v_fK_f\overline{e}_f  \label{eqn:47}\\
\epsilon = S_{dr}\sigma = v_mS_m\overline{s}_m + v_fS_f\overline{s}_f, \label{eqn:48}
\end{gather}

\noindent
where $\overline{e}_\alpha$ and $\overline{s}_\alpha$ are the average microscopic volumetric strain and mean stress fields of continuum $\alpha$ respectively, and $S_{dr}$ is the drained compressibility of continuum $\alpha$.   

Fundamental to the micromechanics approach is understanding how macroscopic stress and strain distribute between individual constituents. Due to the linearity of the material behaviour, the following relations hold (\citealt{Hashin1963}) 

\begin{gather}
\overline{e}_m = A_m\epsilon, \quad \overline{e}_f = A_f\epsilon, \label{eqn:49}  \\
\overline{s}_m = B_m\sigma, \quad \overline{s}_f = B_f\sigma, \label{eqn:50}
\end{gather}

\noindent
where $A_\alpha$ and $B_\alpha$ are concentration factors mapping macroscopic volumetric strain and mean stress to the equivalent averaged microscopic fields. Importantly $A_\alpha$ and $B_\alpha$ encode the geometrical information of the problem.

Additionally, $A_\alpha$ (resp. $B_\alpha$) admit the following compatibility relations 

\begin{equation}
v_mA_m + v_fA_f = 1, \quad v_mB_m + v_fB_f = 1.  \label{eqn:51}  \\
\end{equation}

\noindent
Use of \crefrange{eqn:49}{eqn:51} in \cref{eqn:47} and \cref{eqn:48} allows us to develop the following fundamental relationships for effective moduli of two phase composites,

\begin{gather}
K_{dr} = K_m + v_f(K_f - K_m)A_f, \label{eqn:52}  \\
S_{dr} = S_m + v_f(S_f - S_m)B_f. \label{eqn:53}
\end{gather}

\noindent
Finding effective moduli then amounts to determining $A_\alpha$ or $B_\alpha$. 

For given underlying constituent properties, $K_{dr}$ (resp. $S_{dr}$) can take on a range of values, existing between certain well-defined lower and upper bounds, due to the geometrical variability of the problem. As a result, the effective coefficients also exhibit a bounded range of values due to their dependence on $K_{dr}$ (\cref{tab:1}). This dependence is further explored in the following section.   

\subsection{Physical implications of explicit decoupling}
In the following we show that explicit decoupling assumptions are coincident with effective coefficient bounds. Bounds are attractive as they provide useful estimates of effective properties of interest, as well as a means to verify the values of these properties (\citealt{Torquato1991}).    

We use arguments from micromechanics to show heuristically that the inverse to explicit decoupling is to assume that \cref{eqn:52} or \cref{eqn:53} can be calculated directly by considering certain limiting behaviours and thus bounds on $K_{dr}$. Bounds, from which explicit decouplings naturally arise, on effective properties then follow from the bounds on $K_{dr}$.     

\subsubsection{Isostrain: $\frac{1}{Q}=0$}
Whilst this explicit assumption has not been used within literature its consideration remains instructive. We define assumption $\frac{1}{Q}=0$ in terms of $\text{d}\phi_m$ (although converse arguments may be used for $\text{d}\phi_f$). The term $\frac{1}{Q}=0$ is then equivalent to

\begin{equation}
\frac{1}{Q} = \pderiv{\phi_m}{p_f}\Bigr|_{\substack{\epsilon=0\\dp_m=0}} = 0. \label{eqn:54}
\end{equation}

\noindent
Next we define the following local constitutive model for a continuum $\alpha$ (\citealt{Dormieux2006}),

\begin{gather}
\text{d}\overline{s}_\alpha = K_\alpha\overline{e}_\alpha - b^*_\alpha\text{d}p_\alpha, \label{eqn:55} \\
\text{d}\phi_\alpha = v_\alpha \left(b^*_\alpha\overline{e}_\alpha + \frac{1}{N^*_\alpha}\text{d}p_\alpha \right). \label{eqn:56}
\end{gather}

\noindent
Consider the local model given by \crefrange{eqn:55}{eqn:56} for the case $\alpha=m$. Under a drained matrix ($\text{d}p_m=0$), \cref{eqn:56} shows that a zero matrix porosity variation ($\text{d}\phi_m=0$) can only hold if the average matrix strain is zero ($\overline{e}_m=0$). Further, we can see from \cref{eqn:55} that if the matrix is drained and does not deform, then the variation in average matrix stress is also zero ($\text{d}\overline{s}_m=0$). This will be a useful result for discussions on the explicit decoupling assumption $A_{23}=0$. 

Proceeding with \cref{eqn:54} and the macroscopic strain constraint, $\epsilon=0$, the strain partition in \cref{eqn:46} shows that if $\epsilon=\overline{e}_m=0$ then $\overline{e}_f=0$. The condition $\frac{1}{Q}=0$ is therefore consistent with a condition of \textit{isostrain}, 

\begin{equation}
\epsilon = \overline{e}_m = \overline{e}_f, \quad A_m = A_f = 1. \label{eqn:57}
\end{equation}

\begin{figure}[h]
\centering
\includegraphics[scale = 0.5]{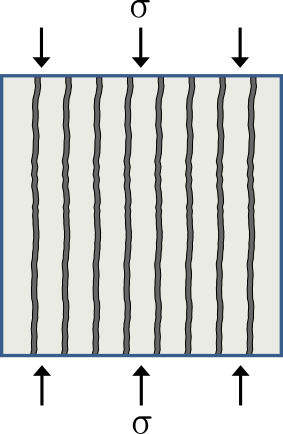}
\caption{Element arrangement in the isostrain condition.} \label{fig:2}
\end{figure}

\noindent
This distribution of strain can be obtained when elements are set in parallel to the direction of loading (\cref{fig:2}). Under isostrain, the bulk modulus of the composite is then the Voigt average of the constituent moduli (\citealt{Voigt1928}), 

\begin{equation}
K^{V}_{dr} = v_mK_m + v_fK_f. \label{eqn:58}
\end{equation}

\noindent
\cite{Hill1963} used variational principles to show that the Voigt average represents an upper bound on $K_{dr}$, and is naturally obtained by substitution of $A_f=1$ into \cref{eqn:52}. As a result, effective coefficients calculated with $K^{V}_{dr}$ correspond to bounds on these parameters. To explore this further we consider the case for the effective Biot coefficients under the the void space assumption.    

When assuming a drained void fracture continuum ($\text{d}p_f=0$), and from \cref{eqn:56}, effective fracture strain is equal to the effective fracture pore strain, $\overline{e}_f=\frac{\text{d}\phi_f}{\phi^0_f}$, and thus in analogy to \cref{eqn:57} isostrain can be summarised in this case as

\begin{equation}
\epsilon = \overline{e}_m = \frac{\text{d}\phi_f}{\phi^0_f}. \label{eqn:59}
\end{equation}

\noindent
Note that \cref{eqn:59}, and isostrain more generally, \cref{eqn:57}, are only equal to zero under the constrained macroscopic strain condition ($\epsilon=0$). However, in general \cref{eqn:57} and \cref{eqn:59} are non-zero due to macroscopic deformation ($\epsilon\not=0$).

From \cref{eqn:19}, with drained matrix and fracture continua, together with the void-space isotrain condition, \cref{eqn:59}, we get the following lower bound on $b_f$ 

\begin{equation}
b_f = \pderiv{\phi_f}{\epsilon}\Bigr|_{\substack{dp_m=0\\dp_f=0}} = \phi^0_f. \label{eqn:60}
\end{equation}   

\noindent
From \cref{eqn:27} the lower bound on $b_f$ corresponds to an upper bound for $b_m$ 

\begin{equation}
b_m = v_m \left(1 - \frac{K_m}{K_s}\right), \label{eqn:61}
\end{equation}

\noindent
where $v_m = 1 - \phi^0_f$, and where we have used $K_{dr} = K^V_{dr} = v_mK_m$. 

From the first equality in \cref{eqn:54} we expect $\frac{1}{Q} \leq 0$ since matrix porosity must reduce in order to accommodate the pressure driven fracture expansion (see also similar arguments in \cite{Berryman1995}). Thus, based on the arguments described in this section we can infer that the explicit decoupling assumption $\frac{1}{Q} = 0$ is concurrent with an \textit{upper bound} on $\frac{1}{Q}$. 

\subsubsection{Incompressible grain isostrain: $\frac{1}{N_\alpha}=\frac{1}{Q}=0$}
We now consider the coefficient models from \cite{Borja2009} under the assumption $\pderiv{\psi_\alpha}{t}\approx0$ made by \cite{Choo2015} and \cite{Choo2016}. 

In Section 3.1.2, we indetified that $\pderiv{\psi_\alpha}{t}\approx0$ amounts to $\frac{1}{N_\alpha}=\frac{1}{Q}=0$ when mapping to the constitutive model shown in \crefrange{eqn:17}{eqn:19}. It is of interest to see under what conditions the result $\frac{1}{N_\alpha}=\frac{1}{Q}=0$ arises when starting from the void space coefficient models built from constituent mechanical properties (i.e. those from \cite{Khalili1996}). 

The set of explicit assumptions: $\frac{1}{N_\alpha}=\frac{1}{Q}=0$, can easily be derived from the microscale by first considering isostrain (and thus $\frac{1}{Q}=0$). With the resulting bounds arising from isostrain, \crefrange{eqn:60}{eqn:61}, along with the assumption $K_s=\infty$, we obtain $\frac{1}{N_\alpha}=0$ using the coefficient models of \cite{Khalili1996} (\cref{tab:1}, with \cref{eqn:26} to decompose $\frac{1}{M_\alpha}$). We therefore refer to conditions resulting in $\frac{1}{N_\alpha}=\frac{1}{Q}=0$ as \textit{incompressible grain isostrain}

As far as parameters in the balance of mass are concerned, \cref{eqn:33} and \cref{eqn:34} are identical when assuming $\pderiv{\psi_\alpha}{t}\approx0$ in the former and incompressible grain isostrain using the constituent mechanical property void space coefficient models in the latter. However, differences in mass balance behaviour may be introduced through the way in which $b_\alpha$ is modelled. It is therefore of interest see how the effective Biot coefficients calculated using the respective void space coefficient models under incompressible grain isostrain compare to the bounds established in \crefrange{eqn:60}{eqn:61}. 

Under the incompressible grain assumption the upper bound for $b_m$ now reads $b_m=v_m$. When using $K^V_{dr}=v_mK_m$ in the coefficient models of \cite{Khalili1996} we can see that the bounds in \crefrange{eqn:60}{eqn:61} are naturally recovered (see \cref{tab:1}). In contrast, from \cref{tab:1} the effective Biot coefficients calculated using the models of \cite{Borja2009} are equivalent to pore fractions, since $b=1$ for incompressible grains. Due to the differences in effective Biot coefficients (and thus other constitutive parameters), we expect disparity in poroelastic behaviour when using the two sets of void space coefficient models.  

\subsubsection{Isostress: $A_{23}(=A_{32})=0$} 
Finally, we study the pure stiffness setting with the condition $A_{23}=0$ in light of the assumptions made in \cite{Nguyen2010}, \cite{Kim2012}, \cite{Mehrabian2014} and \cite{Mehrabian2018}. We continue to work in terms of $\text{d}\phi_m$. Accordingly, and using similar arguments to those used in the derivations of \crefrange{eqn:23}{eqn:24}, 

\begin{equation}
A_{23} = \pderiv{\xi_m}{p_f}\Bigr|_{\substack{d\sigma=0\\dp_m=0}} \equiv \pderiv{\phi_m}{p_f}\Bigr|_{\substack{d\sigma=0\\dp_m=0}} = 0. \label{eqn:62}
\end{equation}

\noindent
In Section 4.2.1 we established that a drained matrix experiencing no deformation is concurrent with a condition of zero (average) matrix stress. From the macroscopic stress constraint in \cref{eqn:62}, $\text{d}\sigma=0$, and the stress partition in \cref{eqn:45}, if $\text{d}\sigma=\text{d}\overline{s}_m=0$ then $\text{d}\overline{s}_f=0$. The condition $A_{23}$ is therefore consistent with a condition of \textit{isostress},

\begin{equation}
\text{d}\sigma = \text{d}\overline{s}_m = \text{d}\overline{s}_f, \quad B_m = B_f = 1. \label{eqn:63}
\end{equation}

\begin{figure}[h]
\centering
\begin{minipage}{0.5\textwidth}
\centering
\includegraphics[scale = 0.5]{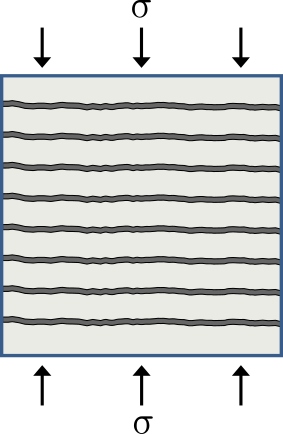}
\subcaption{Transverse}\label{fig:3a}
\end{minipage}%
\begin{minipage}{0.5\textwidth}
\centering
\includegraphics[scale = 0.5]{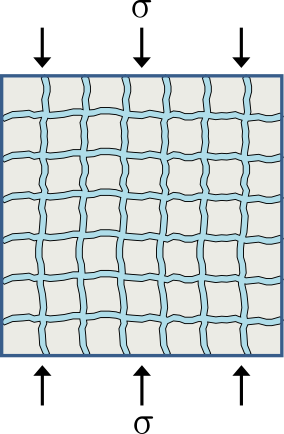}
\subcaption{Solid-fluid suspension}\label{fig:3b}
\end{minipage}
\caption{Typical element arrangements in the isostress condition.} \label{fig:3}
\end{figure}

\noindent
The classical configuration under which an isostress distribution is observed, are elements that are arranged transversely to the direction of applied load (\cref{fig:3a}). For isotropic fracture networks, a more frequent configuration that shows isostress behaviour is within a solid-fluid suspension (\cref{fig:3b}). Such a situation may occur if a network of open fractures totally permeates the solid, thus completely dissociating the matrix material. 

Under isostress, the bulk modulus of the composite is then the Reuss average of the constituent bulk moduli (\citealt{Reuss1929}), 

\begin{equation}
\frac{1}{K^R_{dr}} = \frac{v_m}{K_m} + \frac{v_f}{K_f}. \label{eqn:64}
\end{equation} 

\noindent
\cite{Hill1963} showed that the Reuss average is a lower bound on $K_{dr}$, and is naturally obtained by substitution of $B_f=1$ into \cref{eqn:53}, or from $A_{23}=0$ in \cref{tab:2}. In analogy to $K^V_{dr}$, use of $K^R_{dr}$ for the calculation of the effective constitutive parameters will result in bounds on these parameters.    

Consider the case when $K_f$ is zero (void space fracture phase) we can see from \cref{eqn:64} that $K_{dr}$ is also zero. Using the void space assumption and the isostress condition, \cref{eqn:63}, with \cref{eqn:17} for $d\sigma$ and \cref{eqn:55} for $d\sigma_m$ and $d\sigma_f$ gives us the following 

\begin{equation}
b_m\text{d}p_m - b_f\text{d}p_f = K_m\overline{e}_m - b^*_m\text{d}p_m = -1\text{d}p_f. \label{eqn:65}
\end{equation}

\noindent
From the required isostress equality, $\sigma = \overline{s}_f$, \cref{eqn:65} followed by \cref{eqn:27} allows us to establish the following bounds 

\begin{equation}
b_f = 1, \quad b_m = 0. \label{eqn:66}
\end{equation}   

\noindent
\cref{eqn:66} shows that use of the Reuss bound corresponds to an upper bound on $b_f$ and a lower bound on $b_m$. Interestingly the bounds on $b_f$ from this section and Section 4.2.1, i.e. $\phi^0_f \leq b_f \leq 1$ are very similar to those established on $b$ for the single-porosity model (see \cite{Dormieux2006}). 

\subsection{On moduli upscaling}    
In \cite{Nguyen2010}, \cite{Kim2012}, \cite{Kim2013}, \cite{Mehrabian2014, Mehrabian2015} and more recently \cite{Mehrabian2018}, isostress is implicitly assumed. Upscaling of constituent moduli is then admitted through the Reuss average. This raises the question as to whether this is a reasonable approach to upscaling or not?   

The Reuss and Voigt average represent lower and upper bounds for effective moduli respectively (\citealt{Hill1963}). Moduli bounds represent a minimum and maximum limit, between, or at which, effective moduli arise. For the bulk modulus, bounds by \cite{Hashin1963} have been shown in the same paper to be the best possible for isotropic composites, given only constituent moduli and volume fractions. This result is obtained by solving exactly for the bulk modulus of a specific geometry composite (composite sphere assemblage), and observing that the resultant solution coincides with either bound depending on the stiffness of the inclusion relative to the host material. The lower Hashin-Shtrikman (HS) effective bulk modulus bound is given as

\begin{equation}
K^{HS-}_{dr} = K_f + \frac{v_m}{(K_m - K_f)^{-1} + v_f(K_f + \frac{4}{3}G_f)^{-1}}, \label{eqn:67}
\end{equation}     

\noindent
where $G_f$ denotes the shear modulus of the fracture material. The upper bound, $K^{HS+}_{dr}$ can be determined by swapping subscripts $f$ and $m$ in \cref{eqn:67}.

When  $G_f=0$, such as in a fluid suspension geometry, the HS lower bound and the Reuss bound coincide. It follows that in this geometry, with the stiffness of a void space fracture phase ($K_f=0$), both the Reuss and HS lower bound result in $K^{R}_{dr}=K^{HS+}_{dr}=0$. However, in situations when the fracture phase has an intrinsic stiffness, the Reuss and HS lower bounds may be significantly different (\citealt{Watt1976}). 

Bounds can be used as a first approach to upscaling under certain geometries. If a fracture network completely percolates a matrix then it will have the maximum effect of weakening the rock (\citealt{Watt1976}). The effective bulk modulus of the composite will then coincide with the HS lower bound (\citealt{Boucher1974}; \citealt{Watt1976}). In the general case fractures are likely to have an associated stiffness (\citealt{Bandis1983}). We thus recommend using the HS lower bound over the Reuss average as a first approach to upscaling for such geometries. This procedure is also in line with the assumptions built into the continuum approach: The continuum assumption is linked to ones ability to define an REV over which properties can be averaged. For a fractured system, such an REV cannot be justified if the system is poorly connected (\citealt{Berkowitz2002}). Use of the HS lower bound, as a first approach to moduli averaging, thus supports the notion of a well-connected isotropic dense fracture network, over which an REV could be defined. 

When the underlying composite geometry precludes the use of bounds as methods for upscaling, one must use other methods of averaging. 
Comprehensive summaries of such approaches can be found in the works of \cite{Aboudi1992}, \cite{Nemat1993} and \cite{Torquato2002} for example.
               
\section{Qualitative analysis using the Mandel problem}  
We use solutions to the double-porosity Mandel-problem (\citealt{Nguyen2010}), to investigate the physical impacts of different coefficient modelling concepts and assumptions, on the poromechanical response of a dual-continuum material. We consider implicit assumptions, where the constitutive model starts with no pressure coupling, and explicit assumptions, where the full constitutive model, \cref{eqn:17} and \crefrange{eqn:23}{eqn:24}, is the starting point but pressure coupling is neglected leading to bounds being used for the calculation of $K_{dr}$ (and thus the calculation of the effective constitutive coefficients).

We first investigate the effects of considering the fractured dual-medium as a void space inclusion composite or stiff inclusion composite as assumed by the coefficient models of \cite{Khalili1996} and \cite{Berryman2002a} respectively. Second, we study the effects of decoupling assumptions. 

\subsection{Double-porosity Mandel problem}
\begin{figure}[h]
\centering
\includegraphics[scale = 0.6]{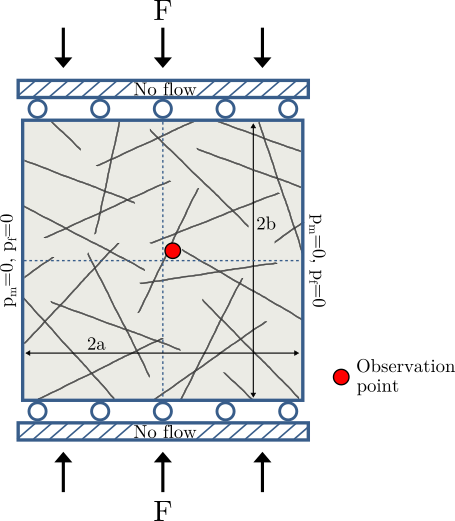}
\caption{Dual-continuum Mandel problem setup.} \label{fig:4}
\end{figure}

The problem geometry is described as an infinitely long (rectangular) cuboid domain such that the plane-strain condition holds (i.e. $u_y=0$ and $q_{m,y}=q_{f,y}=0$) (\cref{fig:4}). The domain is sandwiched between two impermeable, rigid plates, and is free to displace both laterally and vertically. A constant compressive force, $\int^a_{-a} \sigma_{zz} \text{ } dx= -2Fa$, is applied at the rigid plate boundaries, $\Gamma_N$ and $\Gamma_S$ (north and south boundaries respectively). The east and west boundaries, $\Gamma_E$ and $\Gamma_W$ respectively, are then free to drain such that $p_m=p_f=0$ at these boundaries. 

In summary the boundary conditions are,

\begin{equation}
\begin{split}
\sigma_{xx} = \sigma_{xz} = p_m = p_f = 0 \quad &\text{on } \Gamma_E \cup \Gamma_W, \\
\sigma_{xz} = q_{m,z} = q_{f,z} = 0 \quad &\text{on } \Gamma_N \cup \Gamma_S, \\
\int^a_{-a} \sigma_{zz} \text{ } dx = -2Fa \quad &\text{on } \Gamma_N \cup \Gamma_S. 
\end{split} \label{eqn:68}
\end{equation}

\noindent
Due to the symmetry of the problem only a quarter of the domain need be considered.

For an isotropic double-porosity material \cref{eqn:17} and \cref{eqn:20} can be re-written and extended to

\begin{equation}
\text{d}\bm{\sigma} = 2G\bm{\epsilon} + \lambda\epsilon\bm{1} -\sum\limits_{\alpha=m,f}b_\alpha \text{d}p_\alpha\bm{1} \label{eqn:69}
\end{equation}

\noindent
where $\lambda=\frac{2Gv}{1-2v}$ is the Lam{\'{e}} constant, in which $v$ is Poisson's ratio. Substitution of \cref{eqn:69} and \cref{eqn:2} in \cref{eqn:1} yields  

\begin{equation}
G\nabla^2\bm{u} + (\lambda + G)\nabla\epsilon = \sum\limits_{\alpha=m,f}b_\alpha\nabla p_\alpha\bm{1} - \rho\bm{g}, \label{eqn:70}
\end{equation}   

\noindent
where we assume $\overline{\gamma}\approx0$. It can be shown that $\nabla^2\bm{u}=\nabla\epsilon$, with which, and in the absence of body forces, \cref{eqn:70} reduces to  

\begin{equation}
\nabla\epsilon = \sum\limits_{\alpha=m,f}c_\alpha\nabla p_\alpha, \label{eqn:71}
\end{equation}

\noindent
where $c_\alpha=\frac{b_\alpha(1-2v)}{2G(1-v)}$ is the consolidation coefficient belonging to material $\alpha$.

Integration of \cref{eqn:71} leads to 

\begin{equation}
\epsilon = \sum\limits_{i=m,f} c_\alpha p_\alpha + f(t), \label{eqn:72}
\end{equation}

\noindent
where $f(t)$ is an integration function. Use of \cref{eqn:72} in \cref{eqn:34} leads to a set of diffusion equations written entirely in terms of continuum pressures. Solutions to the resulting system of equations are presented in \cite{Nguyen2010}, and \cite{Mehrabian2014}. Further, these solutions allow for calculation of vertical stress and strain (see \citealt{Nguyen2010} or \citealt{Mehrabian2014}).    

\subsection{Data for analysis}
For the qualitative analysis we use a quarter of a $2\text{ m}\times 2\text{ m}$ deformable porous domain. The studied domain is subjected to a constant top boundary force, $\int^1_0 -2\times 10^6 \text{ Pa }dx = -2 \text{ MPa.m}$. Where possible we use values for material properties that are typically encountered in naturally fractured carbonate reservoirs. The mechanical properties are then $K_{m}=20 \text{ GPa}$, $K_{s}=70 \text{ GPa}$, and $v=0.2$ (\citealt{Wang2000}). Values for $K_{dr}$ and $K_f$ are problem dependent. In cases where no rigorous justification is used, we choose values for $K_{dr}$ and $K_f$ arbitrarily (denoted by a superscript $\dagger$). Fluid properties are for that of water; $\rho^0_l = 1000 \text{ kg m}^{\text{-3}}$, $\mu_l = 1 \text{ cp}$ and $K_l = 2.3 \text{ GPa}$. Rock properties related to fluid storage and flow are $\phi^0_m=0.05$, $k_m=0.01 \text{ md}$ and $\phi^0_f=0.01$ (\citealt{Nelson2001}). Effective permeability of the fracture is assumed to be $k_f\approx1000 \text{ md}$. With the cubic law (\citealt{Witherspoon1980}), this corresponds to an aperture, $a_f=7\times10^{-5}\text{ m}$, and fracture spacing, $d=2.8\times 10^{-2} \text{ m}$. 

\subsection{Test cases}
We consider one test case to investigate the differences between void space and stiff inclusion composite coefficient models, and three test cases to investigate implicit and explicit decoupling assumptions. In each case, analytical solutions to the double-porosity Mandel problem are compared. Differences in solutions for each case then arise due the parameter permutations described in the case descriptions that follow.  

\subsubsection{Case 1: Intrinsic fracture properties}
In Case 1 we are interested in comparing the differences that arise when considering intrinsic fracture properties. In particular, it is of interest to investigate if coefficient models from \cite{Khalili1996} could still be used even when a fracture has an associated phase stiffness.  

From \crefrange{tab:1}{tab:2}, we hypothesise that provided that $\phi^*_f\approx1$ and the fracture phase stiffness is orders of magnitude lower than the grain stiffness ($K_f\ll K^f_s$), effects arising due to deviations of intrinsic fracture coefficients under the void space assumption (i.e. deviations from: $b^*_f=1$, $\frac{1}{M^*_\alpha} = \frac{1}{K_l}$ and $B^*_\alpha = 1$) are  negligible. Void space coefficient models could then be used in place of coefficient models that include intrinsic fracture properties. When $\phi^*_f=1$ we use coefficient models from \cite{Khalili1996}, and when $\phi^*_f<1$ we use coefficient models from \cite{Berryman2002a} with $K^f_s=K_s$. 

Strictly speaking, use of void space coefficient models implies $K_f=0$. However, the aim of this test case is to highlight the effect of missing physics by not including intrinsic poromechanical parameters within coefficient model formulations. To do this, and to ensure non-zero values of $K_{dr}$, we use the same upscaled bulk modulus for both sets of coefficient models, which is calculated with a non-zero $K_f$. We consider a composite medium with a network of fractures that completely dissociates the matrix. Upscaling is then done through the HS lower bound. To test our hypothesis we use combinations of various values of $\phi^*_f$ and $K_f$. 

We compare results from coefficient models calculated using a fracture modulus several orders of magnitude lower than the solid grain modulus, ($K^{\dagger}_f=\frac{K^f_s}{1750}$), versus ones that use a fracture modulus only an order of magnitude lower than the solid modulus, ($K^{\dagger}_f=\frac{K^f_s}{35}$). 

\subsubsection{Case 2: Implicit decoupling assumptions}
For Case 2 we investigate the impact of implicit decoupling assumptions. This is particularly poignant considering the use of such decoupled constitutive models in the recent works of \cite{Alberto2019} and \cite{Hajiabadi2019}. To mimic implicit assumptions we consider $\frac{1}{Q}=0$ and $A_{23}=0$ and make no acknowledgement of these assumptions with respect to relations between mechanical properties. When considering $\frac{1}{Q}=0$ and $A_{23}=0$ we use coefficient models from \cite{Khalili1996} and \cite{Berryman2002a}, with $K_f=0$, respectively. We reference these results against ones for which no decoupling is made with coefficient models from \cite{Khalili1996}.

Use of the void space models implies $K_f=0$. Since we do not enforce relations on $K_{dr}$ (a fundamental difference between the implicit assumptions considered here and explicit assumptions) we take an arbitrary value of $K^{\dagger}_{dr}=10 \text{ GPa}$. As a result bounds on the effective constitutive coefficients are not enforced.  

\subsubsection{Case 3: Explicit decoupling assumption - isostrain}
We investigate the effect of assuming isostrain at the microscale whilst making use of coefficient models from \cite{Khalili1996}. Under isostrain the composite and constituent bulk moduli are linked by the Voigt average ($K^V_{dr}= 19.8 \text{ GPa}$).  This leads naturally to bounds on the effective constitutive coefficients, with $\frac{1}{Q}=0$ representing an upper bound. We compare the isostrain results to those computed when using coefficient models with a composite bulk modulus coinciding with the HS upper bound and an arbitrary value ($K^{HS+}_{dr}=19.5 \text{ GPa}$ and $K^{\dagger}_{dr}=10 \text{ GPa}$ respectively). 

Further, we investigate the disparity between results when using the void space coefficient models of \cite{Borja2009} under the assumption of $\pderiv{\psi_\alpha}{t}=0$, and \cite{Khalili1996} under the assumption of incompressible grain isostrain. We use $K^V_{dr}= 19.8 \text{ GPa}$ and $K_s=\infty$ for both sets of coefficient models in this latter isostrain investigation.   

\subsubsection{Case 4: Explicit decoupling assumption - isostress}
For Case 4 we study the effect of assuming isostress at the microscale. In previous works the coefficient models of \cite{Berryman2002a} have been used with an explicit decoupling assumption ($A_{23}=0$) that implies isostress (\citealt{Kim2012}; \citealt{Mehrabian2014}; \citealt{Mehrabian2018}). 

To avoid cases where $K_{dr}=0$ we consider the fracture phase to have the following properties: $\phi^*_f=0.7$ with $K^f_s=K_s$ and $K^{\dagger}_f=\frac{K_m}{500}$. Coefficient models from \cite{Berryman2002a} are then used. 

We compare results arising when calculating the composite bulk modulus with the Reuss average ($K^R_{dr}=3.3 \text{ GPa}$), the HS lower bound ($K^{HS-}_{dr}=5.7 \text{ GPa}$), and an arithmetic average of the HS bounds ($K^{AHS}_{dr}=12.7 \text{ GPa}$). The latter modulus is tested in analogy to a dual system with inclusions that do not have the maximum weakening effect on the host material. One example would be a fracture system composed of a network of open and closed fractures. Another would be aggregate material. 
    
\section{Results and discussions}
In the following, we show results of the test cases described above for the double-porosity Mandel problem. Results are given in terms in evolutions of matrix and fracture pressures, and vertical strain with time. 

To aid in our analysis for pressure and vertical strain we introduce the following notions of the \textit{instantaneous} problem and the \textit{time-dependent} problem. In both cases, mechanical equilibrium is governed, for this system, by \cref{eqn:70}. In the instantaneous problem, fluid pressure for continuum $\alpha$ can be shown to be a state function of total stress and fluid pressure $\beta$. In the time-dependent problem, continuum pressures are governed by way of the diffusion equation, \cref{eqn:34}. 

The instantaneous problem considers the change of the system from an unloaded state, $t(0)$, to a loaded state, $t(0^+)$, upon application of instantaneous loading. Under such conditions the rate of loading is infinitely faster than the rate of inter-continuum fluid transfer (\citealt{Coussy2004}). Consequently, each continuum is undrained, $m^{t(0^+)}_{l,\alpha}=m^{t(0)}_{l,\alpha}$. From \cref{eqn:23} and \cref{eqn:24}, together with the undrained condition ($\text{d}m_{l,\alpha}=0$) we can recover

\begin{equation}
\text{d}p_\alpha = \left(\frac{1}{M_\alpha} - \frac{M_\beta}{(Q)^2}\right)^{-1} \left(\frac{M_\beta b_\beta}{Q}-b_\alpha\right)\epsilon. \label{eqn:73}
\end{equation} 

\noindent
We note that $\epsilon_{zz} \propto \epsilon$. Substitution of \cref{eqn:17} into \cref{eqn:73} leads to   

\begin{equation}
\text{d}p_\alpha = -\frac{M_\alpha b_\alpha}{\{K_{dr} + M_\alpha(b_\alpha)^2\}}\left\lbrace \text{d}\sigma + \left( b_\beta +\frac{K_{dr}}{b_\alpha Q} \right) \text{d}p_\beta \right\rbrace. \label{eqn:74}
\end{equation}   

\subsection{Case 1: Intrinsic fracture properties}
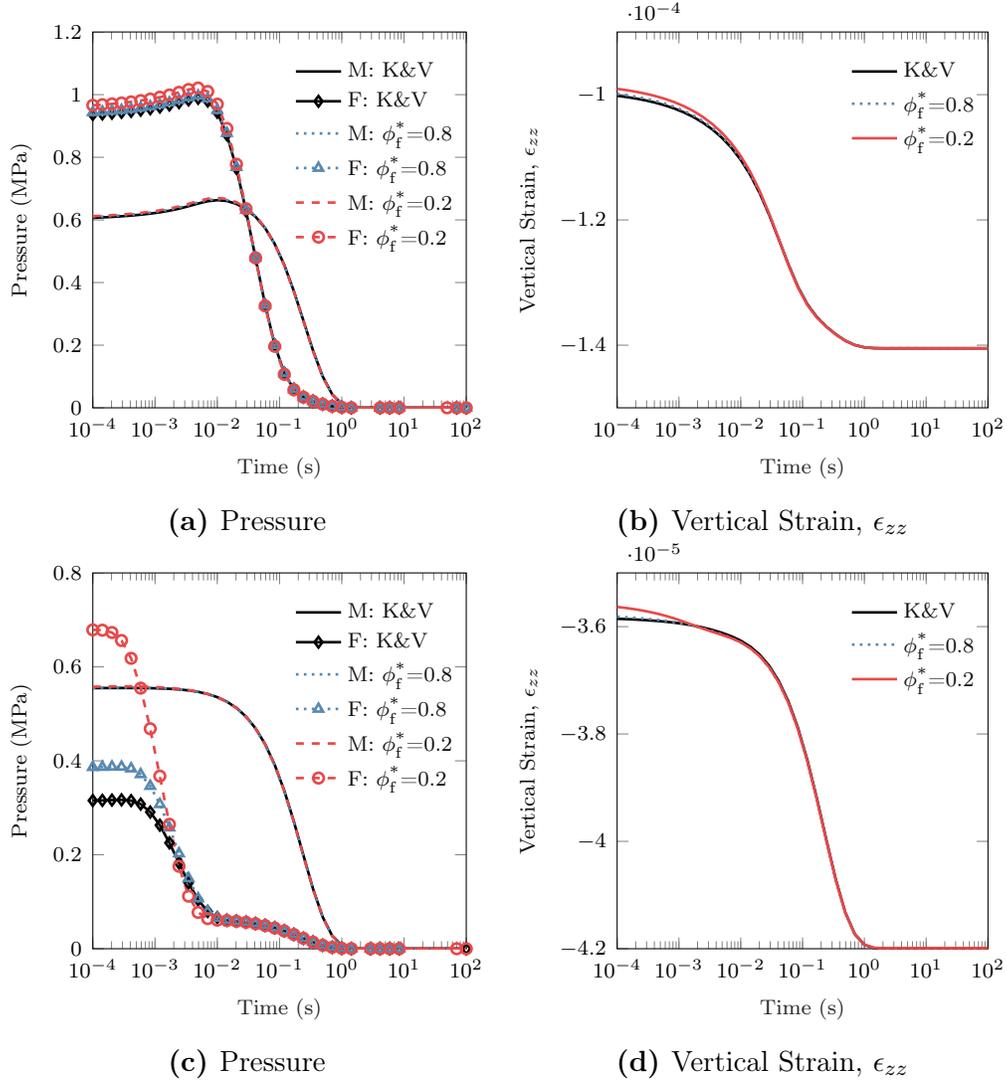
\begin{figure}[ht!]
\centering
\begin{minipage}[b]{0.5\textwidth}
\centering
\setlength\figureheight{5cm}
\setlength\figurewidth{5cm}
\definecolor{mycolor1}{rgb}{0.34667,0.53600,0.69067}%
\definecolor{mycolor2}{rgb}{0.91529,0.28157,0.28784}%
\begin{tikzpicture}

\begin{axis}[%
width=0.993\figurewidth,
height=\figureheight,
at={(0\figurewidth,0\figureheight)},
scale only axis,
xmode=log,
xmin=0.0001,
xmax=100,
xminorticks=true,
xlabel style={font=\fontsize{8}{144}\selectfont\color{white!15!black}},
xlabel={Time (s)},
ymin=0,
ymax=1.2,
ylabel style={font=\fontsize{8}{144}\selectfont\color{white!5!black}},
ylabel={Pressure (MPa)},
ticklabel style={font=\fontsize{8}{144}},
axis background/.style={fill=white},
legend style={font=\fontsize{8}{144}\selectfont\color{white!5!black}, at={(1,0.95)}, anchor=north east, legend cell align=left, align=left, fill=none, draw=none}
]
\addplot [color=black, line width=1pt]
  table[row sep=crcr]{%
0.0001	0.606802317916914\\
0.0001425102670303	0.608096401840264\\
0.000203091762090473	0.609660583010374\\
0.000289426612471675	0.611555490766611\\
0.000412462638290135	0.613857089183898\\
0.000587801607227491	0.616661590081352\\
0.000837677640068292	0.620091531721799\\
0.00119377664171444	0.624298310971091\\
0.00170125427985259	0.629475863452012\\
0.00242446201708233	0.635897341189309\\
0.00345510729459222	0.643731921331288\\
0.00492388263170674	0.652344445580647\\
0.00701703828670383	0.659715122797999\\
0.01	0.663084837403345\\
0.01425102670303	0.66044092607879\\
0.0203091762090473	0.651108017291987\\
0.0289426612471675	0.634730095144185\\
0.0412462638290135	0.609903525581246\\
0.0587801607227491	0.573863480920987\\
0.0837677640068292	0.523313064660928\\
0.119377664171444	0.45596144189249\\
0.170125427985259	0.372424427857229\\
0.242446201708233	0.277885303340767\\
0.345510729459222	0.182694370850875\\
0.492388263170674	0.100556406417325\\
0.701703828670383	0.0430325409452169\\
1	0.0126511009256802\\
1.425102670303	0.00183625130262449\\
2.03091762090473	-0.000111618590252783\\
2.89426612471675	2.3214426068051e-05\\
4.12462638290135	0.000118906682983421\\
5.87801607227491	6.50922376824674e-05\\
8.37677640068292	1.31592675304516e-05\\
11.9377664171444	-4.71035354096015e-06\\
17.0125427985259	-5.97136200419977e-06\\
24.2446201708233	-3.93893366668888e-06\\
34.5510729459222	-1.59524198771144e-06\\
49.2388263170674	-1.04739117638748e-06\\
70.1703828670382	-3.4347846627055e-07\\
100	4.12068375768083e-08\\
};
\addlegendentry{$\text{M: }\text{K\&V}$}

\addplot [color=black, line width=1pt, mark size=2pt, mark=diamond, mark options={solid, black}]
  table[row sep=crcr]{%
0.0001	0.936644360013544\\
0.0001425102670303	0.93851963651818\\
0.000203091762090473	0.940759393570696\\
0.000289426612471675	0.943434653269694\\
0.000412462638290135	0.946630406258331\\
0.000587801607227491	0.950450026660674\\
0.000837677640068292	0.955018559647716\\
0.00119377664171444	0.960462590026157\\
0.00170125427985259	0.966926579749137\\
0.00242446201708233	0.974671351673465\\
0.00345510729459222	0.983216089286014\\
0.00492388263170674	0.988302495634242\\
0.00701703828670383	0.979553331424453\\
0.01	0.943728720380892\\
0.01425102670303	0.872095969117007\\
0.0203091762090473	0.764787153766549\\
0.0289426612471675	0.629338038000321\\
0.0412462638290135	0.478162215954929\\
0.0587801607227491	0.328501772590595\\
0.0837677640068292	0.200809917126263\\
0.119377664171444	0.110739880023332\\
0.170125427985259	0.0595398140665441\\
0.242446201708233	0.0346042759520285\\
0.345510729459222	0.0212027584293762\\
0.492388263170674	0.01177982510368\\
0.701703828670383	0.00514021560654755\\
1	0.00152749421383989\\
1.425102670303	0.000217972126835761\\
2.03091762090473	-2.20863328087223e-05\\
2.89426612471675	-5.44489628766133e-06\\
4.12462638290135	9.08188794390194e-06\\
5.87801607227491	5.45736221413732e-06\\
8.37677640068292	7.7940250148997e-07\\
11.9377664171444	-7.63250479372666e-07\\
17.0125427985259	-7.06658043269764e-07\\
24.2446201708233	-4.42458851202061e-07\\
34.5510729459222	-1.59035225716338e-07\\
49.2388263170674	-8.41080998587008e-08\\
70.1703828670382	-3.05211029926722e-08\\
100	3.70886254352126e-08\\
};
\addlegendentry{$\text{F: }\text{K\&V}$}

\addplot [color=mycolor1, dotted, line width=1pt]
  table[row sep=crcr]{%
0.0001	0.609757965243391\\
0.0001425102670303	0.611075821316171\\
0.000203091762090473	0.612668751420063\\
0.000289426612471675	0.614598502544649\\
0.000412462638290135	0.616942453728741\\
0.000587801607227491	0.619798632314265\\
0.000837677640068292	0.623291833379932\\
0.00119377664171444	0.627576047922735\\
0.00170125427985259	0.63284917521493\\
0.00242446201708233	0.639389377387162\\
0.00345510729459222	0.647356009460245\\
0.00492388263170674	0.656063515445244\\
0.00701703828670383	0.663410258751225\\
0.01	0.666576585078975\\
0.01425102670303	0.663548129413817\\
0.0203091762090473	0.653686848693296\\
0.0289426612471675	0.636671246337198\\
0.0412462638290135	0.611127626302579\\
0.0587801607227491	0.574341886739815\\
0.0837677640068292	0.523101887227152\\
0.119377664171444	0.455207223543833\\
0.170125427985259	0.371328232604537\\
0.242446201708233	0.276662079920756\\
0.345510729459222	0.181546151792828\\
0.492388263170674	0.0996679665748382\\
0.701703828670383	0.042494120230939\\
1	0.0124198266561414\\
1.425102670303	0.00177826909450366\\
2.03091762090473	-0.000113847241103709\\
2.89426612471675	2.65498244518816e-05\\
4.12462638290135	0.000118992449001025\\
5.87801607227491	6.50149336033607e-05\\
8.37677640068292	1.29560562637201e-05\\
11.9377664171444	-5.67933220160298e-06\\
17.0125427985259	-6.05091199556773e-06\\
24.2446201708233	-3.27099681656012e-06\\
34.5510729459222	-1.11362808007232e-06\\
49.2388263170674	-8.16410895475293e-07\\
70.1703828670382	2.74343339115993e-07\\
100	2.5343557359388e-08\\
};
\addlegendentry{$\text{M: }\phi{}^\text{*}_\text{f}\text{=0.8}$}

\addplot [color=mycolor1, dotted, line width=1pt, mark size=2pt, mark=triangle, mark options={solid, mycolor1}]
  table[row sep=crcr]{%
0.0001	0.943539671561516\\
0.0001425102670303	0.945454723663178\\
0.000203091762090473	0.947742142492042\\
0.000289426612471675	0.95047454159764\\
0.000412462638290135	0.953738868230124\\
0.000587801607227491	0.957640962122487\\
0.000837677640068292	0.962308635764501\\
0.00119377664171444	0.96787086913345\\
0.00170125427985259	0.974477621897408\\
0.00242446201708233	0.982394517403813\\
0.00345510729459222	0.991088875902925\\
0.00492388263170674	0.996128614790577\\
0.00701703828670383	0.986915730716067\\
0.01	0.950127658835314\\
0.01425102670303	0.87715476060825\\
0.0203091762090473	0.768334659321792\\
0.0289426612471675	0.631367817326198\\
0.0412462638290135	0.478823154720874\\
0.0587801607227491	0.328150555849665\\
0.0837677640068292	0.199978258626438\\
0.119377664171444	0.109924274547317\\
0.170125427985259	0.0589749969028124\\
0.242446201708233	0.0342642680056269\\
0.345510729459222	0.0209874953416345\\
0.492388263170674	0.0116390914639274\\
0.701703828670383	0.00506098009635363\\
1	0.00149512393557153\\
1.425102670303	0.000210131386482231\\
2.03091762090473	-2.25228142907528e-05\\
2.89426612471675	-5.11935869973785e-06\\
4.12462638290135	9.05911477070062e-06\\
5.87801607227491	5.32271457019848e-06\\
8.37677640068292	7.12056281884202e-07\\
11.9377664171444	-8.04446043199991e-07\\
17.0125427985259	-6.84543484026135e-07\\
24.2446201708233	-3.41231462775549e-07\\
34.5510729459222	-1.41732449169663e-07\\
49.2388263170674	-1.18586531339227e-07\\
70.1703828670382	6.5298236281224e-08\\
100	7.46683069137476e-08\\
};
\addlegendentry{$\text{F: }\phi{}^\text{*}_\text{f}\text{=0.8}$}

\addplot [color=mycolor2, dashed, line width=1pt]
  table[row sep=crcr]{%
0.0001	0.612002916683659\\
0.0001425102670303	0.613379445848091\\
0.000203091762090473	0.615043729157239\\
0.000289426612471675	0.617060538619289\\
0.000412462638290135	0.619511098659274\\
0.000587801607227491	0.622498507676134\\
0.000837677640068292	0.626153831226809\\
0.00119377664171444	0.63063853220457\\
0.00170125427985259	0.636162793522515\\
0.00242446201708233	0.643018001512987\\
0.00345510729459222	0.651329534795222\\
0.00492388263170674	0.660280605991457\\
0.00701703828670383	0.667608993852572\\
0.01	0.670453035882391\\
0.01425102670303	0.666909614788159\\
0.0203091762090473	0.656476544320742\\
0.0289426612471675	0.638875777788901\\
0.0412462638290135	0.612712060378388\\
0.0587801607227491	0.575283446518822\\
0.0837677640068292	0.523452924278351\\
0.119377664171444	0.455122763122387\\
0.170125427985259	0.371022481378664\\
0.242446201708233	0.276316228375616\\
0.345510729459222	0.181266651001559\\
0.492388263170674	0.0994839501506136\\
0.701703828670383	0.0424001442488052\\
1	0.0123861112350273\\
1.425102670303	0.00177093447299403\\
2.03091762090473	-0.000115478396354704\\
2.89426612471675	2.56862244034212e-05\\
4.12462638290135	0.000118295425989188\\
5.87801607227491	6.42883752134464e-05\\
8.37677640068292	1.32900061120682e-05\\
11.9377664171444	-5.65733981573415e-06\\
17.0125427985259	-6.44393393477877e-06\\
24.2446201708233	-3.66613067224766e-06\\
34.5510729459222	-1.58987472801022e-06\\
49.2388263170674	-5.88245309717682e-07\\
70.1703828670382	-7.22638191273132e-08\\
100	-1.85877163849376e-08\\
};
\addlegendentry{$\text{M: }\phi{}^\text{*}_\text{f}\text{=0.2}$}

\addplot [color=mycolor2, dashed, line width=1pt, mark size=2pt, mark=o, mark options={solid, mycolor2}]
  table[row sep=crcr]{%
0.0001	0.965983245265953\\
0.0001425102670303	0.968021620772473\\
0.000203091762090473	0.970456531914383\\
0.000289426612471675	0.973365384240101\\
0.000412462638290135	0.97684090483693\\
0.000587801607227491	0.98099628082832\\
0.000837677640068292	0.98596709333171\\
0.00119377664171444	0.991888430993198\\
0.00170125427985259	0.99892758570496\\
0.00242446201708233	1.00735858224756\\
0.00345510729459222	1.01644219113443\\
0.00492388263170674	1.0211184420644\\
0.00701703828670383	1.01004408663432\\
0.01	0.969680977053261\\
0.01425102670303	0.891918354610913\\
0.0203091762090473	0.777844388854002\\
0.0289426612471675	0.63576378768724\\
0.0412462638290135	0.478842745961797\\
0.0587801607227491	0.325295828269648\\
0.0837677640068292	0.196286311632895\\
0.119377664171444	0.107086048486902\\
0.170125427985259	0.0575288773341831\\
0.242446201708233	0.0337708589103904\\
0.345510729459222	0.0208562396100738\\
0.492388263170674	0.0115863240338469\\
0.701703828670383	0.00502930380771323\\
1	0.00148062769107307\\
1.425102670303	0.000205890794830852\\
2.03091762090473	-2.31682839442147e-05\\
2.89426612471675	-5.19229507010087e-06\\
4.12462638290135	9.05512543370314e-06\\
5.87801607227491	5.45424107924804e-06\\
8.37677640068292	7.66792576876389e-07\\
11.9377664171444	-7.85084275356393e-07\\
17.0125427985259	-6.81842896200177e-07\\
24.2446201708233	-3.70561365156208e-07\\
34.5510729459222	-1.78876050723424e-07\\
49.2388263170674	3.31588472041195e-09\\
70.1703828670382	-4.40558749415926e-08\\
100	4.38085638558718e-08\\
};
\addlegendentry{$\text{F: }\phi{}^\text{*}_\text{f}\text{=0.2}$}

\end{axis}
\end{tikzpicture}
\subcaption{Pressure}\label{fig:5a}
\end{minipage}%
\begin{minipage}[b]{0.5\textwidth}
\centering
\setlength\figureheight{5cm}
\setlength\figurewidth{5cm}
\definecolor{mycolor1}{rgb}{0.34667,0.53600,0.69067}%
\definecolor{mycolor2}{rgb}{0.91529,0.28157,0.28784}%
\begin{tikzpicture}

\begin{axis}[%
width=0.986\figurewidth,
height=\figureheight,
at={(0\figurewidth,0\figureheight)},
scale only axis,
xmode=log,
xmin=0.0001,
xmax=100,
xminorticks=true,
xlabel style={font=\fontsize{8}{144}\selectfont\color{white!15!black}},
xlabel={Time (s)},
ymin=-0.00015,
ymax=-9e-05,
ylabel style={font=\fontsize{8}{144}\selectfont\color{white!5!black}},
ylabel={$\text{Vertical Strain, }\epsilon_{zz}$},
ticklabel style={font=\fontsize{8}{144}},
axis background/.style={fill=white},
legend style={font=\fontsize{8}{144}\selectfont\color{white!5!black}, at={(1,0.95)}, anchor=north east, legend cell align=left, align=left, fill=none, draw=none}
]
\addplot [color=black, line width=1pt]
  table[row sep=crcr]{%
0.0001	-0.000100175284971938\\
0.0001425102670303	-0.000100379344256491\\
0.000203091762090473	-0.000100623880830242\\
0.000289426612471675	-0.00010091713730982\\
0.000412462638290135	-0.000101269123586019\\
0.000587801607227491	-0.000101692036089564\\
0.000837677640068292	-0.000102200781728247\\
0.00119377664171444	-0.000102813655838594\\
0.00170125427985259	-0.000103553214406843\\
0.00242446201708233	-0.000104447419053629\\
0.00345510729459222	-0.00010553115283328\\
0.00492388263170674	-0.000106848098643118\\
0.00701703828670383	-0.000108453051596924\\
0.01	-0.000110415923394478\\
0.01425102670303	-0.000112823598038625\\
0.0203091762090473	-0.000115758474509381\\
0.0289426612471675	-0.00011923433123999\\
0.0412462638290135	-0.000123112678275139\\
0.0587801607227491	-0.000127061526244125\\
0.0837677640068292	-0.00013062829397076\\
0.119377664171444	-0.000133456054296233\\
0.170125427985259	-0.000135506162418797\\
0.242446201708233	-0.00013702770534771\\
0.345510729459222	-0.00013827019608386\\
0.492388263170674	-0.000139281130475125\\
0.701703828670383	-0.000139981657679139\\
1	-0.000140348891989312\\
1.425102670303	-0.000140477836760748\\
2.03091762090473	-0.000140500378044133\\
2.89426612471675	-0.000140498561869509\\
4.12462638290135	-0.000140497420558101\\
5.87801607227491	-0.000140498046226726\\
8.37677640068292	-0.000140498638659803\\
11.9377664171444	-0.000140498847803837\\
17.0125427985259	-0.000140498836694099\\
24.2446201708233	-0.000140498819727223\\
34.5510729459222	-0.000140498788844835\\
49.2388263170674	-0.000140498779424399\\
70.1703828670382	-0.000140498777664904\\
100	-0.000140498778396201\\
};
\addlegendentry{$\text{K\&V}$}

\addplot [color=mycolor1, dotted, line width=1pt]
  table[row sep=crcr]{%
0.0001	-9.98849609174901e-05\\
0.0001425102670303	-0.000100091189339698\\
0.000203091762090473	-0.000100338337692669\\
0.000289426612471675	-0.000100634744516319\\
0.000412462638290135	-0.000100990537302552\\
0.000587801607227491	-0.000101418061668601\\
0.000837677640068292	-0.000101932406775331\\
0.00119377664171444	-0.000102552101248096\\
0.00170125427985259	-0.000103299998315807\\
0.00242446201708233	-0.00010420443979088\\
0.00345510729459222	-0.000105300798024075\\
0.00492388263170674	-0.00010663339947103\\
0.00701703828670383	-0.000108257871219455\\
0.01	-0.000110245259749081\\
0.01425102670303	-0.000112683698843408\\
0.0203091762090473	-0.00011565595413292\\
0.0289426612471675	-0.000119173865350438\\
0.0412462638290135	-0.000123094004267064\\
0.0587801607227491	-0.000127077414807536\\
0.0837677640068292	-0.000130665259211174\\
0.119377664171444	-0.000133499468866193\\
0.170125427985259	-0.000135546210152494\\
0.242446201708233	-0.000137060799392506\\
0.345510729459222	-0.00013829575065367\\
0.492388263170674	-0.000139298951705919\\
0.701703828670383	-0.000139992368774991\\
1	-0.000140354505919707\\
1.425102670303	-0.000140480905892293\\
2.03091762090473	-0.000140502728320742\\
2.89426612471675	-0.000140500858399845\\
4.12462638290135	-0.000140499755883859\\
5.87801607227491	-0.000140500389720136\\
8.37677640068292	-0.000140500974901692\\
11.9377664171444	-0.000140501166670304\\
17.0125427985259	-0.000140501176936476\\
24.2446201708233	-0.000140501140261078\\
34.5510729459222	-0.00014050111330046\\
49.2388263170674	-0.00014050110764946\\
70.1703828670382	-0.000140501093140025\\
100	-0.000140501092451787\\
};
\addlegendentry{$\phi{}^\text{*}_\text{f}\text{=0.8}$}

\addplot [color=mycolor2, line width=1pt]
  table[row sep=crcr]{%
0.0001	-9.90737849989282e-05\\
0.0001425102670303	-9.92865545605994e-05\\
0.000203091762090473	-9.95415862383761e-05\\
0.000289426612471675	-9.9847508706718e-05\\
0.000412462638290135	-0.000100214811468652\\
0.000587801607227491	-0.000100656289604622\\
0.000837677640068292	-0.000101187600383441\\
0.00119377664171444	-0.000101827983621164\\
0.00170125427985259	-0.00010260120334698\\
0.00242446201708233	-0.000103536762815254\\
0.00345510729459222	-0.000104671545645048\\
0.00492388263170674	-0.000106051807851683\\
0.00701703828670383	-0.000107735700755733\\
0.01	-0.000109797659206625\\
0.01425102670303	-0.000112329147642694\\
0.0203091762090473	-0.000115412298699662\\
0.0289426612471675	-0.00011904962620535\\
0.0412462638290135	-0.000123078651106338\\
0.0587801607227491	-0.000127136741730875\\
0.0837677640068292	-0.000130749923596496\\
0.119377664171444	-0.000133567457561257\\
0.170125427985259	-0.000135582791898455\\
0.242446201708233	-0.000137075004692896\\
0.345510729459222	-0.000138300867455917\\
0.492388263170674	-0.000139301534578842\\
0.701703828670383	-0.00013999380957301\\
1	-0.000140355095621846\\
1.425102670303	-0.000140481067362819\\
2.03091762090473	-0.000140502754005741\\
2.89426612471675	-0.000140500873826826\\
4.12462638290135	-0.000140499763946467\\
5.87801607227491	-0.000140500386148281\\
8.37677640068292	-0.000140500971906318\\
11.9377664171444	-0.000140501177084945\\
17.0125427985259	-0.000140501171429055\\
24.2446201708233	-0.000140501141251092\\
34.5510729459222	-0.000140501122163881\\
49.2388263170674	-0.000140501104170396\\
70.1703828670382	-0.000140501097025764\\
100	-0.000140501094384694\\
};
\addlegendentry{$\phi{}^\text{*}_\text{f}\text{=0.2}$}

\end{axis}
\end{tikzpicture}
\subcaption{Vertical Strain, $\epsilon_{zz}$}\label{fig:5b}
\end{minipage}
\begin{minipage}[b]{0.5\textwidth}
\centering
\setlength\figureheight{5cm}
\setlength\figurewidth{5cm}
\definecolor{mycolor1}{rgb}{0.34667,0.53600,0.69067}%
\definecolor{mycolor2}{rgb}{0.91529,0.28157,0.28784}%
\begin{tikzpicture}

\begin{axis}[%
width=0.993\figurewidth,
height=\figureheight,
at={(0\figurewidth,0\figureheight)},
scale only axis,
xmode=log,
xmin=0.0001,
xmax=100,
xminorticks=true,
xlabel style={font=\fontsize{8}{144}\selectfont\color{white!15!black}},
xlabel={Time (s)},
ymin=0,
ymax=0.8,
ylabel style={font=\fontsize{8}{144}\selectfont\color{white!5!black}},
ylabel={Pressure (MPa)},
ticklabel style={font=\fontsize{8}{144}},
ylabel={Pressure (MPa)},
axis background/.style={fill=white},
legend style={font=\fontsize{8}{144}\selectfont\color{white!5!black}, at={(1,0.95)}, anchor=north east, legend cell align=left, align=left, fill=none, draw=none}
]
\addplot [color=black, line width=1pt]
  table[row sep=crcr]{%
0.0001	0.554845051481069\\
0.0001425102670303	0.554898883544811\\
0.000203091762090473	0.554952720328124\\
0.000289426612471675	0.554998035841847\\
0.000412462638290135	0.555006018844913\\
0.000587801607227491	0.554919472881055\\
0.000837677640068292	0.554664725083925\\
0.00119377664171444	0.554163267138041\\
0.00170125427985259	0.553320726145418\\
0.00242446201708233	0.551990341891002\\
0.00345510729459222	0.549940403768927\\
0.00492388263170674	0.546827716838343\\
0.00701703828670383	0.54220669969091\\
0.01	0.535506624516648\\
0.01425102670303	0.52598228063657\\
0.0203091762090473	0.512653253034414\\
0.0289426612471675	0.494219662687583\\
0.0412462638290135	0.469065261538707\\
0.0587801607227491	0.435381629848643\\
0.0837677640068292	0.391457452177167\\
0.119377664171444	0.336310970736486\\
0.170125427985259	0.270728524069719\\
0.242446201708233	0.198559703429793\\
0.345510729459222	0.127517557601164\\
0.492388263170674	0.0678219546763053\\
0.701703828670383	0.0275536339559164\\
1	0.00741926744488852\\
1.425102670303	0.000847077419579141\\
2.03091762090473	-0.000106994276608334\\
2.89426612471675	4.30560504883466e-05\\
4.12462638290135	8.79578985731062e-05\\
5.87801607227491	4.15051488184852e-05\\
8.37677640068292	6.53667028560706e-06\\
11.9377664171444	-4.06079135486957e-06\\
17.0125427985259	-4.47310751478423e-06\\
24.2446201708233	-2.41711571479394e-06\\
34.5510729459222	-1.23774145503941e-06\\
49.2388263170674	-3.52498477033845e-07\\
70.1703828670382	-2.96056221373826e-07\\
100	-6.28680571498382e-08\\
};
\addlegendentry{$\text{M: }\text{K\&V}$}

\addplot [color=black, line width=1pt, mark size=2pt, mark=diamond, mark options={solid, black}]
  table[row sep=crcr]{%
0.0001	0.315416718453746\\
0.0001425102670303	0.315856969504272\\
0.000203091762090473	0.316472501863632\\
0.000289426612471675	0.316896959095085\\
0.000412462638290135	0.315268828055762\\
0.000587801607227491	0.307901615248898\\
0.000837677640068292	0.29096618189006\\
0.00119377664171444	0.262923744297662\\
0.00170125427985259	0.225436694294071\\
0.00242446201708233	0.182603288306261\\
0.00345510729459222	0.140060097439907\\
0.00492388263170674	0.104074144381636\\
0.00701703828670383	0.0793413983560557\\
0.01	0.0662716072536234\\
0.01425102670303	0.0610156500447769\\
0.0203091762090473	0.0587881983144864\\
0.0289426612471675	0.0566969131435001\\
0.0412462638290135	0.053775541884443\\
0.0587801607227491	0.0498056072025683\\
0.0837677640068292	0.0446552639495951\\
0.119377664171444	0.0382329823498595\\
0.170125427985259	0.0306413522910264\\
0.242446201708233	0.0223440496908126\\
0.345510729459222	0.0142448670851584\\
0.492388263170674	0.00750891264115463\\
0.701703828670383	0.0030179475541721\\
1	0.000801405583803521\\
1.425102670303	8.89229138204329e-05\\
2.03091762090473	-1.13798734187774e-05\\
2.89426612471675	5.34472448202335e-06\\
4.12462638290135	9.82949121174075e-06\\
5.87801607227491	4.52911267679185e-06\\
8.37677640068292	6.84793192268503e-07\\
11.9377664171444	-4.64140421426868e-07\\
17.0125427985259	-5.02670240824922e-07\\
24.2446201708233	-2.74778260943124e-07\\
34.5510729459222	-1.14034890164236e-07\\
49.2388263170674	-2.85028386279461e-08\\
70.1703828670382	-2.57415399561104e-08\\
100	7.58545438169255e-09\\
};
\addlegendentry{$\text{F: }\text{K\&V}$}

\addplot [color=mycolor1, dotted, line width=1pt]
  table[row sep=crcr]{%
0.0001	0.555411367055293\\
0.0001425102670303	0.555524931875323\\
0.000203091762090473	0.555653426542416\\
0.000289426612471675	0.555783058735375\\
0.000412462638290135	0.555862753453316\\
0.000587801607227491	0.555800784606306\\
0.000837677640068292	0.555496371590759\\
0.00119377664171444	0.554869739034207\\
0.00170125427985259	0.553845973383957\\
0.00242446201708233	0.552304572229937\\
0.00345510729459222	0.550039467673538\\
0.00492388263170674	0.546735181696407\\
0.00701703828670383	0.541969300206141\\
0.01	0.53517397061063\\
0.01425102670303	0.525588135780291\\
0.0203091762090473	0.512207853672122\\
0.0289426612471675	0.493704503055274\\
0.0412462638290135	0.468460168568812\\
0.0587801607227491	0.434670600503162\\
0.0837677640068292	0.390624994890727\\
0.119377664171444	0.335363200461\\
0.170125427985259	0.269699641508442\\
0.242446201708233	0.197530390887845\\
0.345510729459222	0.126609254230514\\
0.492388263170674	0.0671574105230397\\
0.701703828670383	0.0271781522941584\\
1	0.00727279284664096\\
1.425102670303	0.000815628540142482\\
2.03091762090473	-0.000107750722170921\\
2.89426612471675	4.43992708756642e-05\\
4.12462638290135	8.72190730246157e-05\\
5.87801607227491	4.09877427904567e-05\\
8.37677640068292	5.82090033761078e-06\\
11.9377664171444	-4.50600684703236e-06\\
17.0125427985259	-4.3575438120288e-06\\
24.2446201708233	-2.02181592253547e-06\\
34.5510729459222	-9.59453240953292e-07\\
49.2388263170674	-1.64730577973952e-07\\
70.1703828670382	-2.02832148091967e-07\\
100	2.63216172912849e-08\\
};
\addlegendentry{$\text{M: }\phi{}^\text{*}_\text{f}\text{=0.8}$}

\addplot [color=mycolor1, dotted, line width=1pt, mark size=2pt, mark=triangle, mark options={solid, mycolor1}]
  table[row sep=crcr]{%
0.0001	0.386448856225353\\
0.0001425102670303	0.386834037328476\\
0.000203091762090473	0.387327925250279\\
0.000289426612471675	0.387149628911579\\
0.000412462638290135	0.383413860187746\\
0.000587801607227491	0.371238024175139\\
0.000837677640068292	0.34627405895346\\
0.00119377664171444	0.307493108067479\\
0.00170125427985259	0.2577594083937\\
0.00242446201708233	0.202723519148727\\
0.00345510729459222	0.149829579076255\\
0.00492388263170674	0.106928886395596\\
0.00701703828670383	0.0790688161681468\\
0.01	0.0654314018080318\\
0.01425102670303	0.0604627964805149\\
0.0203091762090473	0.0584504263435867\\
0.0289426612471675	0.0563975344398912\\
0.0412462638290135	0.0534644526036207\\
0.0587801607227491	0.0494915430970254\\
0.0837677640068292	0.0443511877716715\\
0.119377664171444	0.0379479018134713\\
0.170125427985259	0.0303844357467211\\
0.242446201708233	0.0221268538045444\\
0.345510729459222	0.0140797941095087\\
0.492388263170674	0.00740229978543278\\
0.701703828670383	0.00296373834798614\\
1	0.000782126721671124\\
1.425102670303	8.51906182470704e-05\\
2.03091762090473	-1.13268674426622e-05\\
2.89426612471675	5.49209383978827e-06\\
4.12462638290135	9.69506146702544e-06\\
5.87801607227491	4.40191292089883e-06\\
8.37677640068292	6.05485446683206e-07\\
11.9377664171444	-5.22050512621043e-07\\
17.0125427985259	-4.9260618450521e-07\\
24.2446201708233	-2.29227700897259e-07\\
34.5510729459222	-9.99923688409911e-08\\
49.2388263170674	-2.01818740097159e-08\\
70.1703828670382	-2.40896285212314e-08\\
100	-1.8308951852979e-08\\
};
\addlegendentry{$\text{F: }\phi{}^\text{*}_\text{f}\text{=0.8}$}

\addplot [color=mycolor2, dashed, line width=1pt]
  table[row sep=crcr]{%
0.0001	0.557623402830528\\
0.0001425102670303	0.557975405044964\\
0.000203091762090473	0.558332892346376\\
0.000289426612471675	0.558579995589977\\
0.000412462638290135	0.558576392899622\\
0.000587801607227491	0.558234423991266\\
0.000837677640068292	0.557540723219781\\
0.00119377664171444	0.556506127926614\\
0.00170125427985259	0.555099120220313\\
0.00242446201708233	0.553213427242788\\
0.00345510729459222	0.550668096084001\\
0.00492388263170674	0.547171690256571\\
0.00701703828670383	0.542292960611528\\
0.01	0.535459757026754\\
0.01425102670303	0.525844629753474\\
0.0203091762090473	0.512448964252595\\
0.0289426612471675	0.493924307599624\\
0.0412462638290135	0.468655409912211\\
0.0587801607227491	0.434834286783549\\
0.0837677640068292	0.390747293754968\\
0.119377664171444	0.335436822188538\\
0.170125427985259	0.269723950099923\\
0.242446201708233	0.197512432966507\\
0.345510729459222	0.12656853998457\\
0.492388263170674	0.067113576284215\\
0.701703828670383	0.0271476924165618\\
1	0.00726029131679716\\
1.425102670303	0.000812561108251263\\
2.03091762090473	-0.00010712801974215\\
2.89426612471675	4.45071578303553e-05\\
4.12462638290135	8.70251203609625e-05\\
5.87801607227491	4.0226561041181e-05\\
8.37677640068292	5.60922371688437e-06\\
11.9377664171444	-4.59119910175684e-06\\
17.0125427985259	-4.54096626919419e-06\\
24.2446201708233	-2.19976435445707e-06\\
34.5510729459222	-3.47715293663529e-07\\
49.2388263170674	3.52435587105668e-08\\
70.1703828670382	1.99808757922104e-07\\
100	-6.41257455059834e-08\\
};
\addlegendentry{$\text{M: }\phi{}^\text{*}_\text{f}\text{=0.2}$}

\addplot [color=mycolor2, dashed, line width=1pt, mark size=2pt, mark=o, mark options={solid, mycolor2}]
  table[row sep=crcr]{%
0.0001	0.678766905841062\\
0.0001425102670303	0.678112912462313\\
0.000203091762090473	0.673167381345451\\
0.000289426612471675	0.656174897743988\\
0.000412462638290135	0.618232490814479\\
0.000587801607227491	0.554758893057497\\
0.000837677640068292	0.46831501497496\\
0.00119377664171444	0.367478143105035\\
0.00170125427985259	0.264901091941501\\
0.00242446201708233	0.175519709496136\\
0.00345510729459222	0.111887741721303\\
0.00492388263170674	0.0772929857593734\\
0.00701703828670383	0.0641332034528409\\
0.01	0.0608879962425824\\
0.01425102670303	0.0598605769128547\\
0.0203091762090473	0.0584137400104618\\
0.0289426612471675	0.0562061670640584\\
0.0412462638290135	0.0532108198584112\\
0.0587801607227491	0.0492658961906374\\
0.0837677640068292	0.0441640612110563\\
0.119377664171444	0.0377923046258575\\
0.170125427985259	0.0302584490954247\\
0.242446201708233	0.0220318457046865\\
0.345510729459222	0.0140163344143502\\
0.492388263170674	0.00736677339183111\\
0.701703828670383	0.00294828996302704\\
1	0.000777574019901499\\
1.425102670303	8.45184909911577e-05\\
2.03091762090473	-1.12100518343473e-05\\
2.89426612471675	5.47887954669053e-06\\
4.12462638290135	9.64502460233045e-06\\
5.87801607227491	4.36043433318735e-06\\
8.37677640068292	5.69783798972488e-07\\
11.9377664171444	-5.23753682296495e-07\\
17.0125427985259	-5.14853550586013e-07\\
24.2446201708233	-2.24533878526664e-07\\
34.5510729459222	-8.26989610740267e-08\\
49.2388263170674	-1.38996038159206e-08\\
70.1703828670382	2.51343618521597e-08\\
100	1.84876271711119e-08\\
};
\addlegendentry{$\text{F: }\phi{}^\text{*}_\text{f}\text{=0.2}$}

\end{axis}
\end{tikzpicture}
\subcaption{Pressure}\label{fig:5c}
\end{minipage}%
\begin{minipage}[b]{0.5\textwidth}
\centering
\setlength\figureheight{5cm}
\setlength\figurewidth{5cm}
\definecolor{mycolor1}{rgb}{0.34667,0.53600,0.69067}%
\definecolor{mycolor2}{rgb}{0.91529,0.28157,0.28784}%
\begin{tikzpicture}

\begin{axis}[%
width=0.986\figurewidth,
height=\figureheight,
at={(0\figurewidth,0\figureheight)},
scale only axis,
xmode=log,
xmin=0.0001,
xmax=100,
xminorticks=true,
xlabel style={font=\fontsize{8}{144}\selectfont\color{white!15!black}},
xlabel={Time (s)},
ymin=-4.2e-5,
ymax=-3.5e-5,
ylabel style={font=\fontsize{8}{144}\selectfont\color{white!5!black}},
ylabel={$\text{Vertical Strain, }\epsilon_{zz}$},
ticklabel style={font=\fontsize{8}{144}},
axis background/.style={fill=white},
legend style={font=\fontsize{8}{144}\selectfont\color{white!5!black}, at={(1,0.95)}, anchor=north east, legend cell align=left, align=left, fill=none, draw=none}
]
\addplot [color=black, line width=1pt]
  table[row sep=crcr]{%
0.0001	-3.58538064776321e-05\\
0.0001425102670303	-3.58607047810153e-05\\
0.000203091762090473	-3.58690381635245e-05\\
0.000289426612471675	-3.58792007612577e-05\\
0.000412462638290135	-3.58915966905656e-05\\
0.000587801607227491	-3.59068367997308e-05\\
0.000837677640068292	-3.59257862726288e-05\\
0.00119377664171444	-3.59494724603949e-05\\
0.00170125427985259	-3.59792542830435e-05\\
0.00242446201708233	-3.6016070515186e-05\\
0.00345510729459222	-3.6060704453777e-05\\
0.00492388263170674	-3.61142163686163e-05\\
0.00701703828670383	-3.61792829985371e-05\\
0.01	-3.62620548586218e-05\\
0.01425102670303	-3.63726353609826e-05\\
0.0203091762090473	-3.65242443052803e-05\\
0.0289426612471675	-3.67323968081388e-05\\
0.0412462638290135	-3.70150078732272e-05\\
0.0587801607227491	-3.73912311033452e-05\\
0.0837677640068292	-3.78781720438483e-05\\
0.119377664171444	-3.84839079504588e-05\\
0.170125427985259	-3.91958360977476e-05\\
0.242446201708233	-3.99676406179232e-05\\
0.345510729459222	-4.0713328246158e-05\\
0.492388263170674	-4.13261062356662e-05\\
0.701703828670383	-4.17292403464377e-05\\
1	-4.19253400986626e-05\\
1.425102670303	-4.19872887137926e-05\\
2.03091762090473	-4.19957115289766e-05\\
2.89426612471675	-4.19942145643173e-05\\
4.12462638290135	-4.19938711372472e-05\\
5.87801607227491	-4.19943536219892e-05\\
8.37677640068292	-4.19946982304059e-05\\
11.9377664171444	-4.19947968589827e-05\\
17.0125427985259	-4.1994791646194e-05\\
24.2446201708233	-4.19947767381292e-05\\
34.5510729459222	-4.19947606003738e-05\\
49.2388263170674	-4.19947538517902e-05\\
70.1703828670382	-4.19947545569962e-05\\
100	-4.1994747537057e-05\\
};
\addlegendentry{$\text{K\&V}$}

\addplot [color=mycolor1, dotted, line width=1pt]
  table[row sep=crcr]{%
0.0001	-3.58168756476309e-05\\
0.0001425102670303	-3.58265097807751e-05\\
0.000203091762090473	-3.58380940614939e-05\\
0.000289426612471675	-3.58520869882716e-05\\
0.000412462638290135	-3.58690254055442e-05\\
0.000587801607227491	-3.58896595292609e-05\\
0.000837677640068292	-3.59149541729234e-05\\
0.00119377664171444	-3.59460918388983e-05\\
0.00170125427985259	-3.59841791285843e-05\\
0.00242446201708233	-3.60295489204284e-05\\
0.00345510729459222	-3.60819284594693e-05\\
0.00492388263170674	-3.61410549467488e-05\\
0.00701703828670383	-3.62091391029022e-05\\
0.01	-3.62928670640422e-05\\
0.01425102670303	-3.64033355280019e-05\\
0.0203091762090473	-3.65544550212251e-05\\
0.0289426612471675	-3.6762072586608e-05\\
0.0412462638290135	-3.7043839034674e-05\\
0.0587801607227491	-3.74188949541599e-05\\
0.0837677640068292	-3.79040953244809e-05\\
0.119377664171444	-3.85073371720519e-05\\
0.170125427985259	-3.92158421984587e-05\\
0.242446201708233	-3.99830975471723e-05\\
0.345510729459222	-4.07232673828867e-05\\
0.492388263170674	-4.13302194908161e-05\\
0.701703828670383	-4.17283497185252e-05\\
1	-4.19212397464373e-05\\
1.425102670303	-4.19818070467333e-05\\
2.03091762090473	-4.19899025386905e-05\\
2.89426612471675	-4.19884029765423e-05\\
4.12462638290135	-4.19880839044017e-05\\
5.87801607227491	-4.19885637156598e-05\\
8.37677640068292	-4.1988900869849e-05\\
11.9377664171444	-4.19889987023502e-05\\
17.0125427985259	-4.19889933024901e-05\\
24.2446201708233	-4.19889750888513e-05\\
34.5510729459222	-4.19889607472059e-05\\
49.2388263170674	-4.19889539477459e-05\\
70.1703828670382	-4.19889510405223e-05\\
100	-4.19889499369819e-05\\
};
\addlegendentry{$\phi{}^\text{*}_\text{f}\text{=0.8}$}

\addplot [color=mycolor2, line width=1pt]
  table[row sep=crcr]{%
0.0001	-3.56327998588526e-05\\
0.0001425102670303	-3.56545575756732e-05\\
0.000203091762090473	-3.56806217248795e-05\\
0.000289426612471675	-3.57118684735405e-05\\
0.000412462638290135	-3.57492646044283e-05\\
0.000587801607227491	-3.57941279972941e-05\\
0.000837677640068292	-3.584712649347e-05\\
0.00119377664171444	-3.59078187847718e-05\\
0.00170125427985259	-3.59726902191731e-05\\
0.00242446201708233	-3.60368163076187e-05\\
0.00345510729459222	-3.60961787338507e-05\\
0.00492388263170674	-3.61525564428879e-05\\
0.00701703828670383	-3.62142589343643e-05\\
0.01	-3.62932912737625e-05\\
0.01425102670303	-3.64018355246598e-05\\
0.0203091762090473	-3.65529380793513e-05\\
0.0289426612471675	-3.67606449895583e-05\\
0.0412462638290135	-3.70426785196093e-05\\
0.0587801607227491	-3.741786517411e-05\\
0.0837677640068292	-3.79033976093116e-05\\
0.119377664171444	-3.85069112336398e-05\\
0.170125427985259	-3.92157407978784e-05\\
0.242446201708233	-3.99832352396943e-05\\
0.345510729459222	-4.07235597603244e-05\\
0.492388263170674	-4.13305227773414e-05\\
0.701703828670383	-4.17285671897605e-05\\
1	-4.19213326171588e-05\\
1.425102670303	-4.19818317332953e-05\\
2.03091762090473	-4.19898988831705e-05\\
2.89426612471675	-4.19884001054937e-05\\
4.12462638290135	-4.19880805312934e-05\\
5.87801607227491	-4.19885704162074e-05\\
8.37677640068292	-4.19889000033977e-05\\
11.9377664171444	-4.19889985121577e-05\\
17.0125427985259	-4.19889917729244e-05\\
24.2446201708233	-4.19889761423471e-05\\
34.5510729459222	-4.19889567358451e-05\\
49.2388263170674	-4.19889471353524e-05\\
70.1703828670382	-4.1988945813038e-05\\
100	-4.19889541109115e-05\\
};
\addlegendentry{$\phi{}^\text{*}_\text{f}\text{=0.2}$}

\end{axis}
\end{tikzpicture}
\subcaption{Vertical Strain, $\epsilon_{zz}$}\label{fig:5d}
\end{minipage}
\caption{Matrix (`M') and fracture (`F') pressure (a, c), and vertical strain (b, d) evolutions for the double-porosity Mandel problem whilst considering the effects of intrinsic fracture properties. `K$\&$V' denote models from \cite{Khalili1996}. Different values of intrinsic fracture porosity $\phi^*_{f}=1$ (`K$\&$V'), $\phi^*_{f}=0.8$, $\phi^*_{f}=0.2$, and hence different coefficient models, are tested for $K^{\dagger}_f=\frac{K^f_s}{1750}$, (a, b), and $K^{\dagger}_f=\frac{K^f_s}{35}$, (c, d).}\label{fig:5}
\end{figure}

\cref{fig:5a} and \cref{fig:5c} show matrix and fracture pressure evolutions whilst varying intrinsic fracture properties. We find a good match in both matrix and fracture pressure evolutions using coefficient models from \cite{Khalili1996} and \cite{Berryman2002a} when the fracture is almost all void space ($\phi^*_f \approx 1$) and the fracture phase stiffness is orders of magnitude lower than the solid grain stiffness ($K_f\ll K^f_s$) (\cref{fig:5a}). Even when the intrinsic fracture porosity is diminishing ($\phi^*_f=0.2$), provided the fracture phase stiffness is orders of magnitude lower than the solid stiffness, the difference between early time matrix pressures with the different coefficient model formulations is small (\cref{fig:5a}). However, when the fracture bulk modulus is only an order of magnitude lower than the solid modulus ($K_f \not\ll K^f_s$), early time fracture pressure differences become measurable as intrinsic porosity decreases (\cref{fig:5c}). This phenomenon can be explained by considering the intrinsic Skempton coefficient, $B^*_\alpha$, written in the following form for the fracture phase (\citealt{Cheng2016}),

\begin{equation}
B^*_f = 1 - \frac{\phi^*_fK_f(K^f_s-K_l)}{K_l(K^f_s-K_f)+\phi^*_fK_f(K^f_s-K_l)}.\label{eqn:75}
\end{equation} 

\noindent
Use of \cref{eqn:75} allows us to cast our observations as a bounding problem such that

\begin{equation}
B^*_f(\phi^*_f=1,K_f) < B^*_f(\phi^*_f<1,K_f) \leq 1.\label{eqn:76}
\end{equation} 

\noindent
The lower bound in \cref{eqn:76} is a fictitious one in that technically materials with an intrinsic porosity of one should also have a zero stiffness. However, accepting this contradiction is useful in approaching the bounding problem from a purely quantitative point of view. For the upper bound, \cref{eqn:75} shows that as $\phi^*_f$ approaches zero, $B^*_f$ must asymptotically approach one.

If the lower bound in \cref{eqn:76} for a given $K_f$ is close to one then changes in intrinsic fracture porosity are negligible due to the proximity of the lower and upper bounds. This is the case when $K_f\ll K^f_s$. If the fracture phase stiffness is \textit{not} orders of magnitude lower than the solid stiffness, then the lower bound of $B^*_f$ may be significantly less than one. In this case, changes in $B^*_f$ cannot be captured when using void space coefficient models, and thus early time fracture pressure is underestimated as $\phi^*_f$ decreases (\cref{fig:5c}).  

\cref{fig:5b} and \cref{fig:5d} show the variation in vertical strain for the softer and stiffer fracture phases respectively. In \cref{fig:5b}, the strain evolutions are almost identical when using the more compliant fracture phase for the range of intrinsic fracture porosities. This observation is coupled to the similarity in pressure evolutions. Whilst induced fracture pressures are significantly larger when fracture stiffness and intrinsic porosity are non-negligible, \cref{fig:5d} shows very little differences in vertical strain across the whole intrinsic porosity range. This is a direct consequence of the algebraic coefficient of strain in \cref{eqn:73} for the fracture phase, which scales proportionally with the variation in fracture pressure.   

\subsection{Case 2: Implicit decoupling}
\begin{figure}[h]
\centering
\begin{minipage}[b]{0.5\textwidth}
\centering
\setlength\figureheight{5cm}
\setlength\figurewidth{5cm}
\definecolor{mycolor1}{rgb}{0.34667,0.53600,0.69067}%
\definecolor{mycolor2}{rgb}{0.91529,0.28157,0.28784}%
\begin{tikzpicture}

\begin{axis}[%
width=0.993\figurewidth,
height=\figureheight,
at={(0\figurewidth,0\figureheight)},
scale only axis,
xmode=log,
xmin=0.0001,
xmax=100,
xminorticks=true,
xminorticks=true,
xlabel style={font=\fontsize{8}{144}\selectfont\color{white!15!black}},
xlabel={Time (s)},
ymin=0,
ymax=1,
ylabel style={font=\fontsize{8}{144}\selectfont\color{white!5!black}},
ylabel={Pressure (MPa)},
ticklabel style={font=\fontsize{8}{144}},
axis background/.style={fill=white},
legend style={font=\fontsize{8}{144}\selectfont\color{white!5!black}, at={(1,0.95)}, anchor=north east, legend cell align=left, align=left, fill=none, draw=none}
]
\addplot [color=black, line width=1pt]
  table[row sep=crcr]{%
0.0001	0.588008861859843\\
0.0001425102670303	0.589222865646489\\
0.000203091762090473	0.590688655803066\\
0.000289426612471675	0.592462354946948\\
0.000412462638290135	0.594614167261408\\
0.000587801607227491	0.597228450477437\\
0.000837677640068292	0.600411347651806\\
0.00119377664171444	0.604311408298439\\
0.00170125427985259	0.608985272659317\\
0.00242446201708233	0.613921711275362\\
0.00345510729459222	0.617683653868398\\
0.00492388263170674	0.618403481343753\\
0.00701703828670383	0.614828569268647\\
0.01	0.606744215861856\\
0.01425102670303	0.594345366868841\\
0.0203091762090473	0.577407155146968\\
0.0289426612471675	0.55503844496626\\
0.0412462638290135	0.525827349148726\\
0.0587801607227491	0.487966017853879\\
0.0837677640068292	0.439487888347963\\
0.119377664171444	0.378982229659834\\
0.170125427985259	0.306925006908518\\
0.242446201708233	0.227183643460519\\
0.345510729459222	0.147944861348025\\
0.492388263170674	0.0803595662677382\\
0.701703828670383	0.0337224861593366\\
1	0.00959914163786117\\
1.425102670303	0.00128343649350528\\
2.03091762090473	-0.000106038375891372\\
2.89426612471675	3.05488503866446e-05\\
4.12462638290135	9.88106427448285e-05\\
5.87801607227491	5.06582001732285e-05\\
8.37677640068292	9.51171635823852e-06\\
11.9377664171444	-4.46788727079937e-06\\
17.0125427985259	-5.54507072224645e-06\\
24.2446201708233	-3.09254912986522e-06\\
34.5510729459222	-8.61825089352176e-07\\
49.2388263170674	-3.74090678995285e-07\\
70.1703828670382	-3.47319834213289e-07\\
100	-1.18910461243184e-07\\
};
\addlegendentry{M: Ref.}

\addplot [color=black, line width=1pt, mark size=2pt, mark=diamond, mark options={solid, black}]
  table[row sep=crcr]{%
0.0001	0.865786557487422\\
0.0001425102670303	0.867442539501856\\
0.000203091762090473	0.869412850980963\\
0.000289426612471675	0.87175712555242\\
0.000412462638290135	0.874546247859219\\
0.000587801607227491	0.877840982416009\\
0.000837677640068292	0.881707330182974\\
0.00119377664171444	0.88629849881174\\
0.00170125427985259	0.891017088691976\\
0.00242446201708233	0.891703725774322\\
0.00345510729459222	0.878649089955865\\
0.00492388263170674	0.839857890258826\\
0.00701703828670383	0.767921657838543\\
0.01	0.663941899856068\\
0.01425102670303	0.536219371444875\\
0.0203091762090473	0.398008175429711\\
0.0289426612471675	0.266863721150236\\
0.0412462638290135	0.16150784093398\\
0.0587801607227491	0.0932156462835419\\
0.0837677640068292	0.0585640235534923\\
0.119377664171444	0.0433691030254215\\
0.170125427985259	0.034336905265377\\
0.242446201708233	0.0255370502427955\\
0.345510729459222	0.0165919694263903\\
0.492388263170674	0.00892227442950372\\
0.701703828670383	0.00368922816652005\\
1	0.00102640925029697\\
1.425102670303	0.000127554805884979\\
2.03091762090473	-1.52869583549923e-05\\
2.89426612471675	2.65807219910275e-06\\
4.12462638290135	1.05054930764106e-05\\
5.87801607227491	5.45179725767616e-06\\
8.37677640068292	1.01589386162804e-06\\
11.9377664171444	-5.15105318060718e-07\\
17.0125427985259	-6.11655287798253e-07\\
24.2446201708233	-3.20398720678881e-07\\
34.5510729459222	-1.19180702285402e-07\\
49.2388263170674	-6.98902605208538e-08\\
70.1703828670382	-4.39924588196007e-08\\
100	-3.72367455730148e-08\\
};
\addlegendentry{F: Ref.}

\addplot [color=mycolor1, dotted, line width=1pt]
  table[row sep=crcr]{%
0.0001	0.342333649831068\\
0.0001425102670303	0.342879723310509\\
0.000203091762090473	0.343557007924162\\
0.000289426612471675	0.344401466087118\\
0.000412462638290135	0.345460451216488\\
0.000587801607227491	0.346800242376333\\
0.000837677640068292	0.348509197525733\\
0.00119377664171444	0.350693504336434\\
0.00170125427985259	0.353614678261324\\
0.00242446201708233	0.358109175346911\\
0.00345510729459222	0.365833962403288\\
0.00492388263170674	0.378658816948361\\
0.00701703828670383	0.39746754465189\\
0.01	0.421330334314535\\
0.01425102670303	0.447364477305312\\
0.0203091762090473	0.470805420024309\\
0.0289426612471675	0.485360581199083\\
0.0412462638290135	0.484783928108412\\
0.0587801607227491	0.465511419680522\\
0.0837677640068292	0.427896378378742\\
0.119377664171444	0.374593284514084\\
0.170125427985259	0.30867967125976\\
0.242446201708233	0.233875232890337\\
0.345510729459222	0.15743922964912\\
0.492388263170674	0.0895137933233008\\
0.701703828670383	0.0400218105635539\\
1	0.0125454112122516\\
1.425102670303	0.00208486752401081\\
2.03091762090473	-5.96443921500017e-05\\
2.89426612471675	-8.9342696198631e-06\\
4.12462638290135	9.58605409365952e-05\\
5.87801607227491	6.06169042903563e-05\\
8.37677640068292	1.51269805506031e-05\\
11.9377664171444	-3.00830008579588e-06\\
17.0125427985259	-5.10978718835763e-06\\
24.2446201708233	-3.11401942993204e-06\\
34.5510729459222	-1.02578054706301e-06\\
49.2388263170674	-6.83573364776318e-07\\
70.1703828670382	-4.32988832853354e-07\\
100	8.38690266929083e-10\\
};
\addlegendentry{M: 1/Q=0}

\addplot [color=mycolor1, dotted, line width=1pt, mark size=2pt, mark=triangle, mark options={solid, mycolor1}]
  table[row sep=crcr]{%
0.0001	0.724013086268892\\
0.0001425102670303	0.724833624758598\\
0.000203091762090473	0.725788904164875\\
0.000289426612471675	0.726895983089687\\
0.000412462638290135	0.728171373160192\\
0.000587801607227491	0.729609815306678\\
0.000837677640068292	0.731194059278226\\
0.00119377664171444	0.732955308072347\\
0.00170125427985259	0.734252863918393\\
0.00242446201708233	0.731446115515708\\
0.00345510729459222	0.716381898821808\\
0.00492388263170674	0.679336906823594\\
0.00701703828670383	0.614860508114505\\
0.01	0.525017385089096\\
0.01425102670303	0.418263313277511\\
0.0203091762090473	0.307338004744887\\
0.0289426612471675	0.20765710218226\\
0.0412462638290135	0.132995109187254\\
0.0587801607227491	0.0881629464554402\\
0.0837677640068292	0.0661171947834672\\
0.119377664171444	0.0545334386139943\\
0.170125427985259	0.0446614358036184\\
0.242446201708233	0.0337505790242466\\
0.345510729459222	0.0224996626671717\\
0.492388263170674	0.0126111447569169\\
0.701703828670383	0.00554191837940727\\
1	0.00169882999959053\\
1.425102670303	0.000270143723754211\\
2.03091762090473	-1.13002903555242e-05\\
2.89426612471675	-5.00296076698537e-07\\
4.12462638290135	1.36132272248307e-05\\
5.87801607227491	8.3526399195143e-06\\
8.37677640068292	1.99376419824059e-06\\
11.9377664171444	-4.54499666034176e-07\\
17.0125427985259	-7.518194639569e-07\\
24.2446201708233	-4.53633115589549e-07\\
34.5510729459222	-1.68552762755918e-07\\
49.2388263170674	-1.09474258009088e-07\\
70.1703828670382	-5.01962144918602e-08\\
100	-1.81275770263713e-08\\
};
\addlegendentry{F: 1/Q=0}

\addplot [color=mycolor2, line width=1pt]
  table[row sep=crcr]{%
0.0001	0.596482182375459\\
0.0001425102670303	0.597726893880011\\
0.000203091762090473	0.599229371249253\\
0.000289426612471675	0.60104687560698\\
0.000412462638290135	0.603251063389678\\
0.000587801607227491	0.605927716653697\\
0.000837677640068292	0.609184846634957\\
0.00119377664171444	0.613173743132453\\
0.00170125427985259	0.617944348879096\\
0.00242446201708233	0.622941609937176\\
0.00345510729459222	0.626634620361626\\
0.00492388263170674	0.627050067759745\\
0.00701703828670383	0.622878900082558\\
0.01	0.613922182636764\\
0.01425102670303	0.60044143964012\\
0.0203091762090473	0.582306760013082\\
0.0289426612471675	0.558751973466319\\
0.0412462638290135	0.528496273316519\\
0.0587801607227491	0.489812888270382\\
0.0837677640068292	0.440703624488934\\
0.119377664171444	0.379667125260485\\
0.170125427985259	0.307118446630342\\
0.242446201708233	0.226953643802503\\
0.345510729459222	0.147449834360916\\
0.492388263170674	0.0798302984103937\\
0.701703828670383	0.033342759686569\\
1	0.00941965355978137\\
1.425102670303	0.00123553482327968\\
2.03091762090473	-0.000109216123159908\\
2.89426612471675	3.36547512751152e-05\\
4.12462638290135	9.79852674788339e-05\\
5.87801607227491	4.99668627505737e-05\\
8.37677640068292	8.94474155180283e-06\\
11.9377664171444	-4.3977224874732e-06\\
17.0125427985259	-5.38708454220222e-06\\
24.2446201708233	-2.86518382738287e-06\\
34.5510729459222	-8.94168044478933e-07\\
49.2388263170674	-1.97553851126475e-08\\
70.1703828670382	-2.09834623779851e-07\\
100	-2.32384206615578e-07\\
};
\addlegendentry{$\text{M: A}_{\text{23}}\text{=0}$}

\addplot [color=mycolor2, line width=1pt, mark size=2pt, mark=o, mark options={solid, mycolor2}]
  table[row sep=crcr]{%
0.0001	0.870811498205689\\
0.0001425102670303	0.872499987868874\\
0.000203091762090473	0.874509604918572\\
0.000289426612471675	0.876901472563236\\
0.000412462638290135	0.879748402605491\\
0.000587801607227491	0.883113081526632\\
0.000837677640068292	0.887064881219634\\
0.00119377664171444	0.891760683221726\\
0.00170125427985259	0.896585907826532\\
0.00242446201708233	0.897312977864369\\
0.00345510729459222	0.884130998658283\\
0.00492388263170674	0.844978688406397\\
0.00701703828670383	0.772457845921634\\
0.01	0.667725967982024\\
0.01425102670303	0.539142770022192\\
0.0203091762090473	0.400027314494691\\
0.0289426612471675	0.268034143885607\\
0.0412462638290135	0.162003366135748\\
0.0587801607227491	0.0932848197266971\\
0.0837677640068292	0.058434538540842\\
0.119377664171444	0.0431788823368563\\
0.170125427985259	0.0341405525813552\\
0.242446201708233	0.0253534247896114\\
0.345510729459222	0.0164363600846825\\
0.492388263170674	0.00881045567303239\\
0.701703828670383	0.00362570849210298\\
1	0.00100103159427022\\
1.425102670303	0.000121751821394463\\
2.03091762090473	-1.54598716154777e-05\\
2.89426612471675	2.88950277484323e-06\\
4.12462638290135	1.04382800921847e-05\\
5.87801607227491	5.30598204105871e-06\\
8.37677640068292	1.00643311861751e-06\\
11.9377664171444	-4.66283790164956e-07\\
17.0125427985259	-5.64326192005973e-07\\
24.2446201708233	-3.06360973463822e-07\\
34.5510729459222	-1.10972460281784e-07\\
49.2388263170674	-2.84080766493666e-08\\
70.1703828670382	-1.4658803693692e-08\\
100	-2.19713713652194e-08\\
};
\addlegendentry{$\text{F: A}_{\text{23}}\text{=0}$}

\end{axis}
\end{tikzpicture}
\subcaption{Pressure}\label{fig:6a}
\end{minipage}%
\begin{minipage}[b]{0.5\textwidth}
\centering
\setlength\figureheight{5cm}
\setlength\figurewidth{5cm}
\definecolor{mycolor1}{rgb}{0.34667,0.53600,0.69067}%
\definecolor{mycolor2}{rgb}{0.91529,0.28157,0.28784}%
\begin{tikzpicture}

\begin{axis}[%
width=0.986\figurewidth,
height=\figureheight,
at={(0\figurewidth,0\figureheight)},
scale only axis,
xmode=log,
xmin=0.0001,
xmax=100,
xminorticks=true,
xlabel style={font=\fontsize{8}{144}\selectfont\color{white!15!black}},
xlabel={Time (s)},
ymin=-8e-5,
ymax=-6e-5,
ylabel style={font=\fontsize{8}{144}\selectfont\color{white!5!black}},
ylabel={Vertical Strain, $\epsilon_{zz}$},
ticklabel style={font=\fontsize{8}{144}},
axis background/.style={fill=white},
legend style={font=\fontsize{8}{144}\selectfont\color{white!5!black}, at={(1,0.95)}, anchor=north east, legend cell align=left, align=left, fill=none, draw=none}
]
\addplot [color=black, line width=1.5pt]
  table[row sep=crcr]{%
0.0001	-6.1530397252663e-05\\
0.0001425102670303	-6.16524065082058e-05\\
0.000203091762090473	-6.17986096031236e-05\\
0.000289426612471675	-6.19739264396517e-05\\
0.000412462638290135	-6.21843295200385e-05\\
0.000587801607227491	-6.24370866783922e-05\\
0.000837677640068292	-6.27410787068519e-05\\
0.00119377664171444	-6.31071793497088e-05\\
0.00170125427985259	-6.35487888656813e-05\\
0.00242446201708233	-6.40824263479163e-05\\
0.00345510729459222	-6.47284467718723e-05\\
0.00492388263170674	-6.5512325407753e-05\\
0.00701703828670383	-6.64647638308645e-05\\
0.01	-6.76119339628764e-05\\
0.01425102670303	-6.89492632503371e-05\\
0.0203091762090473	-7.0411044723691e-05\\
0.0289426612471675	-7.18636333108818e-05\\
0.0412462638290135	-7.31500630673961e-05\\
0.0587801607227491	-7.41828830008914e-05\\
0.0837677640068292	-7.50129228365083e-05\\
0.119377664171444	-7.57846684229355e-05\\
0.170125427985259	-7.66139035048521e-05\\
0.242446201708233	-7.75124766745683e-05\\
0.345510729459222	-7.83972099301119e-05\\
0.492388263170674	-7.91416599913366e-05\\
0.701703828670383	-7.96462780365657e-05\\
1	-7.99018167182423e-05\\
1.425102670303	-7.99875728557295e-05\\
2.03091762090473	-8.0001154772083e-05\\
2.89426612471675	-7.99995664923653e-05\\
4.12462638290135	-7.99989347639379e-05\\
5.87801607227491	-7.99994727365716e-05\\
8.37677640068292	-7.99999044628163e-05\\
11.9377664171444	-8.00000518028585e-05\\
17.0125427985259	-8.00000576820653e-05\\
24.2446201708233	-8.00000321386134e-05\\
34.5510729459222	-8.00000107947198e-05\\
49.2388263170674	-8.00000024870309e-05\\
70.1703828670382	-8.00000043942109e-05\\
100	-8.00000039208186e-05\\
};
\addlegendentry{Ref.}

\addplot [color=mycolor1, dotted, line width=1.5pt]
  table[row sep=crcr]{%
0.0001	-6.60047752848498e-05\\
0.0001425102670303	-6.60904523606348e-05\\
0.000203091762090473	-6.61930566485765e-05\\
0.000289426612471675	-6.63159994290428e-05\\
0.000412462638290135	-6.64634002605193e-05\\
0.000587801607227491	-6.66402430271716e-05\\
0.000837677640068292	-6.68525646700278e-05\\
0.00119377664171444	-6.71076809598591e-05\\
0.00170125427985259	-6.74144744821104e-05\\
0.00242446201708233	-6.77836586984461e-05\\
0.00345510729459222	-6.8228089633317e-05\\
0.00492388263170674	-6.8763278511808e-05\\
0.00701703828670383	-6.9406873239253e-05\\
0.01	-7.01712195835336e-05\\
0.01425102670303	-7.10458547913747e-05\\
0.0203091762090473	-7.19807353351397e-05\\
0.0289426612471675	-7.28909377706689e-05\\
0.0412462638290135	-7.36971520485078e-05\\
0.0587801607227491	-7.43845168872047e-05\\
0.0837677640068292	-7.50221641050708e-05\\
0.119377664171444	-7.57099757047993e-05\\
0.170125427985259	-7.6504253935564e-05\\
0.242446201708233	-7.73871614109539e-05\\
0.345510729459222	-7.82736044528978e-05\\
0.492388263170674	-7.90411087306807e-05\\
0.701703828670383	-7.95832743729098e-05\\
1	-7.98739960127988e-05\\
1.425102670303	-7.99803030776507e-05\\
2.03091762090473	-8.00007474644805e-05\\
2.89426612471675	-7.9999880187587e-05\\
4.12462638290135	-7.99989206720425e-05\\
5.87801607227491	-7.99993696683809e-05\\
8.37677640068292	-7.99998569783912e-05\\
11.9377664171444	-8.00000432656985e-05\\
17.0125427985259	-8.00000608309775e-05\\
24.2446201708233	-8.00000317964203e-05\\
34.5510729459222	-8.00000102080017e-05\\
49.2388263170674	-8.00000055154837e-05\\
70.1703828670382	-8.00000008201062e-05\\
100	-8.00000016718177e-05\\
};
\addlegendentry{1/Q=0}

\addplot [color=mycolor2, line width=1.5pt]
  table[row sep=crcr]{%
0.0001	-6.13744509132846e-05\\
0.0001425102670303	-6.14976601712121e-05\\
0.000203091762090473	-6.16453077005195e-05\\
0.000289426612471675	-6.18223605158131e-05\\
0.000412462638290135	-6.20348591471652e-05\\
0.000587801607227491	-6.22901483393183e-05\\
0.000837677640068292	-6.25972087831814e-05\\
0.00119377664171444	-6.29670369984834e-05\\
0.00170125427985259	-6.34131966352077e-05\\
0.00242446201708233	-6.39524190981132e-05\\
0.00345510729459222	-6.46053182619675e-05\\
0.00492388263170674	-6.53977324000396e-05\\
0.00701703828670383	-6.63608372058336e-05\\
0.01	-6.75211074463943e-05\\
0.01425102670303	-6.88738351722793e-05\\
0.0203091762090473	-7.03523569822981e-05\\
0.0289426612471675	-7.18213380798446e-05\\
0.0412462638290135	-7.31217844040649e-05\\
0.0587801607227491	-7.41650549030259e-05\\
0.0837677640068292	-7.5002355138617e-05\\
0.119377664171444	-7.57795866778076e-05\\
0.170125427985259	-7.66135971400706e-05\\
0.242446201708233	-7.75160827710249e-05\\
0.345510729459222	-7.84030540221335e-05\\
0.492388263170674	-7.9147428725592e-05\\
0.701703828670383	-7.96502463229815e-05\\
1	-7.99036334973079e-05\\
1.425102670303	-7.99880340521231e-05\\
2.03091762090473	-8.00011690035761e-05\\
2.89426612471675	-7.99995374203255e-05\\
4.12462638290135	-7.99989367264723e-05\\
5.87801607227491	-7.99994789427403e-05\\
8.37677640068292	-7.99999087962485e-05\\
11.9377664171444	-8.00000554361654e-05\\
17.0125427985259	-8.00000556146112e-05\\
24.2446201708233	-8.00000273359749e-05\\
34.5510729459222	-8.0000013364213e-05\\
49.2388263170674	-8.00000002044135e-05\\
70.1703828670382	-8.00000012233352e-05\\
100	-8.00000097525884e-05\\
};
\addlegendentry{$\text{A}_{\text{23}}\text{=0}$}

\end{axis}
\end{tikzpicture}
\subcaption{Vertical strain, $\epsilon_{zz}$}\label{fig:6b}
\end{minipage}
\caption{Matrix (`M') and fracture (`F') pressure (a), and vertical strain (b) evolutions for the double-porosity Mandel problem whilst considering macroscale assumptions $\frac{1}{Q}=0$ and $A_{23}=0$. Results are referenced against the set of coefficient models from \cite{Khalili1996} for which no assumptions have been made.} \label{fig:6}
\end{figure}
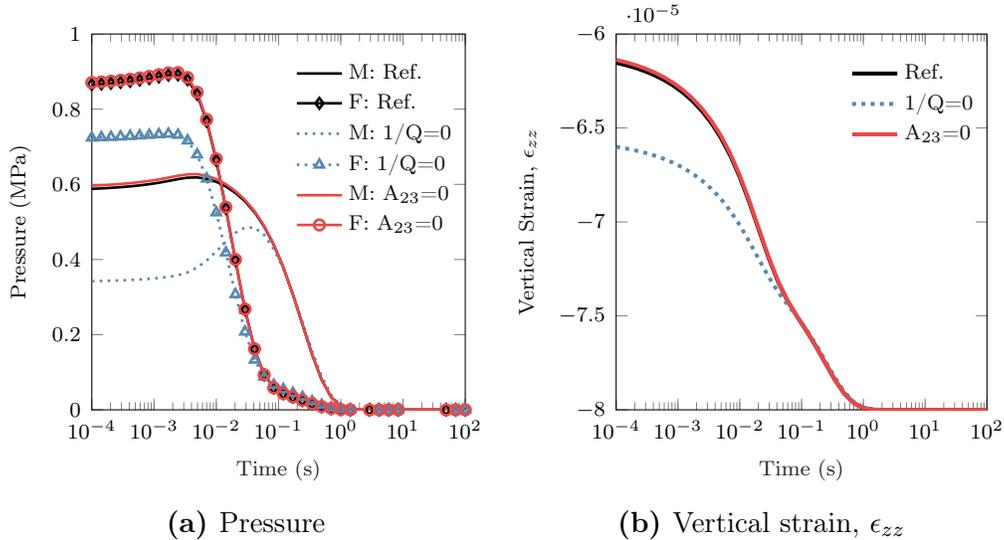

Pressure and vertical strain results for the implicit decoupling assumptions test case are presented in \cref{fig:6}. When assuming $A_{23}=0$, the matrix and fracture pressure evolutions are almost identical to the reference case (\cref{fig:6a}), with vertical strains also being correspondingly very similar (\cref{fig:6b}). When assuming $\frac{1}{Q}=0$, the early time matrix and fracture pressures are measurably lower than the reference case (\cref{fig:6a}). The early time vertical strain is greater when $\frac{1}{Q}=0$ than the strain in the reference case (\cref{fig:6b}).    

The results in \cref{fig:6a} suggest that the assumption $\frac{1}{Q}=0$ has the most noticeable effect on the poromechanical behaviour of the dual-medium. As discussed in Section 4.2.1 we would expect $\frac{1}{Q}<0$. From \cref{eqn:74}, setting $\frac{1}{Q}=0$ thus has the effect of removing a pressure source from continuum $\alpha$. Removal of this poromechanical coupling explains the lower than expected induced matrix and fracture pressures. From \cref{eqn:73} we can see $\text{d}p_\alpha \propto \epsilon^{-1}$. Underestimated pressures therefore explain the over estimated strain when taking $\frac{1}{Q}=0$.

In contrast, \cref{eqn:38} shows that although $A_{23}=0$, $\frac{1}{Q}\not=0$. The pressures in each continuum therefore remain poromechanically coupled with respect to the mixed compliance setting, hence the similarities in pressures and vertical strain when $A_{23}=0$ versus the reference case.   

The current results suggest that assuming $A_{23}=0$ is a reasonable implicit assumption to make. However, we advise caution when interpreting this result. Based on the results in this section it would be easy, and incorrect, to use them as a justification for assumptions made at the microscale. In Section 4.2 it was shown that explicit assumptions affect all of the constitutive coefficients due to bounds on bulk moduli. The remainder of this results section aims at qualitatively supporting these findings.  

\subsection{Case 3: Explicit decoupling - isostrain}
\begin{figure}[h]
\centering
\begin{minipage}[b]{0.5\textwidth}
\centering
\setlength\figureheight{5cm}
\setlength\figurewidth{5cm}
\definecolor{mycolor1}{rgb}{0.34667,0.53600,0.69067}%
\definecolor{mycolor2}{rgb}{0.91529,0.28157,0.28784}%
\begin{tikzpicture}

\begin{axis}[%
width=0.993\figurewidth,
height=\figureheight,
at={(0\figurewidth,0\figureheight)},
scale only axis,
xmode=log,
xmin=0.0001,
xmax=100,
xminorticks=true,
xlabel style={font=\fontsize{8}{144}\selectfont\color{white!15!black}},
xlabel={Time (s)},
ymin=0,
ymax=1,
ylabel style={font=\fontsize{8}{144}\selectfont\color{white!5!black}},
ylabel={Pressure (MPa)},
ticklabel style={font=\fontsize{8}{144}},
ylabel={Pressure (MPa)},
axis background/.style={fill=white},
legend style={font=\fontsize{8}{144}\selectfont\color{white!5!black}, at={(1,0.95)}, anchor=north east, legend cell align=left, align=left, fill=none, draw=none}
]
\addplot [color=black, line width=1pt]
  table[row sep=crcr]{%
0.0001	0.556070357755134\\
0.0001425102670303	0.555967691252713\\
0.000203091762090473	0.555821060712745\\
0.000289426612471675	0.555613583034257\\
0.000412462638290135	0.555322193455421\\
0.000587801607227491	0.554912352916848\\
0.000837677640068292	0.554333511017722\\
0.00119377664171444	0.553509832294357\\
0.00170125427985259	0.552334444928862\\
0.00242446201708233	0.550655316199869\\
0.00345510729459222	0.548214672643822\\
0.00492388263170674	0.544875474549839\\
0.00701703828670383	0.540077465525184\\
0.01	0.53329449371502\\
0.01425102670303	0.523766547338706\\
0.0203091762090473	0.510472355564413\\
0.0289426612471675	0.492087290859658\\
0.0412462638290135	0.467007510371078\\
0.0587801607227491	0.433417624649561\\
0.0837677640068292	0.389621215602182\\
0.119377664171444	0.334639898461863\\
0.170125427985259	0.269267688380597\\
0.242446201708233	0.19735338165153\\
0.345510729459222	0.126607250245783\\
0.492388263170674	0.0672261786358334\\
0.701703828670383	0.0272394342141211\\
1	0.00730076856986096\\
1.425102670303	0.00082150822377182\\
2.03091762090473	-0.000107717651909025\\
2.89426612471675	4.44093798782598e-05\\
4.12462638290135	8.71485594257561e-05\\
5.87801607227491	4.14001045544648e-05\\
8.37677640068292	6.04692847341569e-06\\
11.9377664171444	-4.49657470807831e-06\\
17.0125427985259	-4.33160033018285e-06\\
24.2446201708233	-2.25771976038337e-06\\
34.5510729459222	-8.84634068130679e-07\\
49.2388263170674	-3.99145270547179e-07\\
70.1703828670382	-6.93264659377529e-08\\
100	2.25225586086452e-07\\
};
\addlegendentry{$\text{M: K}^{\text{V}}_{\text{dr}}$}

\addplot [color=black, line width=1pt, mark size=2pt, mark=diamond, mark options={solid, black}]
  table[row sep=crcr]{%
0.0001	0.0591821310063018\\
0.0001425102670303	0.0603893527677845\\
0.000203091762090473	0.0620185756523956\\
0.000289426612471675	0.0639064473528922\\
0.000412462638290135	0.0656003078388308\\
0.000587801607227491	0.0666189184585039\\
0.000837677640068292	0.0667977382300337\\
0.00119377664171444	0.0663354988515799\\
0.00170125427985259	0.0655515329641567\\
0.00242446201708233	0.0646784266742365\\
0.00345510729459222	0.0638473292369657\\
0.00492388263170674	0.0630599180322714\\
0.00701703828670383	0.0623070448107497\\
0.01	0.0614450627983187\\
0.01425102670303	0.0603112502388959\\
0.0203091762090473	0.0587441037642695\\
0.0289426612471675	0.0565781186119566\\
0.0412462638290135	0.0536260554186753\\
0.0587801607227491	0.0496813078414826\\
0.0837677640068292	0.0445520235183748\\
0.119377664171444	0.03813746572784\\
0.170125427985259	0.0305493760106613\\
0.242446201708233	0.0222588240893845\\
0.345510729459222	0.0141734383848083\\
0.492388263170674	0.00745803070468701\\
0.701703828670383	0.00298944962294718\\
1	0.000790088510298673\\
1.425102670303	8.63979142643554e-05\\
2.03091762090473	-1.14232872206289e-05\\
2.89426612471675	5.49294843775176e-06\\
4.12462638290135	9.72077205242711e-06\\
5.87801607227491	4.4635465420627e-06\\
8.37677640068292	6.05444308343204e-07\\
11.9377664171444	-5.04498903098763e-07\\
17.0125427985259	-4.97246159654542e-07\\
24.2446201708233	-2.71506916603144e-07\\
34.5510729459222	-9.20430196422747e-08\\
49.2388263170674	-7.12599003201988e-08\\
70.1703828670382	1.06271656001834e-08\\
100	-1.25412474880662e-08\\
};
\addlegendentry{$\text{F: K}^{\text{V}}_{\text{dr}}$}

\addplot [color=mycolor1, dotted, line width=1pt]
  table[row sep=crcr]{%
0.0001	0.555364424334278\\
0.0001425102670303	0.555303902511923\\
0.000203091762090473	0.555201278993454\\
0.000289426612471675	0.555061971493542\\
0.000412462638290135	0.554847977536069\\
0.000587801607227491	0.5545422339771\\
0.000837677640068292	0.554090011043663\\
0.00119377664171444	0.553415652029263\\
0.00170125427985259	0.55240773543111\\
0.00242446201708233	0.550906491136302\\
0.00345510729459222	0.54866856406609\\
0.00492388263170674	0.54537873489643\\
0.00701703828670383	0.54063732773046\\
0.01	0.533873392131044\\
0.01425102670303	0.524340251560336\\
0.0203091762090473	0.511036595103963\\
0.0289426612471675	0.492638349467984\\
0.0412462638290135	0.46753758523086\\
0.0587801607227491	0.433923379643397\\
0.0837677640068292	0.390097385490813\\
0.119377664171444	0.335074047739491\\
0.170125427985259	0.269646915531054\\
0.242446201708233	0.197667366575088\\
0.345510729459222	0.126844233447495\\
0.492388263170674	0.0673818851603997\\
0.701703828670383	0.0273215259271478\\
1	0.0073325599184992\\
1.425102670303	0.000828200724231805\\
2.03091762090473	-0.000107752858665584\\
2.89426612471675	4.36135512126576e-05\\
4.12462638290135	8.75117884632167e-05\\
5.87801607227491	4.11588821056527e-05\\
8.37677640068292	6.37723617056961e-06\\
11.9377664171444	-4.49583754751774e-06\\
17.0125427985259	-4.6329399107013e-06\\
24.2446201708233	-2.31012442987342e-06\\
34.5510729459222	-1.27751639346684e-06\\
49.2388263170674	-3.84407714991581e-07\\
70.1703828670382	3.64611034607205e-07\\
100	8.28657316863789e-08\\
};
\addlegendentry{$\text{M: K}^{\text{HS+}}_{\text{dr}}$}

\addplot [color=mycolor1, dotted, line width=1pt, mark size=2pt, mark=asterisk, mark options={solid, mycolor1}]
  table[row sep=crcr]{%
0.0001	0.145484159253228\\
0.0001425102670303	0.146377639708059\\
0.000203091762090473	0.147544737734107\\
0.000289426612471675	0.148487303795445\\
0.000412462638290135	0.147851173371981\\
0.000587801607227491	0.143935154984271\\
0.000837677640068292	0.135756929620762\\
0.00119377664171444	0.123663440673153\\
0.00170125427985259	0.109062192480374\\
0.00242446201708233	0.093922367924293\\
0.00345510729459222	0.0804535242764361\\
0.00492388263170674	0.0705202999335686\\
0.00701703828670383	0.0646833478663473\\
0.01	0.0618724999387668\\
0.01425102670303	0.0603285049339706\\
0.0203091762090473	0.0587661172370444\\
0.0289426612471675	0.0566190023197385\\
0.0412462638290135	0.0536608285375569\\
0.0587801607227491	0.0497079997862685\\
0.0837677640068292	0.0445757614686726\\
0.119377664171444	0.0381608896803827\\
0.170125427985259	0.0305726251371278\\
0.242446201708233	0.0222804578068157\\
0.345510729459222	0.0141918651277884\\
0.492388263170674	0.00747125272507322\\
0.701703828670383	0.00299685742814725\\
1	0.00079312235810389\\
1.425102670303	8.70471729374693e-05\\
2.03091762090473	-1.14483881238919e-05\\
2.89426612471675	5.43072733092688e-06\\
4.12462638290135	9.78724071206339e-06\\
5.87801607227491	4.47431860998466e-06\\
8.37677640068292	6.34050744040452e-07\\
11.9377664171444	-5.13204468225363e-07\\
17.0125427985259	-5.27376353546759e-07\\
24.2446201708233	-2.85097708988411e-07\\
34.5510729459222	-1.34468190343127e-07\\
49.2388263170674	-3.22542976886741e-08\\
70.1703828670382	1.26396689425239e-08\\
100	1.30506568012052e-09\\
};
\addlegendentry{$\text{F: K}^{\text{HS+}}_{\text{dr}}$}

\addplot [color=mycolor2, dashed, line width=1pt]
  table[row sep=crcr]{%
0.0001	0.588008861859843\\
0.0001425102670303	0.589222865646489\\
0.000203091762090473	0.590688655803066\\
0.000289426612471675	0.592462354946948\\
0.000412462638290135	0.594614167261408\\
0.000587801607227491	0.597228450477437\\
0.000837677640068292	0.600411347651806\\
0.00119377664171444	0.604311408298439\\
0.00170125427985259	0.608985272659317\\
0.00242446201708233	0.613921711275362\\
0.00345510729459222	0.617683653868398\\
0.00492388263170674	0.618403481343753\\
0.00701703828670383	0.614828569268647\\
0.01	0.606744215861856\\
0.01425102670303	0.594345366868841\\
0.0203091762090473	0.577407155146968\\
0.0289426612471675	0.55503844496626\\
0.0412462638290135	0.525827349148726\\
0.0587801607227491	0.487966017853879\\
0.0837677640068292	0.439487888347963\\
0.119377664171444	0.378982229659834\\
0.170125427985259	0.306925006908518\\
0.242446201708233	0.227183643460519\\
0.345510729459222	0.147944861348025\\
0.492388263170674	0.0803595662677382\\
0.701703828670383	0.0337224861593366\\
1	0.00959914163786117\\
1.425102670303	0.00128343649350528\\
2.03091762090473	-0.000106038375891372\\
2.89426612471675	3.05488503866446e-05\\
4.12462638290135	9.88106427448285e-05\\
5.87801607227491	5.06582001732285e-05\\
8.37677640068292	9.51171635823852e-06\\
11.9377664171444	-4.46788727079937e-06\\
17.0125427985259	-5.54507072224645e-06\\
24.2446201708233	-3.09254912986522e-06\\
34.5510729459222	-8.61825089352176e-07\\
49.2388263170674	-3.74090678995285e-07\\
70.1703828670382	-3.47319834213289e-07\\
100	-1.18910461243184e-07\\
};
\addlegendentry{$\text{M: K}^\dagger_{\text{dr}}$}

\addplot [color=mycolor2, dashed, line width=1pt, mark size=2pt, mark=o, mark options={solid, mycolor2}]
  table[row sep=crcr]{%
0.0001	0.865786557487422\\
0.0001425102670303	0.867442539501856\\
0.000203091762090473	0.869412850980963\\
0.000289426612471675	0.87175712555242\\
0.000412462638290135	0.874546247859219\\
0.000587801607227491	0.877840982416009\\
0.000837677640068292	0.881707330182974\\
0.00119377664171444	0.88629849881174\\
0.00170125427985259	0.891017088691976\\
0.00242446201708233	0.891703725774322\\
0.00345510729459222	0.878649089955865\\
0.00492388263170674	0.839857890258826\\
0.00701703828670383	0.767921657838543\\
0.01	0.663941899856068\\
0.01425102670303	0.536219371444875\\
0.0203091762090473	0.398008175429711\\
0.0289426612471675	0.266863721150236\\
0.0412462638290135	0.16150784093398\\
0.0587801607227491	0.0932156462835419\\
0.0837677640068292	0.0585640235534923\\
0.119377664171444	0.0433691030254215\\
0.170125427985259	0.034336905265377\\
0.242446201708233	0.0255370502427955\\
0.345510729459222	0.0165919694263903\\
0.492388263170674	0.00892227442950372\\
0.701703828670383	0.00368922816652005\\
1	0.00102640925029697\\
1.425102670303	0.000127554805884979\\
2.03091762090473	-1.52869583549923e-05\\
2.89426612471675	2.65807219910275e-06\\
4.12462638290135	1.05054930764106e-05\\
5.87801607227491	5.45179725767616e-06\\
8.37677640068292	1.01589386162804e-06\\
11.9377664171444	-5.15105318060718e-07\\
17.0125427985259	-6.11655287798253e-07\\
24.2446201708233	-3.20398720678881e-07\\
34.5510729459222	-1.19180702285402e-07\\
49.2388263170674	-6.98902605208538e-08\\
70.1703828670382	-4.39924588196007e-08\\
100	-3.72367455730148e-08\\
};
\addlegendentry{$\text{F: K}^\dagger_{\text{dr}}$}

\end{axis}
\end{tikzpicture}
\subcaption{Pressure}\label{fig:7a}
\end{minipage}%
\begin{minipage}[b]{0.5\textwidth}
\centering
\setlength\figureheight{5cm}
\setlength\figurewidth{5cm}
\definecolor{mycolor1}{rgb}{0.34667,0.53600,0.69067}%
\definecolor{mycolor2}{rgb}{0.91529,0.28157,0.28784}%
\begin{tikzpicture}

\begin{axis}[%
width=0.993\figurewidth,
height=\figureheight,
at={(0\figurewidth,0\figureheight)},
scale only axis,
xmode=log,
xmin=0.0001,
xmax=100,
xminorticks=true,
xlabel style={font=\fontsize{8}{144}\selectfont\color{white!15!black}},
xlabel={Time (s)},
ymin=-9e-5,
ymax=-2e-5,
ylabel style={font=\fontsize{8}{144}\selectfont\color{white!5!black}},
ylabel={Vertical Strain, $\epsilon_{zz}$},
ticklabel style={font=\fontsize{8}{144}},
axis background/.style={fill=white},
legend style={font=\fontsize{8}{144}\selectfont\color{white!5!black}, at={(1,0.95)}, anchor=north east, legend cell align=left, align=left, fill=none, draw=none}
]
\addplot [color=black, line width=1.5pt]
  table[row sep=crcr]{%
0.0001	-3.4439291604396e-05\\
0.0001425102670303	-3.44406569358793e-05\\
0.000203091762090473	-3.44425268367956e-05\\
0.000289426612471675	-3.44451133918904e-05\\
0.000412462638290135	-3.44487230989443e-05\\
0.000587801607227491	-3.44537447118939e-05\\
0.000837677640068292	-3.44608097057718e-05\\
0.00119377664171444	-3.44707180752393e-05\\
0.00170125427985259	-3.44847743340963e-05\\
0.00242446201708233	-3.45048865995503e-05\\
0.00345510729459222	-3.45392636052847e-05\\
0.00492388263170674	-3.45714723607955e-05\\
0.00701703828670383	-3.46271520922399e-05\\
0.01	-3.47048967202394e-05\\
0.01425102670303	-3.4813674971168e-05\\
0.0203091762090473	-3.49646109231999e-05\\
0.0289426612471675	-3.51723078047356e-05\\
0.0412462638290135	-3.54542912883572e-05\\
0.0587801607227491	-3.58294215280801e-05\\
0.0837677640068292	-3.63148620707579e-05\\
0.119377664171444	-3.69184341958033e-05\\
0.170125427985259	-3.76273557648085e-05\\
0.242446201708233	-3.8395165829408e-05\\
0.345510729459222	-3.91360592289729e-05\\
0.492388263170674	-3.97437996985631e-05\\
0.701703828670383	-4.01426592111545e-05\\
1	-4.03360472590734e-05\\
1.425102670303	-4.03968393778533e-05\\
2.03091762090473	-4.04049909889341e-05\\
2.89426612471675	-4.04034918978184e-05\\
4.12462638290135	-4.04031726786469e-05\\
5.87801607227491	-4.04036501474854e-05\\
8.37677640068292	-4.04039901049535e-05\\
11.9377664171444	-4.04040884825744e-05\\
17.0125427985259	-4.04040845074405e-05\\
24.2446201708233	-4.04040657988564e-05\\
34.5510729459222	-4.04040488197786e-05\\
49.2388263170674	-4.04040455951453e-05\\
70.1703828670382	-4.04040400523953e-05\\
100	-4.0404037474257e-05\\
};
\addlegendentry{$\text{K}^{\text{V}}_{\text{dr}}$}

\addplot [color=mycolor1, dotted, line width=1.5pt]
  table[row sep=crcr]{%
0.0001	-3.48295824355617e-05\\
0.0001425102670303	-3.48319669184576e-05\\
0.000203091762090473	-3.48343140093747e-05\\
0.000289426612471675	-3.48380612437957e-05\\
0.000412462638290135	-3.48423971030907e-05\\
0.000587801607227491	-3.48486276272059e-05\\
0.000837677640068292	-3.48570746096949e-05\\
0.00119377664171444	-3.4868473001827e-05\\
0.00170125427985259	-3.48840316942636e-05\\
0.00242446201708233	-3.49054234882082e-05\\
0.00345510729459222	-3.49347362988331e-05\\
0.00492388263170674	-3.4975074533518e-05\\
0.00701703828670383	-3.50310606142059e-05\\
0.01	-3.51090546034997e-05\\
0.01425102670303	-3.52178258010957e-05\\
0.0203091762090473	-3.53689214438816e-05\\
0.0289426612471675	-3.55767556628488e-05\\
0.0412462638290135	-3.5858878538109e-05\\
0.0587801607227491	-3.6234272652084e-05\\
0.0837677640068292	-3.67200898379223e-05\\
0.119377664171444	-3.73242054733791e-05\\
0.170125427985259	-3.80338887733018e-05\\
0.242446201708233	-3.88027308792576e-05\\
0.345510729459222	-3.95448368304816e-05\\
0.492388263170674	-4.01538904761502e-05\\
0.701703828670383	-4.05538671849681e-05\\
1	-4.07479563603601e-05\\
1.425102670303	-4.08090589512278e-05\\
2.03091762090473	-4.08172852097339e-05\\
2.89426612471675	-4.08157832160923e-05\\
4.12462638290135	-4.08154520190776e-05\\
5.87801607227491	-4.08159355432587e-05\\
8.37677640068292	-4.08162743291878e-05\\
11.9377664171444	-4.08163782719205e-05\\
17.0125427985259	-4.08163714119219e-05\\
24.2446201708233	-4.08163524235603e-05\\
34.5510729459222	-4.08163382154578e-05\\
49.2388263170674	-4.08163279301896e-05\\
70.1703828670382	-4.08163268727539e-05\\
100	-4.08163221713187e-05\\
};
\addlegendentry{$\text{K}^{\text{HS+}}_{\text{dr}}$}

\addplot [color=mycolor2, line width=1.5pt]
  table[row sep=crcr]{%
0.0001	-6.1530397252663e-05\\
0.0001425102670303	-6.16524065082058e-05\\
0.000203091762090473	-6.17986096031236e-05\\
0.000289426612471675	-6.19739264396517e-05\\
0.000412462638290135	-6.21843295200385e-05\\
0.000587801607227491	-6.24370866783922e-05\\
0.000837677640068292	-6.27410787068519e-05\\
0.00119377664171444	-6.31071793497088e-05\\
0.00170125427985259	-6.35487888656813e-05\\
0.00242446201708233	-6.40824263479163e-05\\
0.00345510729459222	-6.47284467718723e-05\\
0.00492388263170674	-6.5512325407753e-05\\
0.00701703828670383	-6.64647638308645e-05\\
0.01	-6.76119339628764e-05\\
0.01425102670303	-6.89492632503371e-05\\
0.0203091762090473	-7.0411044723691e-05\\
0.0289426612471675	-7.18636333108818e-05\\
0.0412462638290135	-7.31500630673961e-05\\
0.0587801607227491	-7.41828830008914e-05\\
0.0837677640068292	-7.50129228365083e-05\\
0.119377664171444	-7.57846684229355e-05\\
0.170125427985259	-7.66139035048521e-05\\
0.242446201708233	-7.75124766745683e-05\\
0.345510729459222	-7.83972099301119e-05\\
0.492388263170674	-7.91416599913366e-05\\
0.701703828670383	-7.96462780365657e-05\\
1	-7.99018167182423e-05\\
1.425102670303	-7.99875728557295e-05\\
2.03091762090473	-8.0001154772083e-05\\
2.89426612471675	-7.99995664923653e-05\\
4.12462638290135	-7.99989347639379e-05\\
5.87801607227491	-7.99994727365716e-05\\
8.37677640068292	-7.99999044628163e-05\\
11.9377664171444	-8.00000518028585e-05\\
17.0125427985259	-8.00000576820653e-05\\
24.2446201708233	-8.00000321386134e-05\\
34.5510729459222	-8.00000107947198e-05\\
49.2388263170674	-8.00000024870309e-05\\
70.1703828670382	-8.00000043942109e-05\\
100	-8.00000039208186e-05\\
};
\addlegendentry{$\text{K}^{\dagger}_{\text{dr}}$}

\end{axis}
\end{tikzpicture}
\subcaption{Vertical strain, $\epsilon_{zz}$}\label{fig:7b}
\end{minipage}
\caption{Matrix (`M') and fracture (`F') pressure (a), and vertical strain (b) evolutions for the double-porosity Mandel problem considering isostrain. `$\text{K}^\text{V}_{dr}$' denotes coefficient models for which isostrain is assumed ($K^V_{dr}=19.8 \text{ GPa}$). `$\text{K}^\text{HS+}_{dr}$' are models calculated using the upper Hashin-Shtrikman bound for $K^{HS+}_{dr}$ ($19.5 \text{ GPa}$). `$\text{K}^{\dagger}_{\text{dr}}$' are coefficient models calculated with an arbitrary bulk modulus of $10 \text{ GPa}$.}  \label{fig:7}
\end{figure}
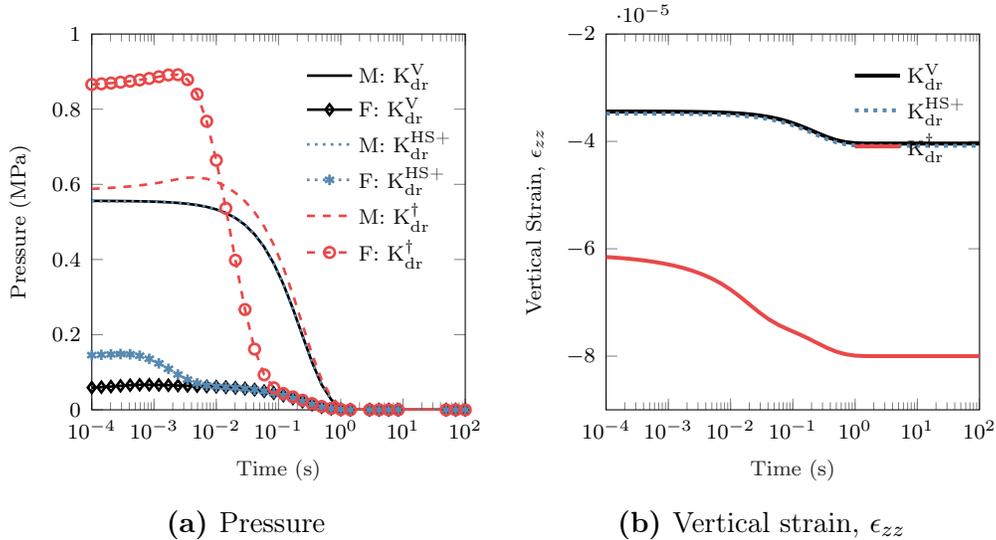

\cref{fig:7a} gives the results for the comparison between coefficient models calculated with different values of $K_{dr}$, pertaining to cases of isostrain (Voigt average), the HS upper bound and an arbitrary value. In the isostrain and HS upper bound cases, matrix pressure evolutions appear to be almost identical. However, for the arbitrary $K_{dr}$ case the fracture pressures are significantly higher. Further, for the arbitrary case the matrix pressures are slightly higher at early to middle times relative to the aforementioned upper bound cases. The contrast in fracture pressures is most pronounced between the isostrain and arbitrary composite bulk modulus cases. For the induced problem, different values of $K_{dr}$, which go on to affect constitutive coefficient calculations, explain the differences in pressure by way of \cref{eqn:74}. 

An alternative, heuristic approach to explaining the pressure distributions in \cref{fig:7a} can be achieved by considering the required geometry that would be necessary to give a $K_{dr}$ that is lower than the Voigt average. Departure from this upper bound occurs when inclusions are arranged so as to weaken the composite. They then take on a greater portion of the distributed stress. We therefore observe a greater induced fracture pressure when using a lower value of $K_{dr}$ compared to the Voigt bound, due to the proportionality between  stress and pressure. 

The early time vertical strains in \cref{fig:7b} can be explained by way of \cref{eqn:73} or through the heuristic argument. With the latter, towards the upper bounds of $K_{dr}$ the matrix supports the majority of deformation. Since the matrix is stiff, deformation is low. In contrast, when fractures are arranged such that they have a greater weakening effect on the solid, deformation is high. 

At later times vertical strain between both the upper bound and arbitrary $K_{dr}$ cases diverge. Towards the upper bounds for $K_{dr}$, $b_f < b_m$. In the case of isostrain this fact is easily seen from the relations in \crefrange{eqn:60}{eqn:61}. The magnitude of $b_m$ means that deformation is more strongly coupled to differences in matrix pressure relative to fracture pressure by way of momentum balance, \cref{eqn:70}. This explains the growth in vertical strain separation at later times shown in \cref{fig:7b}.     

Of further interest, considering these both represent upper bounds on the composite bulk modulus, is the difference in early time fracture pressures between the isostrain and HS upper bound cases. The early time fracture pressure associated with the HS upper bound is over double that of the isostrain case. This highlights the need for caution before making assumptions on the distribution of strain between constituents. 

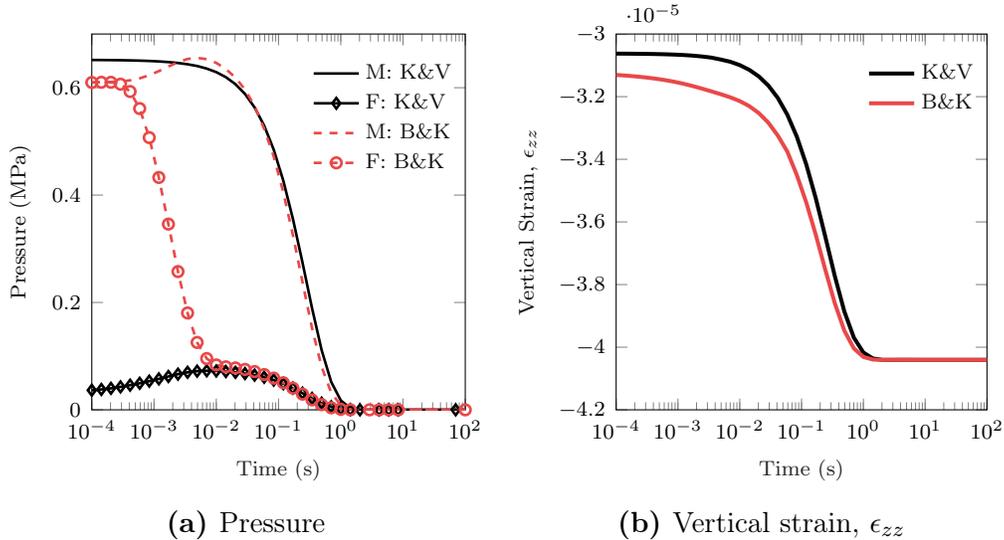
\begin{figure}[h]
\centering
\begin{minipage}[b]{0.5\textwidth}
\centering
\setlength\figureheight{5cm}
\setlength\figurewidth{5cm}
\definecolor{mycolor1}{rgb}{0.91529,0.28157,0.28784}%
\begin{tikzpicture}

\begin{axis}[%
width=0.993\figurewidth,
height=\figureheight,
at={(0\figurewidth,0\figureheight)},
scale only axis,
xmode=log,
xmin=0.0001,
xmax=100,
xminorticks=true,
xlabel style={font=\fontsize{8}{144}\selectfont\color{white!15!black}},
xlabel={Time (s)},
ymin=0,
ymax=0.7,
ylabel style={font=\fontsize{8}{144}\selectfont\color{white!5!black}},
ylabel={Pressure (MPa)},
ticklabel style={font=\fontsize{8}{144}},
ylabel={Pressure (MPa)},
axis background/.style={fill=white},
legend style={font=\fontsize{8}{144}\selectfont\color{white!5!black}, at={(1,0.95)}, anchor=north east, legend cell align=left, align=left, fill=none, draw=none}
]
\addplot [color=black, line width=1pt]
  table[row sep=crcr]{%
0.0001	0.651657723580612\\
0.0001425102670303	0.651549175719571\\
0.000203091762090473	0.651393717366867\\
0.000289426612471675	0.651173502091207\\
0.000412462638290135	0.650863185732571\\
0.000587801607227491	0.650427580061993\\
0.000837677640068292	0.649815120804032\\
0.00119377664171444	0.648955524975898\\
0.00170125427985259	0.647747854736905\\
0.00242446201708233	0.646055159935015\\
0.00345510729459222	0.643676881408856\\
0.00492388263170674	0.640427254180141\\
0.00701703828670383	0.63561829952533\\
0.01	0.629006958109854\\
0.01425102670303	0.619662727695461\\
0.0203091762090473	0.606577603073076\\
0.0289426612471675	0.588370789146657\\
0.0412462638290135	0.563328878928571\\
0.0587801607227491	0.529408097998353\\
0.0837677640068292	0.484465179785256\\
0.119377664171444	0.426779177674387\\
0.170125427985259	0.35598801321149\\
0.242446201708233	0.274570173509219\\
0.345510729459222	0.189288998687616\\
0.492388263170674	0.111200853085715\\
0.701703828670383	0.0520376527297269\\
1	0.0174696243776097\\
1.425102670303	0.00334024232968063\\
2.03091762090473	3.21724682028888e-05\\
2.89426612471675	-4.59890087731049e-05\\
4.12462638290135	0.000103051651502947\\
5.87801607227491	7.83561755710588e-05\\
8.37677640068292	2.30663077409199e-05\\
11.9377664171444	-1.50692751161288e-06\\
17.0125427985259	-5.79985702212852e-06\\
24.2446201708233	-3.90668667791535e-06\\
34.5510729459222	-1.99809522010502e-06\\
49.2388263170674	-5.48815076727766e-07\\
70.1703828670382	-9.24691418782667e-08\\
100	-4.36981247910043e-07\\
};
\addlegendentry{M: K\&V}

\addplot [color=black, line width=1pt, mark size=2pt, mark=diamond, mark options={solid, black}]
  table[row sep=crcr]{%
0.0001	0.036472174131459\\
0.0001425102670303	0.0379712702158186\\
0.000203091762090473	0.040049857534061\\
0.000289426612471675	0.0427434045983444\\
0.000412462638290135	0.0459416006055493\\
0.000587801607227491	0.0495034885083905\\
0.000837677640068292	0.053396642723433\\
0.00119377664171444	0.057651174375148\\
0.00170125427985259	0.0621458074926763\\
0.00242446201708233	0.0664542779220329\\
0.00345510729459222	0.0699323540339156\\
0.00492388263170674	0.0720738583087111\\
0.00701703828670383	0.0727660632506216\\
0.01	0.0723924641404798\\
0.01425102670303	0.0713481290592186\\
0.0203091762090473	0.0697910043922927\\
0.0289426612471675	0.0676372442619251\\
0.0412462638290135	0.0646851108768145\\
0.0587801607227491	0.0606933312510804\\
0.0837677640068292	0.0554179429228873\\
0.119377664171444	0.0486713898534667\\
0.170125427985259	0.0404355703079344\\
0.242446201708233	0.0310270281186269\\
0.345510729459222	0.0212554776376321\\
0.492388263170674	0.0123957957456694\\
0.701703828670383	0.00575421159833375\\
1	0.00191480819369339\\
1.425102670303	0.000362027010230721\\
2.03091762090473	3.19238395765144e-06\\
2.89426612471675	-4.33917014183756e-06\\
4.12462638290135	1.16858898663012e-05\\
5.87801607227491	8.64198640590977e-06\\
8.37677640068292	2.47524166760895e-06\\
11.9377664171444	-2.21187739504089e-07\\
17.0125427985259	-6.66861351455377e-07\\
24.2446201708233	-4.63930995084287e-07\\
34.5510729459222	-2.3783501905498e-07\\
49.2388263170674	-6.39542675476676e-08\\
70.1703828670382	2.44529064348256e-08\\
100	-2.05039104605465e-08\\
};
\addlegendentry{F: K\&V}

\addplot [color=mycolor1, dashed, line width=1pt]
  table[row sep=crcr]{%
0.0001	0.609778338086326\\
0.0001425102670303	0.610193921723257\\
0.000203091762090473	0.610737725033068\\
0.000289426612471675	0.611696293405628\\
0.000412462638290135	0.61368689044973\\
0.000587801607227491	0.617421490023211\\
0.000837677640068292	0.623250826055953\\
0.00119377664171444	0.630890048112198\\
0.00170125427985259	0.639449684155491\\
0.00242446201708233	0.647548011903603\\
0.00345510729459222	0.653467996220269\\
0.00492388263170674	0.655612590581553\\
0.00701703828670383	0.653163020589675\\
0.01	0.646190747992625\\
0.01425102670303	0.634892948029066\\
0.0203091762090473	0.618779021217881\\
0.0289426612471675	0.596482810763303\\
0.0412462638290135	0.566061180615315\\
0.0587801607227491	0.523317127689728\\
0.0837677640068292	0.472040179894624\\
0.119377664171444	0.405205113272063\\
0.170125427985259	0.32563071988974\\
0.242446201708233	0.238065612314434\\
0.345510729459222	0.152028286804417\\
0.492388263170674	0.0800916162197418\\
0.701703828670383	0.0320282029897274\\
1	0.00838212985220713\\
1.425102670303	0.000873161022800995\\
2.03091762090473	-0.000134102807680307\\
2.89426612471675	6.07939975438108e-05\\
4.12462638290135	0.000106330063790241\\
5.87801607227491	4.78446180484676e-05\\
8.37677640068292	6.75339599763343e-06\\
11.9377664171444	-5.25517059647253e-06\\
17.0125427985259	-5.07371159904286e-06\\
24.2446201708233	-2.99774866168618e-06\\
34.5510729459222	-9.54385083878964e-07\\
49.2388263170674	-3.18903394302738e-07\\
70.1703828670382	-1.52315515795329e-07\\
100	4.89009784499141e-07\\
};
\addlegendentry{M: B\&K}

\addplot [color=mycolor1, dashed, line width=1pt, mark size=2pt, mark=o, mark options={solid, mycolor1}]
  table[row sep=crcr]{%
0.0001	0.609738798696342\\
0.0001425102670303	0.610122921426951\\
0.000203091762090473	0.610070857173647\\
0.000289426612471675	0.60657339604029\\
0.000412462638290135	0.592903046052845\\
0.000587801607227491	0.561142042976977\\
0.000837677640068292	0.507072168731907\\
0.00119377664171444	0.432832630332931\\
0.00170125427985259	0.345955166087583\\
0.00242446201708233	0.257515649137109\\
0.00345510729459222	0.180450310519295\\
0.00492388263170674	0.125526134132848\\
0.00701703828670383	0.0954761226932603\\
0.01	0.083647671343247\\
0.01425102670303	0.08006229157458\\
0.0203091762090473	0.0780483692735239\\
0.0289426612471675	0.0752455895420793\\
0.0412462638290135	0.0712496694244265\\
0.0587801607227491	0.0657442922266081\\
0.0837677640068292	0.0590412035777614\\
0.119377664171444	0.0504687401836497\\
0.170125427985259	0.0403393207305322\\
0.242446201708233	0.0292915806644524\\
0.345510729459222	0.0185525578934025\\
0.492388263170674	0.009682757105818\\
0.701703828670383	0.00383223358663367\\
1	0.000990823766005333\\
1.425102670303	0.000101063358712965\\
2.03091762090473	-1.52112620911036e-05\\
2.89426612471675	8.07582516615235e-06\\
4.12462638290135	1.29132438226808e-05\\
5.87801607227491	5.67031575136401e-06\\
8.37677640068292	7.48987019498015e-07\\
11.9377664171444	-6.86402930645727e-07\\
17.0125427985259	-6.15423065877264e-07\\
24.2446201708233	-3.60588399337383e-07\\
34.5510729459222	-1.16642397932348e-07\\
49.2388263170674	-4.23785247711872e-08\\
70.1703828670382	-1.04552735720408e-08\\
100	3.52588825607637e-08\\
};
\addlegendentry{F: B\&K}

\end{axis}
\end{tikzpicture}
\subcaption{Pressure}\label{fig:8a}
\end{minipage}%
\begin{minipage}[b]{0.5\textwidth}
\centering
\setlength\figureheight{5cm}
\setlength\figurewidth{5cm}
\definecolor{mycolor1}{rgb}{0.91529,0.28157,0.28784}%
\begin{tikzpicture}

\begin{axis}[%
width=0.986\figurewidth,
height=\figureheight,
at={(0\figurewidth,0\figureheight)},
scale only axis,
xmode=log,
xmin=0.0001,
xmax=100,
xminorticks=true,
xlabel style={font=\fontsize{8}{144}\selectfont\color{white!15!black}},
xlabel={Time (s)},
ymin=-4.2e-5,
ymax=-3e-5,
ylabel style={font=\fontsize{8}{144}\selectfont\color{white!5!black}},
ylabel={Vertical Strain, $\epsilon_{zz}$},
ticklabel style={font=\fontsize{8}{144}},
axis background/.style={fill=white},
legend style={font=\fontsize{8}{144}\selectfont\color{white!5!black}, at={(1,0.95)}, anchor=north east, legend cell align=left, align=left, fill=none, draw=none}
]
\addplot [color=black, line width=1.5pt]
  table[row sep=crcr]{%
0.0001	-3.06257100034757e-05\\
0.0001425102670303	-3.06275468881251e-05\\
0.000203091762090473	-3.06300847494513e-05\\
0.000289426612471675	-3.06336396802857e-05\\
0.000412462638290135	-3.06386084342997e-05\\
0.000587801607227491	-3.06455833288566e-05\\
0.000837677640068292	-3.0655348613637e-05\\
0.00119377664171444	-3.06691214667839e-05\\
0.00170125427985259	-3.06885022357824e-05\\
0.00242446201708233	-3.07158548624028e-05\\
0.00345510729459222	-3.07527339118044e-05\\
0.00492388263170674	-3.08144764809537e-05\\
0.00701703828670383	-3.08851569460548e-05\\
0.01	-3.09924053779205e-05\\
0.01425102670303	-3.11430208181824e-05\\
0.0203091762090473	-3.13528399846454e-05\\
0.0289426612471675	-3.164311465797e-05\\
0.0412462638290135	-3.20401662293112e-05\\
0.0587801607227491	-3.25742273749555e-05\\
0.0837677640068292	-3.32760662308442e-05\\
0.119377664171444	-3.41680513019634e-05\\
0.170125427985259	-3.52489401524672e-05\\
0.242446201708233	-3.64729021625837e-05\\
0.345510729459222	-3.77309663072936e-05\\
0.492388263170674	-3.88583145941028e-05\\
0.701703828670383	-3.96930509800874e-05\\
1	-4.01697778356941e-05\\
1.425102670303	-4.03602838829574e-05\\
2.03091762090473	-4.04036831455294e-05\\
2.89426612471675	-4.04044644069222e-05\\
4.12462638290135	-4.04025575853192e-05\\
5.87801607227491	-4.04029767616871e-05\\
8.37677640068292	-4.04037468223483e-05\\
11.9377664171444	-4.0404071907845e-05\\
17.0125427985259	-4.04041280152952e-05\\
24.2446201708233	-4.04040973020058e-05\\
34.5510729459222	-4.04040718825328e-05\\
49.2388263170674	-4.04040472192991e-05\\
70.1703828670382	-4.04040391202833e-05\\
100	-4.04040434375518e-05\\
};
\addlegendentry{K\&V}

\addplot [color=mycolor1, line width=1.5pt]
  table[row sep=crcr]{%
0.0001	-3.13055367780748e-05\\
0.0001425102670303	-3.13267418786444e-05\\
0.000203091762090473	-3.1352161139547e-05\\
0.000289426612471675	-3.13826847269107e-05\\
0.000412462638290135	-3.1419404897464e-05\\
0.000587801607227491	-3.14636976591565e-05\\
0.000837677640068292	-3.15172706205925e-05\\
0.00119377664171444	-3.15816327095586e-05\\
0.00170125427985259	-3.16569353143777e-05\\
0.00242446201708233	-3.17408504584298e-05\\
0.00345510729459222	-3.18295485417698e-05\\
0.00492388263170674	-3.19211788225597e-05\\
0.00701703828670383	-3.20206940344532e-05\\
0.01	-3.2141495410055e-05\\
0.01425102670303	-3.23022846947737e-05\\
0.0203091762090473	-3.25237313482201e-05\\
0.0289426612471675	-3.28282562394013e-05\\
0.0412462638290135	-3.32413757082435e-05\\
0.0587801607227491	-3.37532375061756e-05\\
0.0837677640068292	-3.45002595688468e-05\\
0.119377664171444	-3.53822136969621e-05\\
0.170125427985259	-3.64149557130923e-05\\
0.242446201708233	-3.75296008385054e-05\\
0.345510729459222	-3.85998532454584e-05\\
0.492388263170674	-3.94719806657236e-05\\
0.701703828670383	-4.00392478480635e-05\\
1	-4.03109408963814e-05\\
1.425102670303	-4.03947458773664e-05\\
2.03091762090473	-4.04054001489073e-05\\
2.89426612471675	-4.04031864103978e-05\\
4.12462638290135	-4.04027869417947e-05\\
5.87801607227491	-4.04035103256642e-05\\
8.37677640068292	-4.04039748758203e-05\\
11.9377664171444	-4.04041149303241e-05\\
17.0125427985259	-4.04040994792575e-05\\
24.2446201708233	-4.04040750739511e-05\\
34.5510729459222	-4.0404051844937e-05\\
49.2388263170674	-4.04040441161027e-05\\
70.1703828670382	-4.04040518986096e-05\\
100	-4.04040438111212e-05\\
};
\addlegendentry{B\&K}

\end{axis}
\end{tikzpicture}
\subcaption{Vertical strain, $\epsilon_{zz}$}\label{fig:8b}
\end{minipage}
\caption{Matrix (`M') and fracture (`F') pressure (a), and vertical strain (b) evolutions for the double-porosity Mandel problem considering incompressible grain isostrain. `K$\&$V' and `B$\&$K' denote coefficient models associated with \cite{Khalili1996} and \cite{Borja2009} respectively. The latter includes the assumption $\pderiv{\psi_\alpha}{t}\approx0$. This allows us to map the effective stress model from \cite{Borja2009} onto the constitutive model \crefrange{eqn:17}{eqn:19}.}  \label{fig:8}
\end{figure}

\cref{fig:8a} displays the results for the incompressible grain isostrain investigation. For the case of coefficient models from \cite{Khalili1996}, the induced matrix pressure is significantly larger than the induced fracture pressure. Conversely, when using coefficient models from \cite{Borja2009} with the assumption $\pderiv{\psi_\alpha}{t} \approx 0$, induced matrix and fracture pressures are equal. This can be explained by considering \cref{eqn:73} which allows us to equate induced variations in matrix and fracture pressures such that

\begin{equation}
\begin{split}
\left(\frac{1}{M_m} - \frac{M_f}{(Q)^2}\right) & \left(\frac{M_fb_f}{Q}-b_m\right)^{-1} \text{d}p_m \\
&= \left(\frac{1}{M_f} - \frac{M_m}{(Q)^2}\right) \left(\frac{M_mb_m}{Q}-b_f\right)^{-1} \text{d}p_f.
\end{split}
\label{eqn:77}
\end{equation}

\noindent
Assuming $\pderiv{\psi_\alpha}{t} \approx 0$ together with the effective Biot coefficient expressions from \cite{Borja2009} (\cref{tab:1}), \cref{eqn:77} reduces to

\begin{equation}
\left(\frac{\phi^0_m}{K_l} \frac{1-\phi_s}{\phi^0_m}\right)\text{d}p_m =
\left(\frac{\phi^0_f}{K_l} \frac{1-\phi_s}{\phi^0_f}\right)\text{d}p_f,
\label{eqn:78}
\end{equation}

\noindent
from which it is easy to see that $\text{d}p_m=\text{d}p_f$. 

Under isostrain, we would expect the distribution of stress required to maintain strain uniformity between matrix and fractures to lead to disparate matrix and fracture pressures. The result $\text{d}p_m=\text{d}p_f$ therefore suggests that the closure condition $\pderiv{\psi_\alpha}{t}\approx 0$ may be an even stronger assumption than incompressible grain isostrain alone. 

\cref{fig:8b} shows vertical strain is lower at early times when using coefficient models from \cite{Khalili1996} under incompressible grain isostrain. This can be explained by \cref{eqn:73}, which is affected by differences in $b_\alpha$ arising from each set of coefficient models.  

In light of the discussions in Section 4.2.2 and the results presented herein, assuming $\pderiv{\psi_\alpha}{t}\approx 0$ appears to be a strong closure assumption to make. Thus, we suggest further development of a constitutive model for $\psi_\alpha$, along with its relationship to the constitutive model shown in \crefrange{eqn:17}{eqn:19}.

\subsection{Case 4: Explicit decoupling - isostress}
\begin{figure}[h]
\centering
\begin{minipage}[b]{0.5\textwidth}
\centering
\setlength\figureheight{5cm}
\setlength\figurewidth{5cm}
\definecolor{mycolor1}{rgb}{0.34667,0.53600,0.69067}%
\definecolor{mycolor2}{rgb}{0.91529,0.28157,0.28784}%
\begin{tikzpicture}

\begin{axis}[%
width=0.993\figurewidth,
height=\figureheight,
at={(0\figurewidth,0\figureheight)},
scale only axis,
xmode=log,
xmin=0.0001,
xmax=100,
xminorticks=true,
xlabel style={font=\fontsize{8}{144}\selectfont\color{white!15!black}},
xlabel={Time (s)},
ymin=0,
ymax=1.2,
ylabel style={font=\fontsize{8}{144}\selectfont\color{white!5!black}},
ylabel={Pressure (MPa)},
ticklabel style={font=\fontsize{8}{144}},
ylabel={Pressure (MPa)},
axis background/.style={fill=white},
legend style={font=\fontsize{8}{144}\selectfont\color{white!5!black}, at={(1,0.95)}, anchor=north east, legend cell align=left, align=left, fill=none, draw=none}
]
\addplot [color=black, line width=1pt]
  table[row sep=crcr]{%
0.0001	0.619787658109028\\
0.0001425102670303	0.620969850326927\\
0.000203091762090473	0.622400584773017\\
0.000289426612471675	0.624136374239443\\
0.000412462638290135	0.626248289665474\\
0.000587801607227491	0.628826323544799\\
0.000837677640068292	0.631985406898223\\
0.00119377664171444	0.635874134013779\\
0.00170125427985259	0.640683852969082\\
0.00242446201708233	0.646655596644854\\
0.00345510729459222	0.654119293501936\\
0.00492388263170674	0.663501957930073\\
0.00701703828670383	0.674877052282382\\
0.01	0.686996718801293\\
0.01425102670303	0.697006732127003\\
0.0203091762090473	0.701694573420746\\
0.0289426612471675	0.698948175009504\\
0.0412462638290135	0.687479568600161\\
0.0587801607227491	0.664976987934012\\
0.0837677640068292	0.626787781339311\\
0.119377664171444	0.566739386793357\\
0.170125427985259	0.480444809549927\\
0.242446201708233	0.370429187411695\\
0.345510729459222	0.249940818180098\\
0.492388263170674	0.140423527266033\\
0.701703828670383	0.0612740934158795\\
1	0.0184728128522968\\
1.425102670303	0.00282681207101768\\
2.03091762090473	-0.000139713722272247\\
2.89426612471675	1.49829081071301e-05\\
4.12462638290135	0.000162078898123463\\
5.87801607227491	9.4106811092813e-05\\
8.37677640068292	2.09289459475241e-05\\
11.9377664171444	-5.32896018862166e-06\\
17.0125427985259	-7.47444718780482e-06\\
24.2446201708233	-5.1656643460297e-06\\
34.5510729459222	-1.04872545983992e-06\\
49.2388263170674	-1.04840312091009e-06\\
70.1703828670382	-3.03475660360041e-07\\
100	3.06730887793247e-07\\
};
\addlegendentry{$\text{M: K}^{\text{R}}_{\text{dr}}$}

\addplot [color=black, line width=1pt, mark size=2pt, mark=diamond, mark options={solid, black}]
  table[row sep=crcr]{%
0.0001	0.972909952841446\\
0.0001425102670303	0.974647104200345\\
0.000203091762090473	0.976723586756574\\
0.000289426612471675	0.979206297074611\\
0.000412462638290135	0.982175580601343\\
0.000587801607227491	0.985727988167327\\
0.000837677640068292	0.989980148256814\\
0.00119377664171444	0.995075541995754\\
0.00170125427985259	1.00118273973569\\
0.00242446201708233	1.0084725159954\\
0.00345510729459222	1.01719630023915\\
0.00492388263170674	1.02762414092864\\
0.00701703828670383	1.03827712323416\\
0.01	1.04236780698056\\
0.01425102670303	1.02715520718419\\
0.0203091762090473	0.979619943789992\\
0.0289426612471675	0.893757062755217\\
0.0412462638290135	0.772527809477008\\
0.0587801607227491	0.624969356205751\\
0.0837677640068292	0.464540577023871\\
0.119377664171444	0.309795071806692\\
0.170125427985259	0.181512764242389\\
0.242446201708233	0.0933131180527383\\
0.345510729459222	0.043591983978981\\
0.492388263170674	0.0193508877971128\\
0.701703828670383	0.00787891826051593\\
1	0.00254650548471074\\
1.425102670303	0.000524232089599426\\
2.03091762090473	4.38651625508786e-05\\
2.89426612471675	6.92502707151141e-06\\
4.12462638290135	1.12959909723238e-05\\
5.87801607227491	4.12289233482371e-06\\
8.37677640068292	-1.45422379663939e-06\\
11.9377664171444	-2.34600834812093e-06\\
17.0125427985259	-1.6051222160256e-06\\
24.2446201708233	-8.55560889179639e-07\\
34.5510729459222	-2.00687521715092e-07\\
49.2388263170674	-5.287600431581e-08\\
70.1703828670382	1.90136969141074e-08\\
100	4.85107638801406e-08\\
};
\addlegendentry{$\text{F: K}^{\text{R}}_{\text{dr}}$}

\addplot [color=mycolor1, dotted, line width=1pt]
  table[row sep=crcr]{%
0.0001	0.610124516647809\\
0.0001425102670303	0.611451869750459\\
0.000203091762090473	0.61305635173876\\
0.000289426612471675	0.615000194523449\\
0.000412462638290135	0.617361403095081\\
0.000587801607227491	0.620238824878845\\
0.000837677640068292	0.623758278357768\\
0.00119377664171444	0.62807495292856\\
0.00170125427985259	0.633388720208943\\
0.00242446201708233	0.639979995662234\\
0.00345510729459222	0.648003114901674\\
0.00492388263170674	0.65675166576966\\
0.00701703828670383	0.664097598265508\\
0.01	0.667213345993258\\
0.01425102670303	0.664101959399314\\
0.0203091762090473	0.654147711728837\\
0.0289426612471675	0.637036812839867\\
0.0412462638290135	0.611391807809025\\
0.0587801607227491	0.574500418191349\\
0.0837677640068292	0.523162012676989\\
0.119377664171444	0.45519521726512\\
0.170125427985259	0.371278046325321\\
0.242446201708233	0.276604679400683\\
0.345510729459222	0.181499839049916\\
0.492388263170674	0.0996374737094881\\
0.701703828670383	0.0424788756862022\\
1	0.012414339500529\\
1.425102670303	0.00177597251777151\\
2.03091762090473	-0.000114296515107929\\
2.89426612471675	2.66642334616126e-05\\
4.12462638290135	0.000118809809982987\\
5.87801607227491	6.46693813997079e-05\\
8.37677640068292	1.27534874278261e-05\\
11.9377664171444	-5.0337648304849e-06\\
17.0125427985259	-5.70810796738429e-06\\
24.2446201708233	-3.6421333430643e-06\\
34.5510729459222	-1.04513023995101e-06\\
49.2388263170674	7.71797495985119e-09\\
70.1703828670382	-7.88954879592925e-07\\
100	-7.24507156502626e-08\\
};
\addlegendentry{$\text{M: K}^{\text{HS-}}_{\text{dr}}$}

\addplot [color=mycolor1, dotted, line width=1pt, mark size=2pt, mark=triangle, mark options={solid, mycolor1}]
  table[row sep=crcr]{%
0.0001	0.947207041730001\\
0.0001425102670303	0.949141897041022\\
0.000203091762090473	0.951453000908954\\
0.000289426612471675	0.954213727420648\\
0.000412462638290135	0.957511954372567\\
0.000587801607227491	0.961454696113903\\
0.000837677640068292	0.966171030343889\\
0.00119377664171444	0.971790889459395\\
0.00170125427985259	0.978466923906741\\
0.00242446201708233	0.986466547981559\\
0.00345510729459222	0.995225603124215\\
0.00492388263170674	1.00021193093568\\
0.00701703828670383	0.990704818079342\\
0.01	0.953341925504837\\
0.01425102670303	0.879591522968611\\
0.0203091762090473	0.769914047922862\\
0.0289426612471675	0.632109838091309\\
0.0412462638290135	0.478845056298385\\
0.0587801607227491	0.327694075270342\\
0.0837677640068292	0.199375063035146\\
0.119377664171444	0.109454378385924\\
0.170125427985259	0.0587325085918215\\
0.242446201708233	0.0341804605581883\\
0.345510729459222	0.0209650641881042\\
0.492388263170674	0.0116302913129954\\
0.701703828670383	0.00505578100777378\\
1	0.00149274743753922\\
1.425102670303	0.000209324364206202\\
2.03091762090473	-2.25998818097433e-05\\
2.89426612471675	-5.14232744033971e-06\\
4.12462638290135	8.9550951163194e-06\\
5.87801607227491	5.43869278708053e-06\\
8.37677640068292	7.00658401962945e-07\\
11.9377664171444	-7.72958932407261e-07\\
17.0125427985259	-6.58853419977333e-07\\
24.2446201708233	-4.3634330741272e-07\\
34.5510729459222	-1.29128038521423e-07\\
49.2388263170674	-3.11394085828883e-09\\
70.1703828670382	-6.71027684450949e-08\\
100	2.90361412140317e-08\\
};
\addlegendentry{$\text{F: K}^{\text{HS-}}_{\text{dr}}$}

\addplot [color=mycolor2, dashed, line width=1pt]
  table[row sep=crcr]{%
0.0001	0.580754304940939\\
0.0001425102670303	0.581860166589434\\
0.000203091762090473	0.583195561379632\\
0.000289426612471675	0.584811320849153\\
0.000412462638290135	0.586767340979136\\
0.000587801607227491	0.589141880515042\\
0.000837677640068292	0.592036120335977\\
0.00119377664171444	0.595411627205305\\
0.00170125427985259	0.598710317012416\\
0.00242446201708233	0.600724958910223\\
0.00345510729459222	0.600140169811865\\
0.00492388263170674	0.596274239569277\\
0.00701703828670383	0.589186176686491\\
0.01	0.579113876052928\\
0.01425102670303	0.56596099886219\\
0.0203091762090473	0.549185762875584\\
0.0289426612471675	0.527791070536424\\
0.0412462638290135	0.500200861119922\\
0.0587801607227491	0.464305852799003\\
0.0837677640068292	0.417944855873817\\
0.119377664171444	0.359809480302866\\
0.170125427985259	0.290566169593838\\
0.242446201708233	0.214158940905914\\
0.345510729459222	0.138594525681108\\
0.492388263170674	0.0745973561136305\\
0.701703828670383	0.0308803128368154\\
1	0.00859285882824185\\
1.425102670303	0.00108128284512459\\
2.03091762090473	-0.000107108161473421\\
2.89426612471675	3.58129795915845e-05\\
4.12462638290135	9.33868959735003e-05\\
5.87801607227491	4.64668951167035e-05\\
8.37677640068292	8.1669706360379e-06\\
11.9377664171444	-4.73908549858393e-06\\
17.0125427985259	-4.61748262960621e-06\\
24.2446201708233	-2.41910294564707e-06\\
34.5510729459222	-1.04106811882394e-06\\
49.2388263170674	-2.14009840910168e-07\\
70.1703828670382	1.19412291583102e-07\\
100	-5.63321741420872e-07\\
};
\addlegendentry{$\text{M: K}^{\text{AHS}}_{\text{dr}}$}

\addplot [color=mycolor2, dashed, line width=1pt, mark size=2pt, mark=o, mark options={solid, mycolor2}]
  table[row sep=crcr]{%
0.0001	0.839397065890786\\
0.0001425102670303	0.840838353187843\\
0.000203091762090473	0.842546107011729\\
0.000289426612471675	0.844565393044083\\
0.000412462638290135	0.846920069517188\\
0.000587801607227491	0.849658784049216\\
0.000837677640068292	0.852851028994258\\
0.00119377664171444	0.85529313634984\\
0.00170125427985259	0.851569786304106\\
0.00242446201708233	0.831207116091611\\
0.00345510729459222	0.783257185783527\\
0.00492388263170674	0.702840887269432\\
0.00701703828670383	0.593558814505297\\
0.01	0.46553073395403\\
0.01425102670303	0.333559436831699\\
0.0203091762090473	0.215887549368814\\
0.0289426612471675	0.128992783776874\\
0.0412462638290135	0.0785810613586609\\
0.0587801607227491	0.0561389259044828\\
0.0837677640068292	0.046888536092103\\
0.119377664171444	0.0402634796071543\\
0.170125427985259	0.0325720139046728\\
0.242446201708233	0.0238792431006279\\
0.345510729459222	0.0153165856065768\\
0.492388263170674	0.00815964822111545\\
0.701703828670383	0.0033373711398148\\
1	0.000913070650606584\\
1.425102670303	0.000109884316966678\\
2.03091762090473	-1.23969231379817e-05\\
2.89426612471675	4.34996964040246e-06\\
4.12462638290135	1.03316876904622e-05\\
5.87801607227491	5.04485471713426e-06\\
8.37677640068292	8.86760538382525e-07\\
11.9377664171444	-5.13134884421462e-07\\
17.0125427985259	-4.89332927147288e-07\\
24.2446201708233	-2.57271317088056e-07\\
34.5510729459222	-1.45879617531981e-07\\
49.2388263170674	-2.02503638869353e-08\\
70.1703828670382	3.84725972839653e-08\\
100	6.89044787025709e-10\\
};
\addlegendentry{$\text{F: K}^{\text{AHS}}_{\text{dr}}$}

\end{axis}
\end{tikzpicture}
\subcaption{Pressure}\label{fig:9a}
\end{minipage}%
\begin{minipage}[b]{0.5\textwidth}
\centering
\setlength\figureheight{5cm}
\setlength\figurewidth{5cm}
\definecolor{mycolor1}{rgb}{0.34667,0.53600,0.69067}%
\definecolor{mycolor2}{rgb}{0.91529,0.28157,0.28784}%
\begin{tikzpicture}

\begin{axis}[%
width=0.986\figurewidth,
height=\figureheight,
at={(0\figurewidth,0\figureheight)},
scale only axis,
xmode=log,
xmin=0.0001,
xmax=100,
xminorticks=true,
xlabel style={font=\fontsize{8}{144}\selectfont\color{white!15!black}},
xlabel={Time (s)},
ymin=-0.00025,
ymax=0,
ylabel style={font=\fontsize{8}{144}\selectfont\color{white!5!black}},
ylabel={Vertical Strain, $\epsilon_{zz}$},
ticklabel style={font=\fontsize{8}{144}},
axis background/.style={fill=white},
legend style={font=\fontsize{8}{144}\selectfont\color{white!5!black}, at={(1,0.95)}, anchor=north east, legend cell align=left, align=left, fill=none, draw=none}
]
\addplot [color=black, line width=1.5pt]
  table[row sep=crcr]{%
0.0001	-0.000162328529810051\\
0.0001425102670303	-0.000162621491486381\\
0.000203091762090473	-0.000162972407282117\\
0.000289426612471675	-0.00016339301294524\\
0.000412462638290135	-0.000163897537900363\\
0.000587801607227491	-0.000164503281065599\\
0.000837677640068292	-0.000165231338353445\\
0.00119377664171444	-0.000166107536379478\\
0.00170125427985259	-0.00016716362892315\\
0.00242446201708233	-0.000168438841747622\\
0.00345510729459222	-0.000169981916772528\\
0.00492388263170674	-0.00017185382683399\\
0.00701703828670383	-0.000174131406707673\\
0.01	-0.000176911782260896\\
0.01425102670303	-0.000180318691021374\\
0.0203091762090473	-0.000184512463296516\\
0.0289426612471675	-0.000189685042760458\\
0.0412462638290135	-0.000195989807963725\\
0.0587801607227491	-0.000203390089926606\\
0.0837677640068292	-0.000211499077271475\\
0.119377664171444	-0.000219536060601127\\
0.170125427985259	-0.00022653361792016\\
0.242446201708233	-0.000231801447543482\\
0.345510729459222	-0.000235292566251493\\
0.492388263170674	-0.000237441435459597\\
0.701703828670383	-0.000238692694869565\\
1	-0.000239321208354423\\
1.425102670303	-0.000239551739543246\\
2.03091762090473	-0.000239599185661755\\
2.89426612471675	-0.000239599403692013\\
4.12462638290135	-0.000239598014597036\\
5.87801607227491	-0.000239598989504633\\
8.37677640068292	-0.000239599893323628\\
11.9377664171444	-0.00023960015338658\\
17.0125427985259	-0.000239600149823459\\
24.2446201708233	-0.000239600091180817\\
34.5510729459222	-0.00023960002823471\\
49.2388263170674	-0.000239600010513737\\
70.1703828670382	-0.000239600003113761\\
100	-0.000239600000668982\\
};
\addlegendentry{$\text{K}^{\text{R}}_{\text{dr}}$}

\addplot [color=mycolor1, dotted, line width=1.5pt]
  table[row sep=crcr]{%
0.0001	-9.97523379405765e-05\\
0.0001425102670303	-9.99596332416148e-05\\
0.000203091762090473	-0.000100208067888663\\
0.000289426612471675	-0.000100506026075155\\
0.000412462638290135	-0.000100863695728111\\
0.000587801607227491	-0.000101293493557357\\
0.000837677640068292	-0.000101810603540838\\
0.00119377664171444	-0.000102433667315593\\
0.00170125427985259	-0.000103185686057372\\
0.00242446201708233	-0.000104095192142215\\
0.00345510729459222	-0.00010519780303885\\
0.00492388263170674	-0.000106538146857658\\
0.00701703828670383	-0.000108172269384008\\
0.01	-0.000110171757079403\\
0.01425102670303	-0.000112625288363604\\
0.0203091762090473	-0.000115615580393902\\
0.0289426612471675	-0.000119152976393274\\
0.0412462638290135	-0.000123091022853115\\
0.0587801607227491	-0.000127086819981688\\
0.0837677640068292	-0.000130679072785874\\
0.119377664171444	-0.000133510704501629\\
0.170125427985259	-0.000135552332225843\\
0.242446201708233	-0.000137063198697806\\
0.345510729459222	-0.000138296611383094\\
0.492388263170674	-0.000139299384132242\\
0.701703828670383	-0.000139992606926851\\
1	-0.000140354607054546\\
1.425102670303	-0.000140480944625091\\
2.03091762090473	-0.000140502728059442\\
2.89426612471675	-0.000140500857740498\\
4.12462638290135	-0.000140499762459248\\
5.87801607227491	-0.0001405003794741\\
8.37677640068292	-0.000140500973892994\\
11.9377664171444	-0.000140501167946987\\
17.0125427985259	-0.000140501163938496\\
24.2446201708233	-0.000140501144608424\\
34.5510729459222	-0.000140501108992444\\
49.2388263170674	-0.000140501104385749\\
70.1703828670382	-0.000140501116822944\\
100	-0.00014050109627958\\
};
\addlegendentry{$\text{K}^{\text{HS-}}_{\text{dr}}$}

\addplot [color=mycolor2, line width=1.5pt]
  table[row sep=crcr]{%
0.0001	-5.02960329542481e-05\\
0.0001425102670303	-5.03879491612109e-05\\
0.000203091762090473	-5.04980663283288e-05\\
0.000289426612471675	-5.06300735918237e-05\\
0.000412462638290135	-5.0788446101006e-05\\
0.000587801607227491	-5.0978609235728e-05\\
0.000837677640068292	-5.12072028584058e-05\\
0.00119377664171444	-5.1482314748248e-05\\
0.00170125427985259	-5.18138439591873e-05\\
0.00242446201708233	-5.22138859436004e-05\\
0.00345510729459222	-5.26974212309903e-05\\
0.00492388263170674	-5.32813941258778e-05\\
0.00701703828670383	-5.39762780245555e-05\\
0.01	-5.47690250015133e-05\\
0.01425102670303	-5.56086406605248e-05\\
0.0203091762090473	-5.6412025410243e-05\\
0.0289426612471675	-5.71028455229353e-05\\
0.0412462638290135	-5.76667772835329e-05\\
0.0587801607227491	-5.8170468047083e-05\\
0.0837677640068292	-5.8713873731135e-05\\
0.119377664171444	-5.93647488064031e-05\\
0.170125427985259	-6.01326124944208e-05\\
0.242446201708233	-6.0972228566979e-05\\
0.345510729459222	-6.1791488472027e-05\\
0.492388263170674	-6.24735916159058e-05\\
0.701703828670383	-6.29303303318812e-05\\
1	-6.31579224342753e-05\\
1.425102670303	-6.32324561967244e-05\\
2.03091762090473	-6.3243590665842e-05\\
2.89426612471675	-6.32420381579546e-05\\
4.12462638290135	-6.32415507882562e-05\\
5.87801607227491	-6.32420615693793e-05\\
8.37677640068292	-6.32424559963131e-05\\
11.9377664171444	-6.3242580401455e-05\\
17.0125427985259	-6.32425773662469e-05\\
24.2446201708233	-6.32425506153312e-05\\
34.5510729459222	-6.32425368059153e-05\\
49.2388263170674	-6.32425329376029e-05\\
70.1703828670382	-6.32425249617173e-05\\
100	-6.32425255639564e-05\\
};
\addlegendentry{$\text{K}^{\text{AHS}}_{\text{dr}}$}

\end{axis}
\end{tikzpicture}
\subcaption{Vertical strain, $\epsilon_{zz}$}\label{fig:9b}
\end{minipage}
\caption{Matrix (`M') and fracture (`F') pressure (a), and vertical strain (b) evolutions for the double-porosity Mandel problem whilst considering isostress (Reuss average) and other upscaling approaches. `$\text{K}^\text{R}_{\text{dr}}$' denotes coefficient models for which isostress is assumed ($K^R_{dr} = 3.3 \text{ GPa}$). Notation `$\text{K}^\text{HS-}_{\text{dr}}$' and `$\text{K}^\text{AHS}_{\text{dr}}$' are models calculated using the lower Hashin-Shtrikman bound and the arimthmetic mean of the Hashin-Shtrikman bounds for $K_{dr}$ respectively ($5.7 \text{ GPa}$ and $12.7 \text{ GPa}$ respectively).} \label{fig:9}
\end{figure}
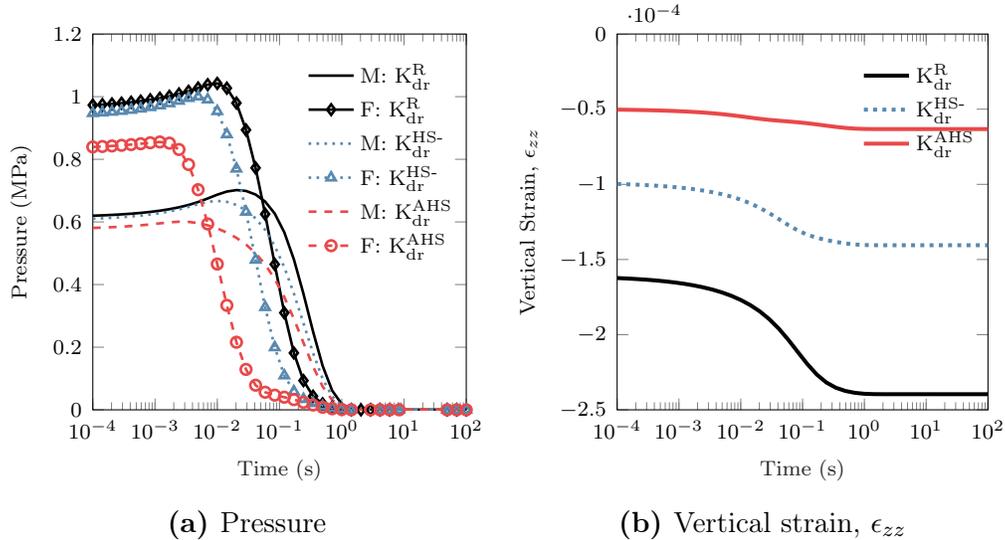

\autoref{fig:9a} shows further impacts of the upscaling method on matrix and fracture pressure evolutions. A stiffer composite bulk modulus leads to a lower induced fracture pressure and earlier onset of pressure diffusion in the same continuum. This is the case when using the arithmetic mean of the HS bounds. Non-monotonic rises in matrix and fracture pressures are more pronounced for more compliant composites. This is the case when assuming isotress (Reuss average) and the HS lower bound. Additionally, more compliant composites exhibit a faster decrease in matrix pressure at later times when compared to the stiffer modulus case.  

Non-monotonic pressure rises are often referred to as the Mandel-Cryer effect within literature (\citealt{Wang2000}; \citealt{Cheng2016}). We predict that such rises lead to greater pressure differentials at middle to late times between the matrix and fracture domains. This would explain the higher rate of matrix pressure diffusion in the compliant material cases. 

\cref{fig:9b} shows pronounced distinctions in vertical strain for the three different upscaling cases. As expected, the stiffer arithmetic mean of the HS bounds shows the lowest deformation. Of more interest, considering that they both correspond to lower bounds, are the difference in strains between the cases of isostress and HS lower bound. Vertical strain at late times is approximately 75\% larger when assuming isostress compared to when using the HS lower bound.       
 
The cause of the difference in vertical strains between the isostress and HS lower bound cases can be explained using similar discussions as to those used in Sections 6.2 and 6.3. First, using the heuristic argument the HS lower bound is higher than the Reuss bound suggesting that the matrix is capable of supporting a greater distribution of strain. This explains the difference in vertical strain at early times. Late time differences can be explained by the differences in fracture pressure. In contrast to Section 6.3, towards the lower bounds for $K_{dr}$, $b_f > b_m$. The magnitude of $b_f$ means that deformation is more strongly coupled to differences in fracture pressure relative to matrix pressure by way of momentum balance, \cref{eqn:70}.   

With a view toward multi-continuum generalisations, based on the results  in Section 4.2.3 and the qualitative results herein we recommend care before assuming isostress. This stress distribution has strong geometrical implications that, without experimental substantiation to prove otherwise, would seem unlikely to hold within a multi-continuum material. 

\section{Conclusion}
The goal of this paper was to formulate a set of recommendations for how and when to use different constitutive modelling concepts. In doing we revisited and reviewed three main modelling approaches, which differ in the parameters, and/or assumptions, used for the calculation of the effective constitutive coefficients. The parameters and assumptions used in these approaches are summarised as follows: 

\begin{enumerate}[(i)]
\item Constituent mechanical properties; assuming the high permeability, low storage continuum is all void space (no intrinsic fracture properties),
\item Constituent pore fractions; assuming the high permeability, low storage continuum is all void space,
\item Constituent mechanical properties, including intrinsic fracture properties.
\end{enumerate}

\noindent
Based on theoretical and qualitative findings in this paper we recommend further work on algebraic closure conditions for models built using continuum pore fractions. Comparing coefficient models (i) and (iii) we found that the effects of intrinsic fracture properties become measurable when there are significant deviations from the intrinsic poromechanical constitutive coefficients of a void space fracture continuum. In this case $\phi^*_f<1$ and $K_f\not\ll K^f_s$, and it is advisable to use coefficient models where intrinsic fracture properties are naturally incorporated. We envisage the aforementioned conditions, and thus real benefit of using models (iii), to be observed when considering non-linear poromechanics, where the internal structure of the high permeability, low storage continuum is evolving (e.g. fracture closure) thus leading to cases where $\phi^*_f<1$ and $K_f\not\ll K^f_s$. However, in the linear case, our results show models of type (i) to give very good matches to models of type (iii) even when $\phi^*_f<1$, provided $K_f\ll K^f_s$. Therefore as a first approach we recommend the use of models (i) for poromechanical dual-continuum modelling given that we expect the high permeability, low storage continuum to be mechanically weaker than solid grains.     

The second set of recommendations is formed on the basis of our investigations into implicit and explicit decoupling assumptions. In both cases mechanical coupling between continuum pressures is neglected. In the former we showed that implicit assumptions can lead to the removal of pressure sources leading to physically unreliable results. We therefore recommend the use of a full constitutive system where possible.  

Even with a full constitutive system, explicit assumptions have been made as a passage to simplifying relations between composite and constituent moduli without considering the physical implications of their use. In this case we showed that explicit decoupling assumptions are coincident with bounds on composite moduli that arise naturally under end-member states of isostrain and isostress. However, for isotropic composite materials it is well known that the bounds obtained under isotrain and isostress can be loose, and that tighter bounds using similar quantities are readily available within literature. Our qualitative investigations showed clear differences in poromechanical behaviour when using these different bounds. 

To conclude, bounds arising from isostrain and isostress states, which are concurrent with explicit decoupling assumptions, can provide a useful means for guiding our intuition into multiscale poromechanical behaviour, given their ease of computation. However, for practical subsurface applications, we recommend against the use of explicit decoupling assumptions, as they have physical and geometrical implications that are unlikely to be justified within isotropic multiscale materials.                     

\subsection*{Acknowledgements}
The authors are very grateful for the funding provided to them by the National Environmental Research Council to carry out this work, as well as to the reviewers who helped to improve this work.  

\newpage
\bibliographystyle{agsm}

\end{document}